# GEANT SIMULATIONS OF NEUTRON CAPTURE EXPERIMENTS

# WITH A 4π $BaF_2$ DETECTOR

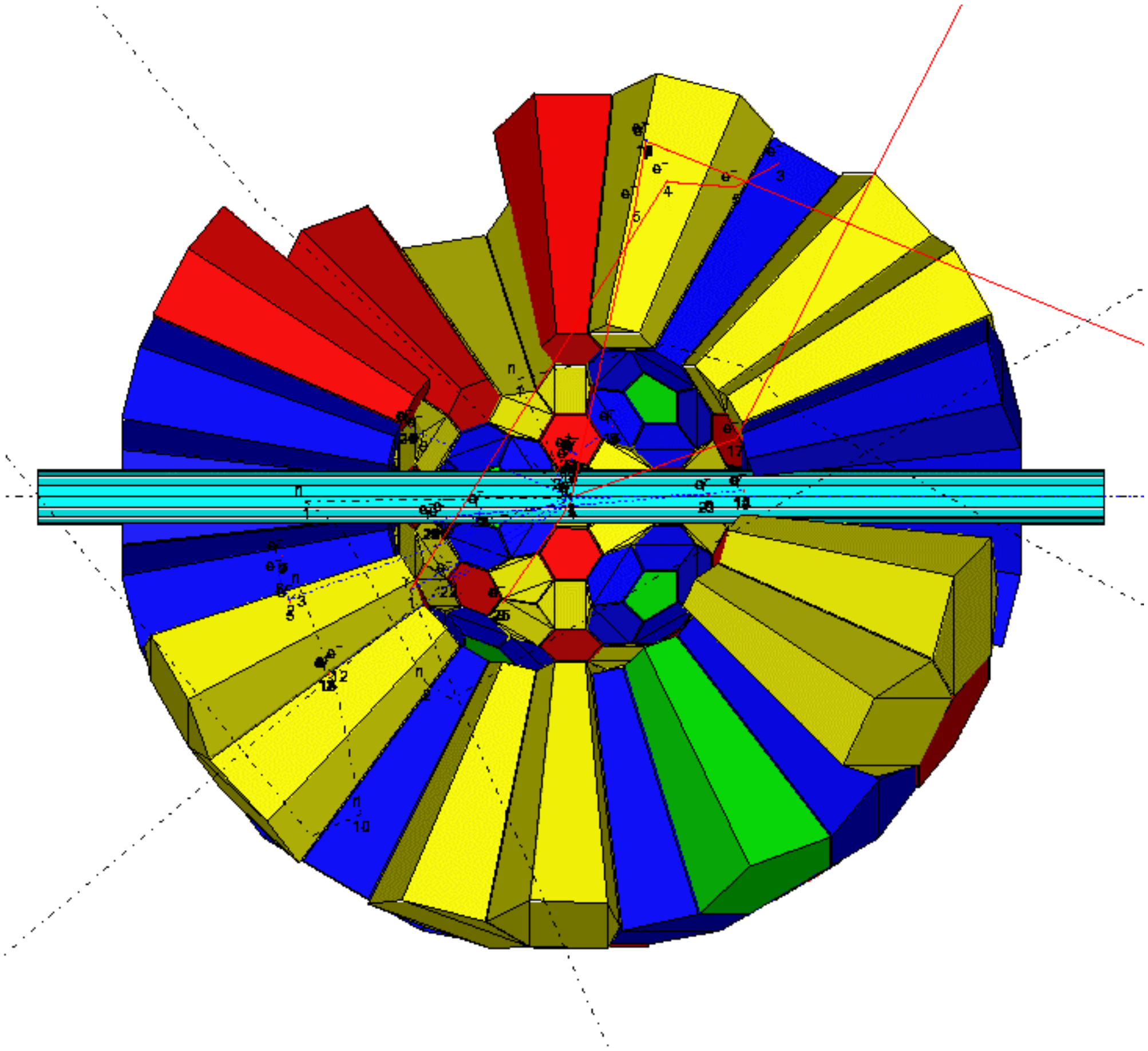

M. Heil, R. Reifarth, F. Kaeppeler, K. Wisshak, F. Voss
Forschungszentrum Karlsruhe

J. L. Ullmann, R. C. Haight, E. H. Seabury,
J. B. Wilhelmy, R. S. Rundberg, M. M. Fowler
Los Alamos National Laboratory

# Table of Contents

## List of Figures

## List of Tables

# 1. Introduction

The goal of this research project is to give quantitative information useful for the design of a γ-ray detector to investigate neutron capture (n, γ) reactions on radioactive nuclei at the Manuel Lujan Jr. Neutron Scattering Center (MLNSC) moderated neutron source at LANSCE. Data for neutron energies from thermal up to approximately 500 keV are desired. The radioactive nuclei can have half-lives as short as a few months. With the sample sizes foreseen, typically 1 mg, the radioactive decay rate can exceed tens of Curies (Ci). Many of the nuclei of interest emit copious quantities of energetic γ-rays which generally have significantly less energy ( < 3 MeV almost always) than the sum energy of gamma rays following neutron capture (~ 6 MeV), but the possibility exists that several gamma rays from unrelated radioactive decays could occur nearly simultaneously and thereby be difficult to separate from the neutron-capture γ-rays.

In addition to the capture γ-rays and radiation from the radioactive decay, the detector will also be subjected to neutrons scattered from the sample and, if the shielding is not perfect, also to background neutrons. The sensitivity of the detector to scattered neutrons must therefore also be assessed.

This design for the DANCE γ-ray detector array is intended to address the experimental challenges discussed above. So that the neutron capture rate is not overwhelmed by the radioactive decay rate, the intense neutron source at MLNSC is chosen. A very reasonable capture rate can then be achieved with a small sample (about 1 mg) of the radioisotope. A design that would accept even smaller samples is a continuing goal. To reduce the sensitivity to scattered neutrons, the materials of the detector should have small neutron capture cross sections. The detector is highly segmented so that the counting rate in a single element is acceptable. For this reason, we choose a 162 element, "soccer-ball" array, which subtends 4π steradians. Finally, we choose a very fast detector so that successive pulses, mostly from the radioactive decay, will be registered separately rather than piling up.

We focus on the well-studied scintillator, $BaF_2$. It has a fast component of 0.6 ns decay time and is the fastest known scintillator used in γ-ray spectroscopy. It has a relatively low neutron capture cross section (compared with NaI and CsI, for example); and it is available at reasonable cost in the size and shapes we require. Arrays of fewer elements have been developed and used for many years for neutron experiments at Karlsruhe and, more recently, at Oak Ridge National Laboratory. The response of some other scintillator materials including $C_6D_6$ and $C_6F_6$ is also calculated here.

To optimize the design of this detector array, information is needed on the following aspects of the performance:

1. The energy response to mono-energetic γ–rays.
2. The multiplicity of hits and the hit pattern for an incident mono-energetic γ–ray.
3. The energy response for typical γ -ray cascades following neutron capture.

4. The multiplicity of hits and the hit pattern for typical γ-ray cascades following neutron capture.
5. The neutron sensitivity of the detector: energy response, multiplicity of hits and hit pattern.

Design variables and considerations include:

1. Inner radius of the spherical detector shell. Clearance from the beam-tube.

2. Thickness of the spherical detector shell, i.e. inside and outside radii. Previous data suggest that a 15 cm thickness is desirable so that the fast light component of BaF2 is not absorbed before it reaches the photomultiplier tube (PMs). Would a thinner or thicker shell thickness be desirable?

3. Shaping the crystals so that they match well the diameter of available PMs. The hexagons and pentagons of the soccer-ball design could transition into circular cylinders that would match the diameter of the PMs. Some efficiency would be lost in this shaping, however. These calculations cover only the detection efficiency of the scintillator, but not the light transport to the PM tube.

4. Maximizing the capture γ-ray sensitivity relative to neutron sensitivity.

5. Assessing the effect of neutron moderating and capture materials ($^{6}Li$, $^{10}B$, $CH_2$, $CD_2$ etc.) placed in thin layers between the elements of the array or between the sample and the inner radius of the shell.

6. Assessing the effects of the structural support, materials of the PMs including the magnetic shields, and features of the experimental cave; analysis of whether a large number of “room-return” events is expected, and what mitigating design steps could be taken to reduce these effects.

7. Assessing the effects of various beam-tube designs.

8. Assessing the effects of collimators, their materials and location.

The calculations described below are intended to provide information on these and other aspects of the design.

# 2. Calculational Approach

With the help of GEANT, a detector description and simulation tool from CERN, the response of a $BaF_2$ detector consisting of 162 crystals was simulated. GEANT tracks photons, electrons and hadrons, which in this case are only neutrons. GEANT offers different interfaces for the tracking of hadrons (see page 312-320 of the GEANT manual). In this work the GHEISHA interface was used for tracking neutrons with an energy of $E_n$ > 200 keV and the GCALOR package was used for neutrons below 200 keV. The GCALOR interface uses MICAP for the tracking of neutrons below an energy of 20 MeV (see GCALOR manual chapter 3-1). The cut-off energy for all particles except for neutrons is 10 keV. If a particle falls below the threshold, its energy will be deposited in the present medium. For the MICAP interface the cut-off energy for neutrons was set to 10 meV. MICAP uses pointwise cross-section data (ENDF/B-V) for various isotopes. Because of the lack of photon data for neutron capture events in barium isotopes and in gold, theoretical γ-cascades calculated by Uhl [Uhl93a,Uhl93b] were used.

The calculations here are only of energy deposited in the scintillators. Light transport to the photocathode of the photomultiplier tubes (PMs) and the response of the PMs are not included here.

GEANT is a simulation tool invented for tracking high energy particles and for some interactions the errors of the calculated cross sections at lower energies can be as much as 20 % (e.g. photoelectric effect; see GEANT manual p. 215). To get an impression what one should expect from a simulation with GEANT, a comparison between simulated and experimental data for the setup in Karlsruhe [Wis90] is shown in Figure 1. Excellent agreement in pulse-height distribution is seen for all multiplicity of hits. We therefore have confidence in this approach of analyzing a similar detector.

For a simple geometry (see Figure 3) with a cylinder of $BaF_2$ a comparison between GEANT and MCNP was done. The cylinder with a diameter of 12.5 cm was 7.5 cm thick and the 1 MeV γ-rays were emitted on the symmetry axis at a distance of 10 cm to the cylinder. The 100,000 simulated γ-rays were randomly distributed into the hemisphere where the cylinder was located. The results of the simulations can be seen in Figure 2. For higher energies one finds rather good agreement, but for small energies (below 0.3 MeV) there are small deviations. The total efficiency (efficiency for any interaction with energy deposition) of the system is the same.

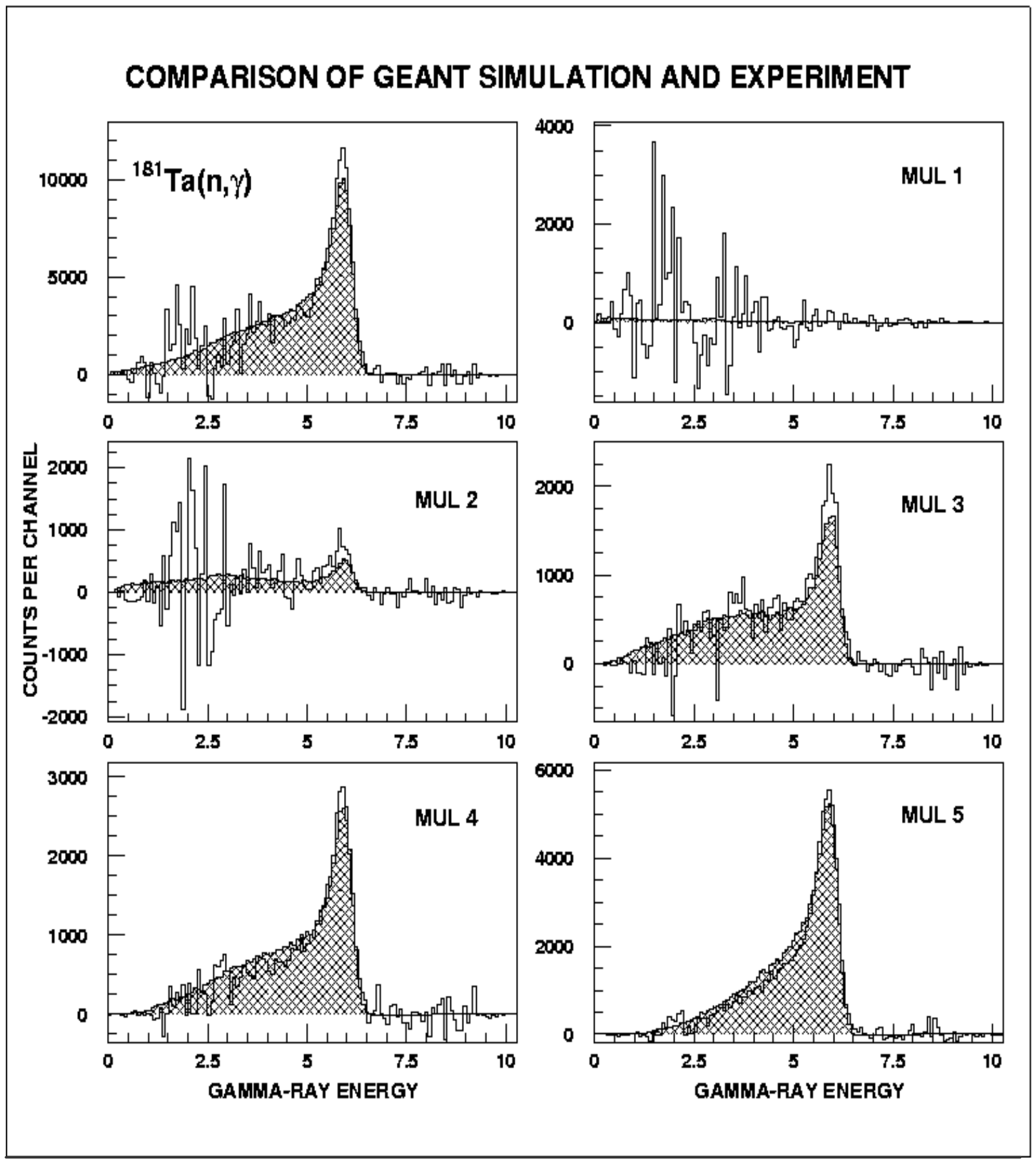

**Figure 1:** Comparison of GEANT simulation and experiment for capture events in $^{181}$Ta viewed by the Karlsruhe 41-detector $BaF_2$ array. The pulse height response is plotted for different multiplicities of detector hits. The top left frame shows the response for all multiplicities.

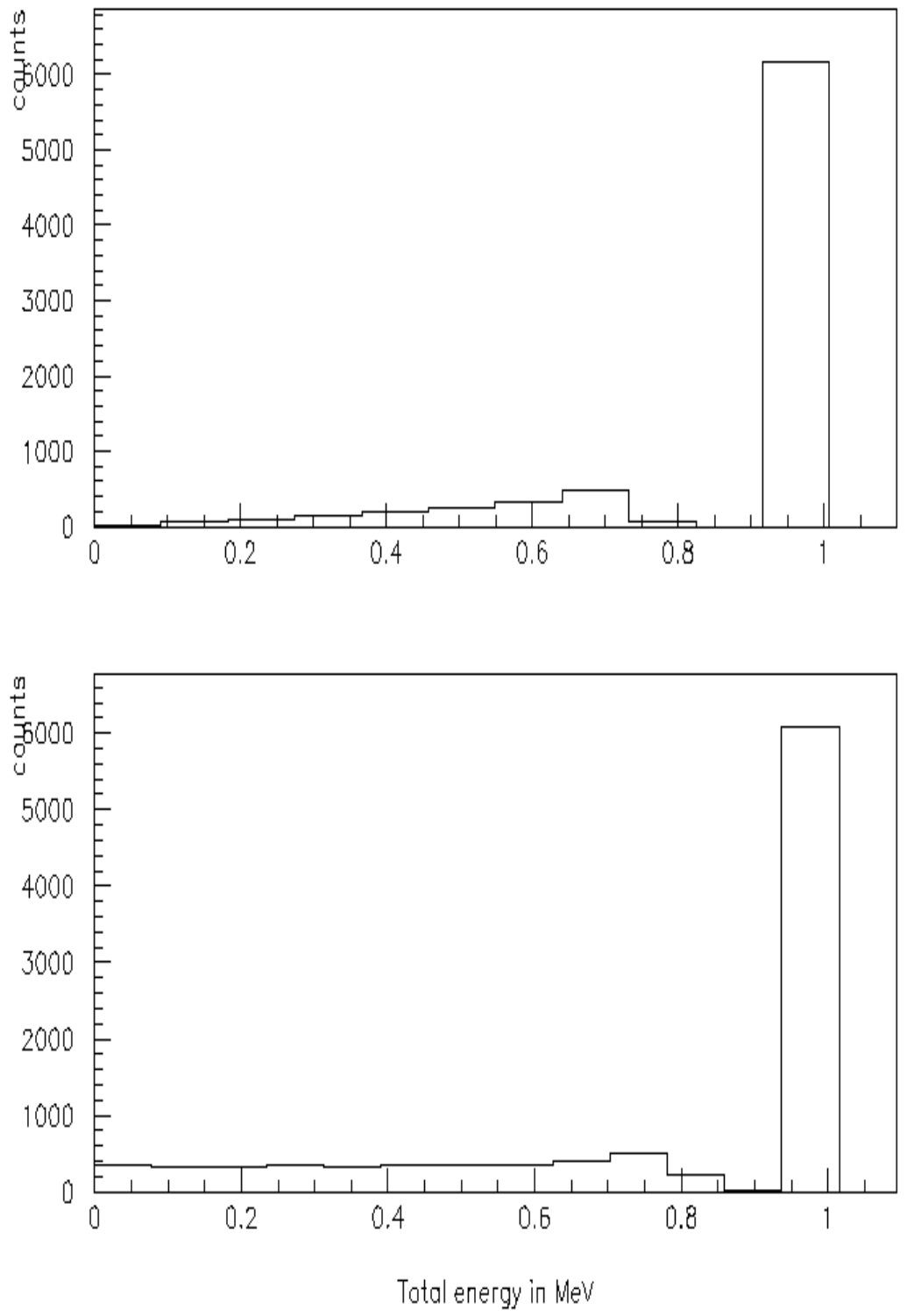

**Figure 3:** Comparison between a simulation with MCNP (top) and GEANT (bottom) for 1 MeV γ-rays and a cylinder of $BaF_2$.

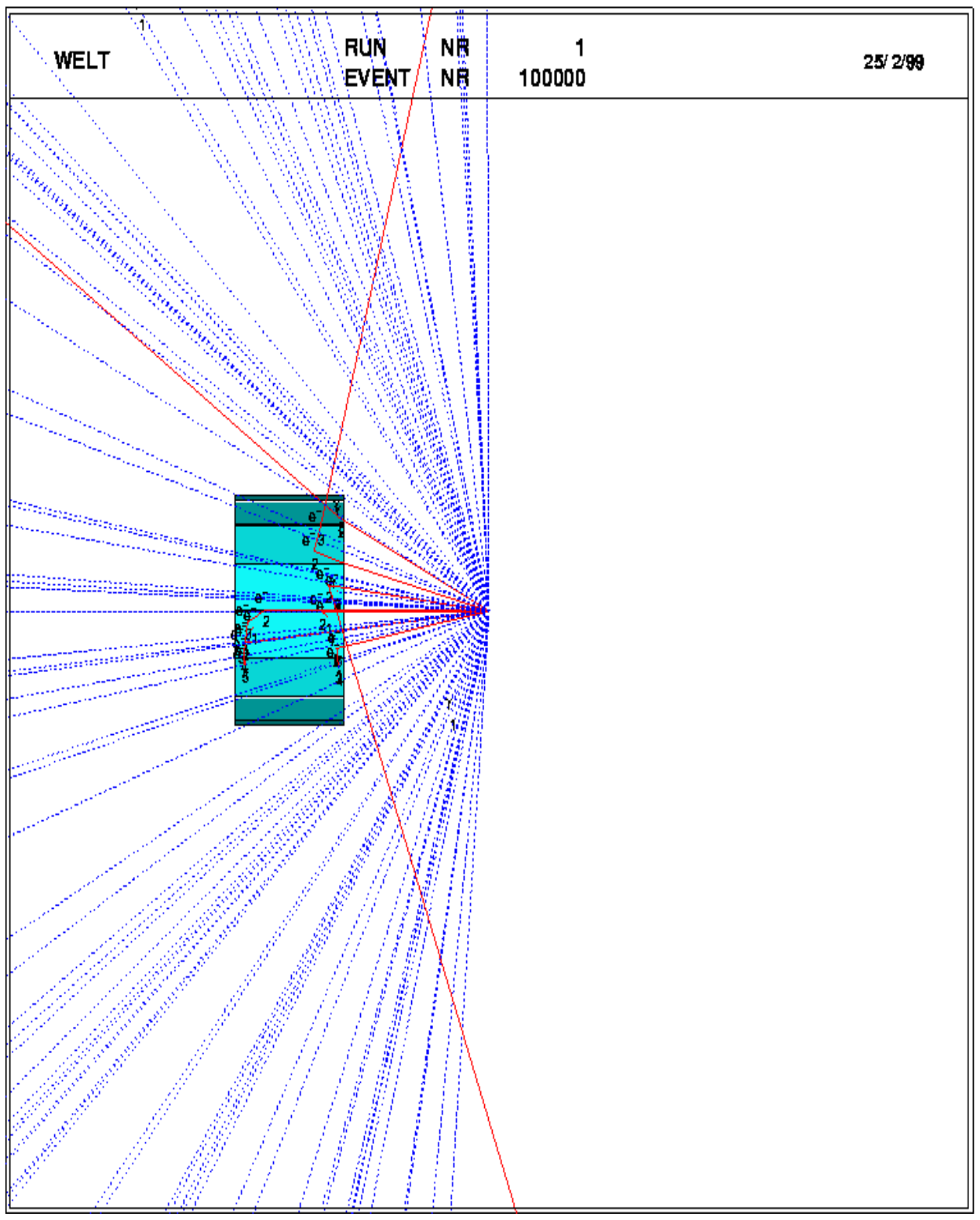

**Figure 2:** Geometry used for the comparison between GEANT and MCNP.

# 3. Geometry

The modeled detector consists of 162 $BaF_2$ crystals of 15 cm length. Covering the full solid angle without any gaps requires four different crystal shapes, which can be seen in Figure 4. The shapes of the crystals are optimized in a way that they all cover the same solid angle, although they have different shapes [Hab79].

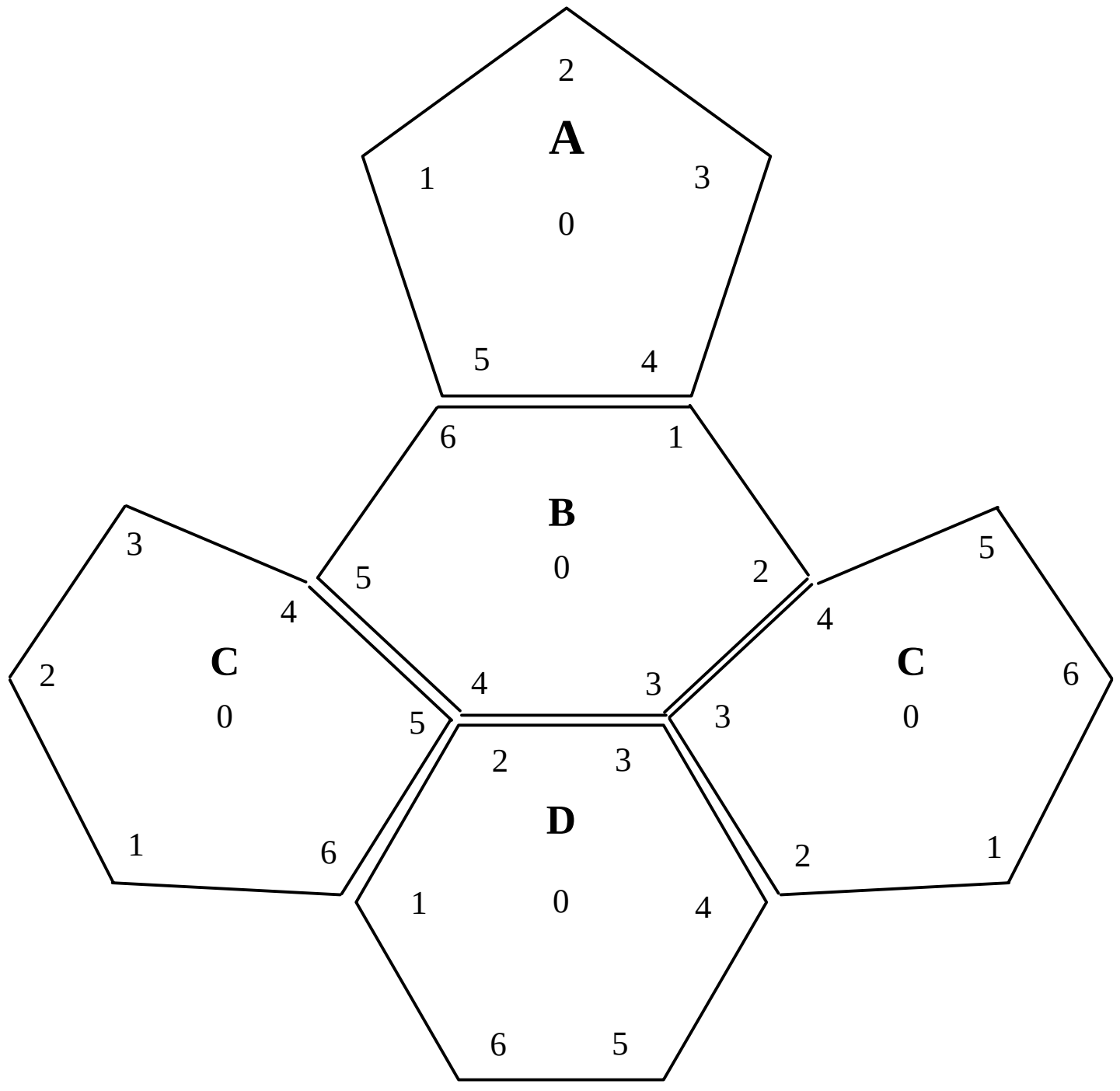

**Figure 4:** The detector is composed of 12 crystals of type A, 60 crystals of type B, 60 crystals of type C and 30 crystal of type D.

Table 1 shows the dimensions of the 4 crystal shapes. See Figure 4 for definitions of the line segments.

| Line Segment | Type | | | |
|---|---|---|---|---|
| | A | B | C | D |
| 12 | 0.21436 | 0.17624 | 0.19423 | 0.17442 |
| 23 | 0.21436 | 0.16630 | 0.17501 | 0.17442 |
| 34 | 0.21436 | 0.17358 | 0.16622 | 0.17442 |
| 45 | 0.21436 | 0.16630 | 0.16622 | 0.17442 |
| 56 | 0.21436 | 0.17624 | 0.17501 | 0.17442 |
| 61 | - | 0.21428 | 0.19423 | 0.17442 |
| 10 | 0.18234 | 0.18000 | 0.15055 | 0.17442 |
| 20 | 0.18234 | 0.20783 | 0.17596 | 0.17442 |
| 30 | 0.18234 | 0.14331 | 0.20455 | 0.17442 |
| 40 | 0.18234 | 0.14331 | 0.15055 | 0.17442 |
| 50 | 0.18234 | 0.20783 | 0.20455 | 0.17442 |
| 60 | - | 0.18000 | 0.17596 | 0.17442 |

**Table 1:** Dimensions of the different crystal shapes for a $4\pi$ array of unit radius.

The volumes described above can be arranged to form a closed sphere with an inner radius of 10 cm and an outer radius of 25 cm. In the present simulation, however, the crystals were moved outwards by 0.335 cm and the resulting gaps of 2 mm between the crystals were filled with a mixture of aluminum, Teflon and air to simulate materials used to reflect the scintillation light at the surfaces and to isolate each crystal from its neighbors. The mixture is equivalent to 0.1 mm aluminum foil, 0.5 mm Teflon and 0.4 mm air surrounding every crystal. The front of the crystal is also covered with aluminum and Teflon. Figure 5 shows the model of the detector with all crystals.

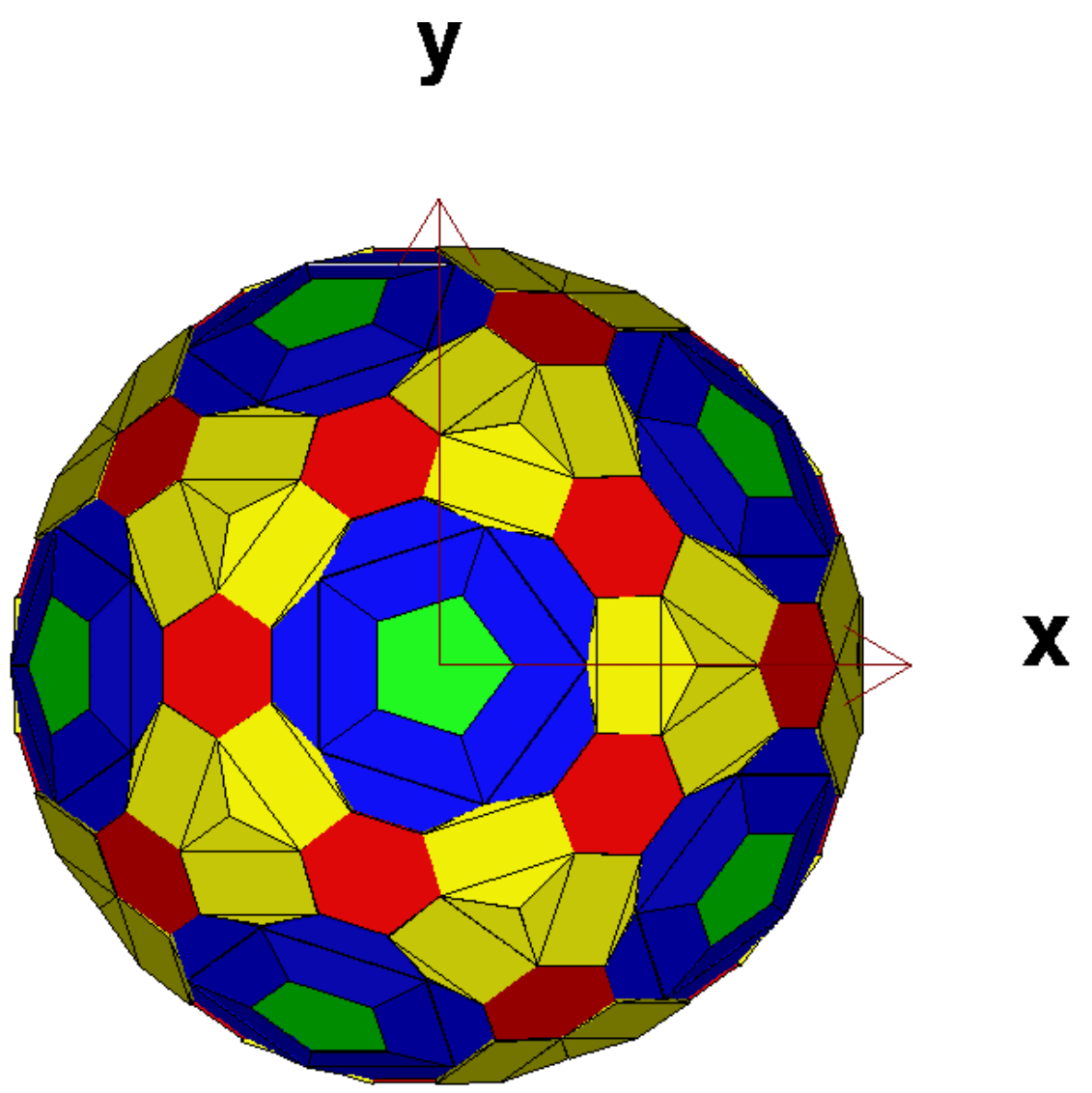

**Figure 5:** Model of the detector with all 162 $BaF_2$ crystals.

# 4. Scintillator material

Various scintillator materials can be used for γ-ray spectroscopy. In this section different scintillator materials are investigated with respect to their neutron sensitivity. Other properties, e.g. decay time of the scintillator light are not considered. It should be mentioned that for all scintillator materials we assumed the same resolution for a given energy deposited. The resolution does depend, however, on several factors including the photon statistics (see Table 2). For $BaF_2$, for example, one could like to use only the very fast component, which has about 10% of the total light output. The effect of decreasing the photon statistics is described in section 6.2.

| Material | Density ($g/cm^3$) | Decay Time (ns) | Wavelength (nm) | Photons / MeV |
|---|---|---|---|---|
| $BaF_2$ | 4.88 | 0.6 ; 630 | 180 ; 310 | 1,800 ; 10,000 |
| $Bi_4Ge_3O_{12}$ (BGO) | 7.13 | 60 ; 300 | 480 | 700 ; 7,500 |
| $CeF_3$ | 6.16 | 3 ; 27 | 300 ; 340 | 200 ; 4,300 |
| $C_6F_6$ | 1.61 | 3.3 | 430 | ≈ 10,000 |

**Table 2** Characteristics of some scintillator materials

The setup used for the calculation was an array with 160 crystals with an inner radius of 10 cm. The thickness of the crystals varies according to the attenuation length of the scintillator. A aluminum beam pipe of 3 mm thickness was also included. We assumed that the sample is 1 mm thick gold, located at the center of the array. This sample was irradiated with neutrons in the energy range from 100 eV up to 20 MeV. The energy spectrum of the neutrons was of the form $1/E_n$ (see section 6.1) and the flight path was 20 m long.

TOF spectra for events from neutron capture in the sample and for events from neutrons which were scattered on the sample and captured in the scintillator material were recorded separately. Various TOF cuts corresponding to neutron energy ranges of 100 eV – 1000 eV, 1 keV - 10 keV, 10 keV – 100 keV, 100 keV – 1000 keV etc. were applied and the ratio between events caused by scattered and captured neutrons was calculated. The neutron scattering-to-capture ratio for these different bins varies from about 1:1 to more than 50:1 (see section 9).

The following scintillator materials were investigated:

1. $BaF_2$: The thickness of the $BaF_2$ crystals was 15 cm. The natural composition of barium was used.

2. $BaF_2$: Again the thickness of the $BaF_2$ crystals was 15 cm but only $^{138}$Ba was used. (At the present time it is not practical to use enriched barium isotopes in the scintillator.)

3. BGO: The thickness of the BGO crystals was 10 cm. Since there are no photon data for neutron capture on germanium in the data base, 2 γ-rays with 25 % and 75 % of the full energy were assumed to be emitted in each capture.

4. $CeF_3$: The thickness of the $CeF_3$ crystals was 13 cm.

5. $C_6F_6$: The thickness of the $C_6F_6$ liquid was 60 cm. No “can” materials were included.

The results are given in Table 3.

| Setup | Ratio between scattered and captured events for different neutron energy regions. | | | |
|---|---|---|---|---|
| | 0.1 .. 1 keV | 1 ..10 keV | 10 .. 100 keV | 0.1 .. 1 MeV |
| 1 $BaF_2$, natural composition (15 cm) | 0.522±0.02 | 1.09±0.04 | 1.34±0.10 | 3.19±0.60 |
| 2 $BaF_2$, only $^{138}Ba$ (15 cm) | 0.011±0.002 | 0.026±0.006 | 0.07±0.02 | 2.53±0.69 |
| 3 BGO (10 cm) | 0.352±0.017 | 0.842±0.052 | 1.24±0.14 | 2.22±0.61 |
| 4 $CeF_3$ (13 cm) | 0.038±0.004 | 0.124±0.014 | 0.75±0.10 | 1.89±0.48 |
| 5 $C_6F_6$ (60 cm) | 0.0349±0.004 | 0.0181±0.0045 | 0.0364±0.014 | 1.28±0.37 |

**Table 3:** Ratio of events from scattered neutrons and true capture for different scintillator materials

The scintillator material $C_6H_6$ was also investigated, but, because of the small atomic number of the material, the efficiency for the $\gamma$-rays was too low to be considered further.

Figure 6 and Figure 7 show a comparison of the detector response for the capture-events in the sample for the scintillators discussed above. The thickness for each detector was chosen in a way that all materials have the same response to the gold cascades.

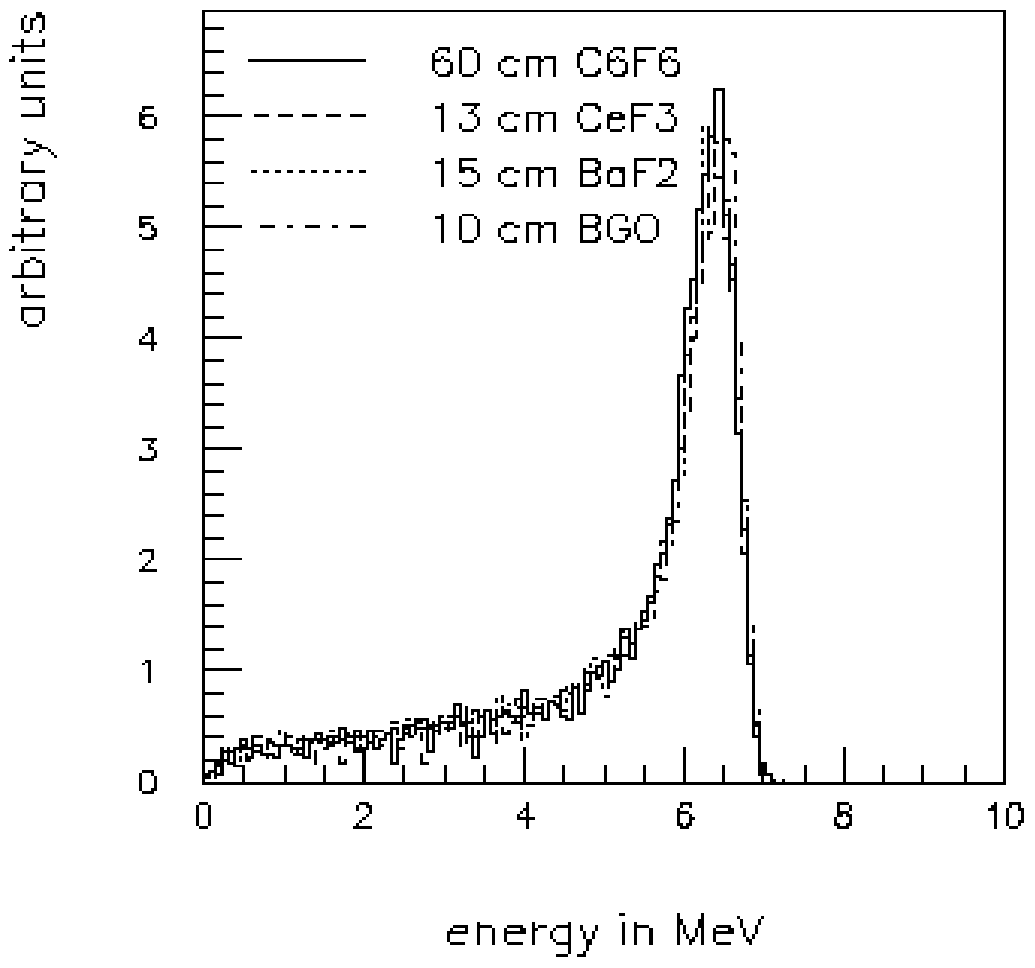

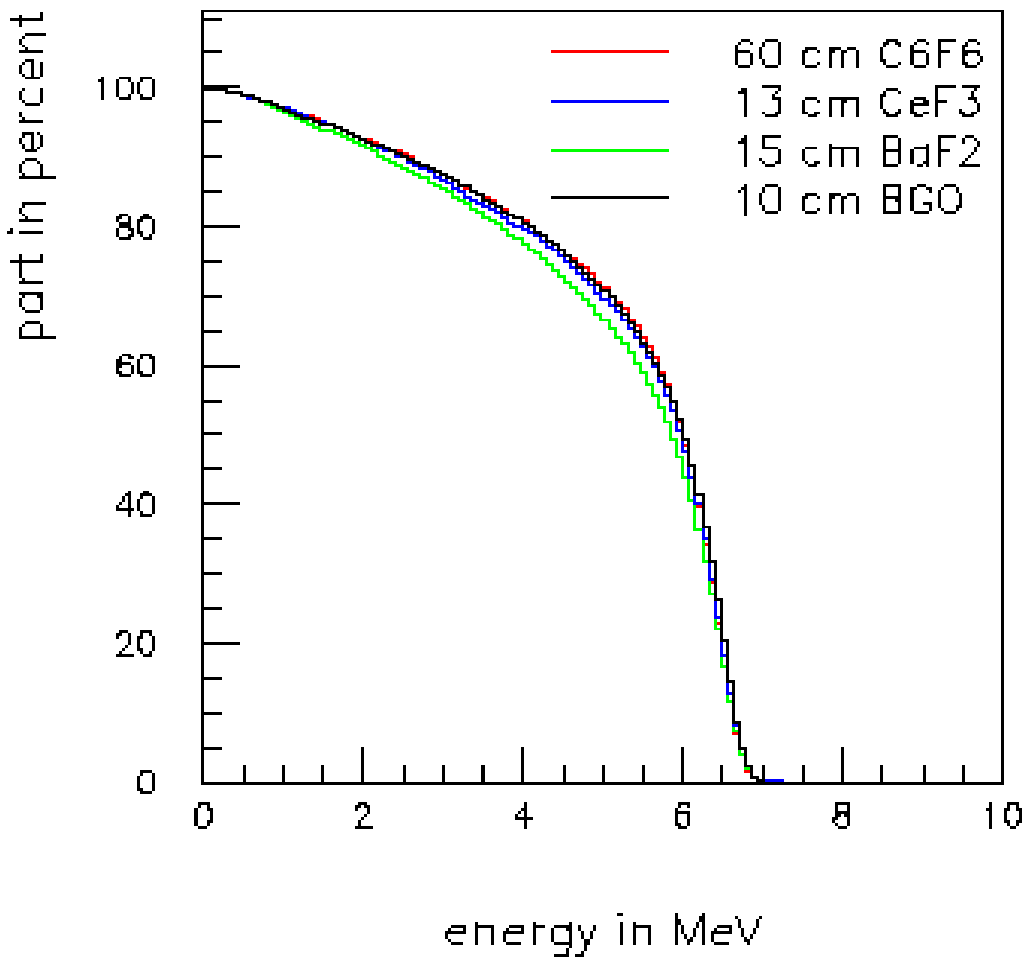

**Figure 6:** Pulse height distribution for various scintillator materials.

**Figure 7:** Percentage of integrated events above a given threshold energy for the pulse height distribution obtained with various scintillator materials.

Because of the rather good (low) neutron sensitivity of $C_6F_6$, possibly this material is an alternative to the solid materials (e.g. $BaF_2$). The following Table 4 as well as Figure 8 show the results for 30, 40 and 60 cm $C_6F_6$ scintillator length. For each run the inner radius of the detectors was 10 cm.

| Setup | Ratio between scattered and captured events for different neutron energy regions. | | | |
|---|---|---|---|---|
| | 0.1 .. 1 keV | 1 ..10 keV | 10 .. 100 keV | 0.1 .. 1 MeV |
| 15 cm $BaF_2$, natural composition | 0.52±0.02 | 1.09±0.04 | 1.3±0.1 | 3.2±0.6 |
| 60 cm $C_6F_6$ | 0.035±0.004 | 0.018±0.005 | 0.04±0.01 | 1.3±0.4 |
| 40 cm $C_6F_6$ | 0.024±0.003 | 0.016±0.004 | 0.03±0.01 | 1.8±0.6 |
| 30 cm $C_6F_6$ | 0.015±0.003 | 0.007±0.003 | 0.03±0.01 | 1.2±0.4 |

**Table 4.** $C_6F_6$ detector responses to scattered neutrons compared with capture on the sample relative to $BaF_2$.

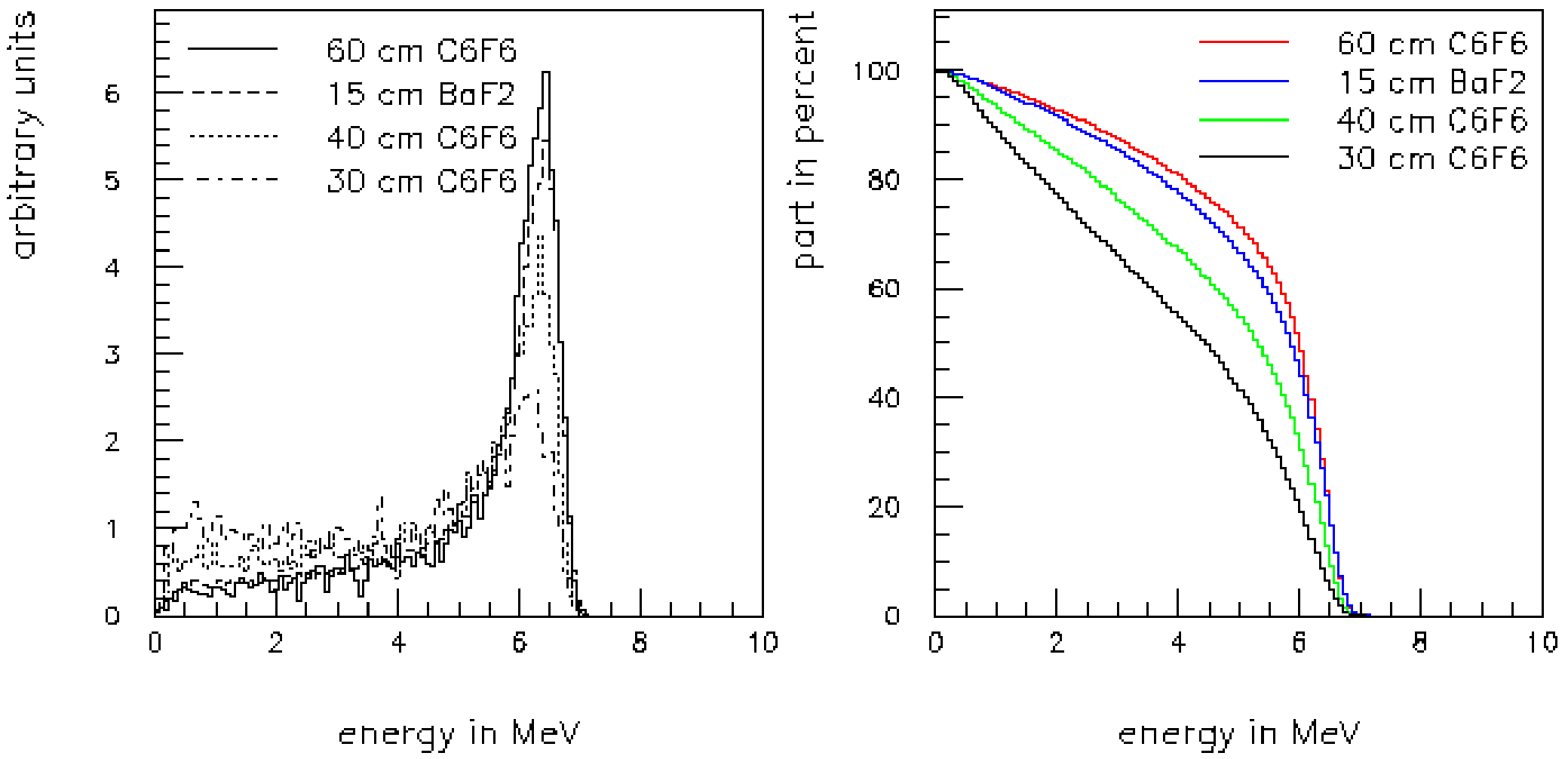

**Figure 8:** Comparison of the energy deposited and the energy deposited above a given energy for different sizes of $C_6F_6$ detectors and 15 cm $BaF_2$ crystals.

# 5. Detector response for γ-rays

## 5.1 Mono-energetic γ's

With the geometry described in section 3 the detector response for randomly distributed γ-rays with energies from 1 MeV up to 9 MeV emitted in the center of the $BaF_2$ crystal ball was investigated. Figure 9 and Figure 10 show the resulting multiplicity distributions and the sum energy spectra for 20,000 initial γ-rays of each energy. For almost all γ-energies multiplicity 2 is the most abundant. This means that having only the multiplicity of an event one cannot unambiguously deduce the energy of the initial γ-ray.

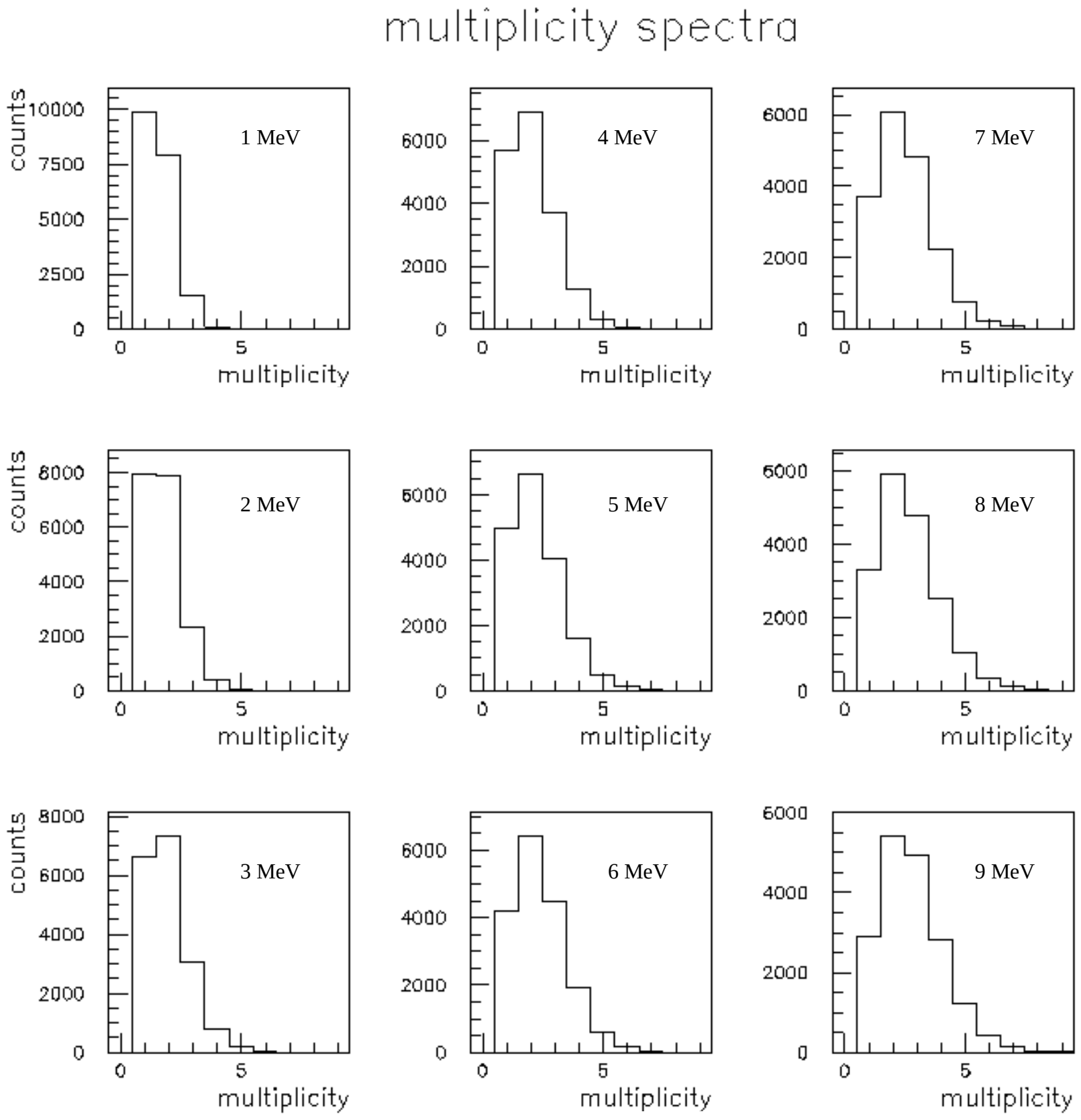

**Figure 9:** Multiplicity distribution for γ-energies from 1 MeV–9 MeV in a $BaF_2$ crystal ball.

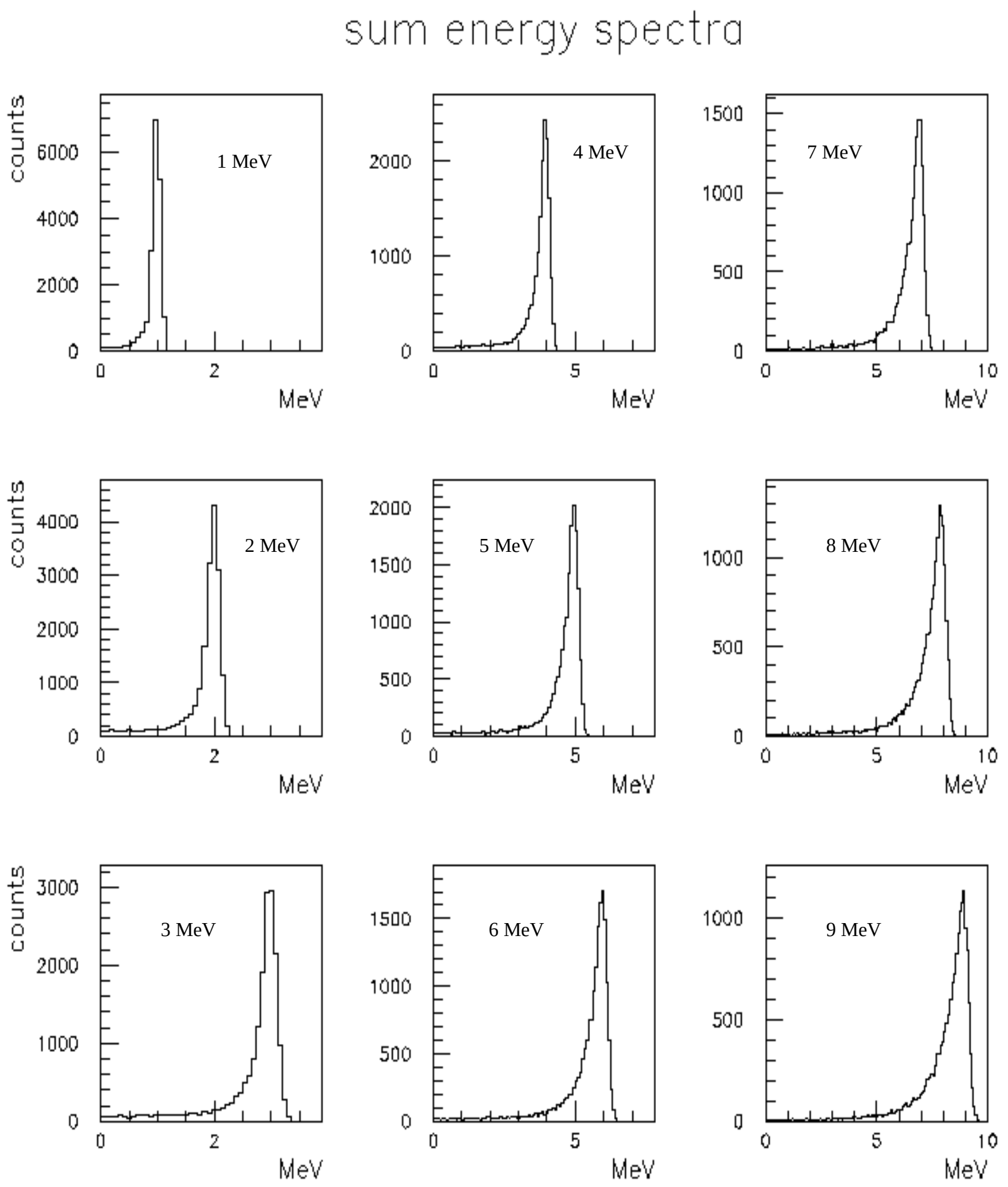

**Figure 10:** Energy deposit summed over all crystals for γ-energies from 1 MeV - 9 MeV. For each calculation, 20,000 initial photons were tracked.

The total efficiency and the average multiplicity as a function of γ-energy are given in Table 5. No energy cut-off was applied for the calculation of the total efficiency.

| γ-energy in MeV | total efficiency in % | averaged multiplicity |
|---|---|---|
| 1.0 | 97 | 1.58 |
| 2.0 | 93 | 1.75 |
| 3.0 | 91 | 1.94 |
| 4.0 | 90 | 2.11 |
| 5.0 | 90 | 2.25 |
| 6.0 | 90 | 2.39 |
| 7.0 | 90 | 2.52 |
| 8.0 | 90 | 2.65 |
| 9.0 | 90 | 2.79 |

**Table 5:** Total γ-ray efficiency as a function of $E_\gamma$.

A linear function fits the dependence of average multiplicity from energy in the energy region up to 10 MeV rather well. Let the linear function be f(x) = mx + n; one finds:

$$m = 0.154\ MeV^{-1} \text{ and } n = 1.44 .$$

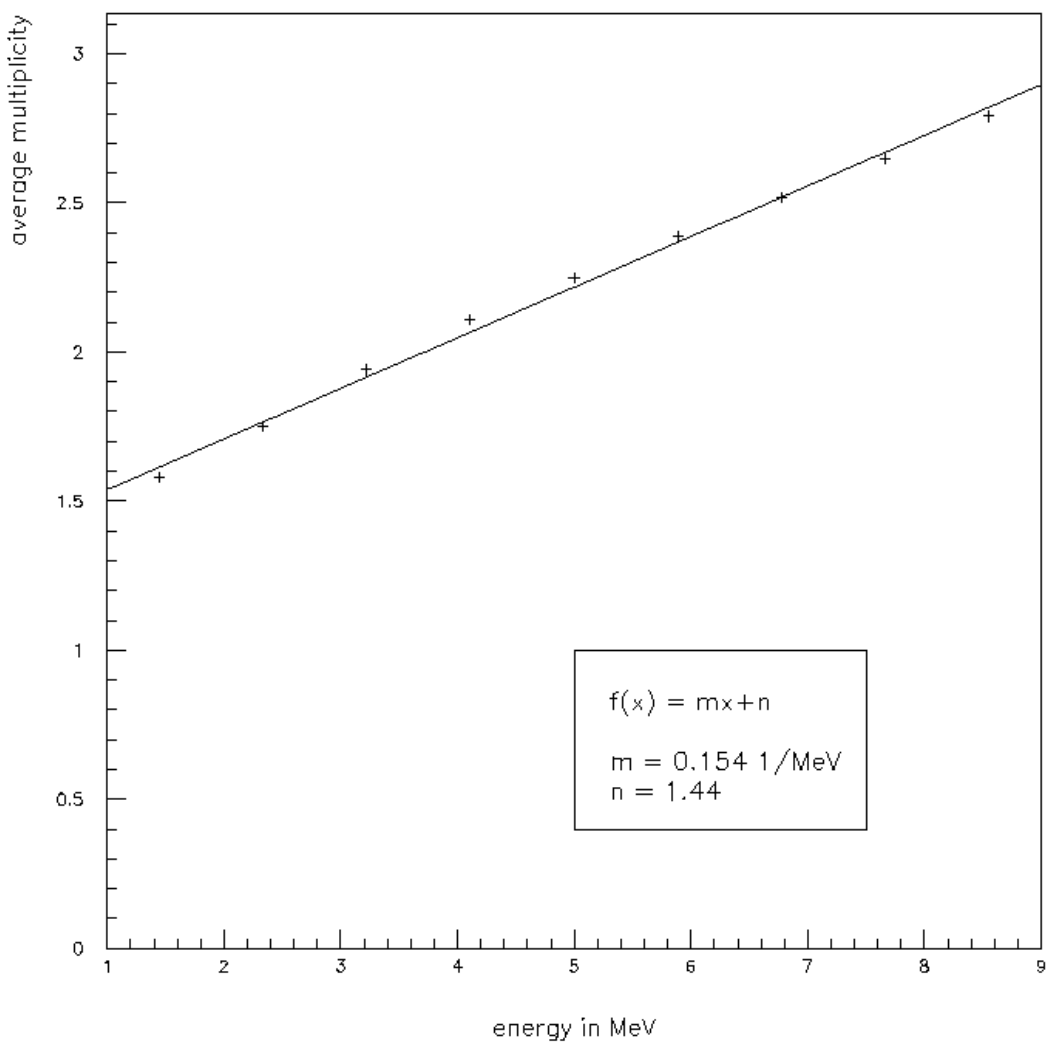

**Figure 11:** Average multiplicity versus primary γ-ray energy.

A simulation with all 162 crystals without any gaps between the crystals for 6 MeV γ-rays gave a total efficiency of 92 % compared with 90 % for the array with gaps. This means that the gaps between the crystals only account for 2 % decrease in the total efficiency. The “no-gap” value of 92 % can be understood easily by noting that the attenuation length for γ-rays in the MeV region in $BaF_2$ is approximately 5 cm. Three e-foldings in 15 cm of $BaF_2$ would yield 95 % efficiency for the detection of a γ-ray. Requiring the full energy of the γ-ray to be deposited in the array lowers the efficiency significantly, however.

In addition, hitting patterns were recorded and Figure 12 shows some illustrative examples of events with different multiplicities for γ-rays with 6 MeV energy. The examples chosen are for multiplicities significantly greater than the average multiplicity. They do show however that for certain events, the detected hit pattern can range over widely separated detection units. To quantify the closeness of the hit pattern,we consider a coherent group of hit modules to form a cluster. For example, the multiplicity 10 event in the upper left corner of Figure 12 consists of 4 clusters. We investigated (section 6.5) whether the concept of clusters could be used to help separate capture events on the sample from capture of scattered neutrons in the scintillator material.

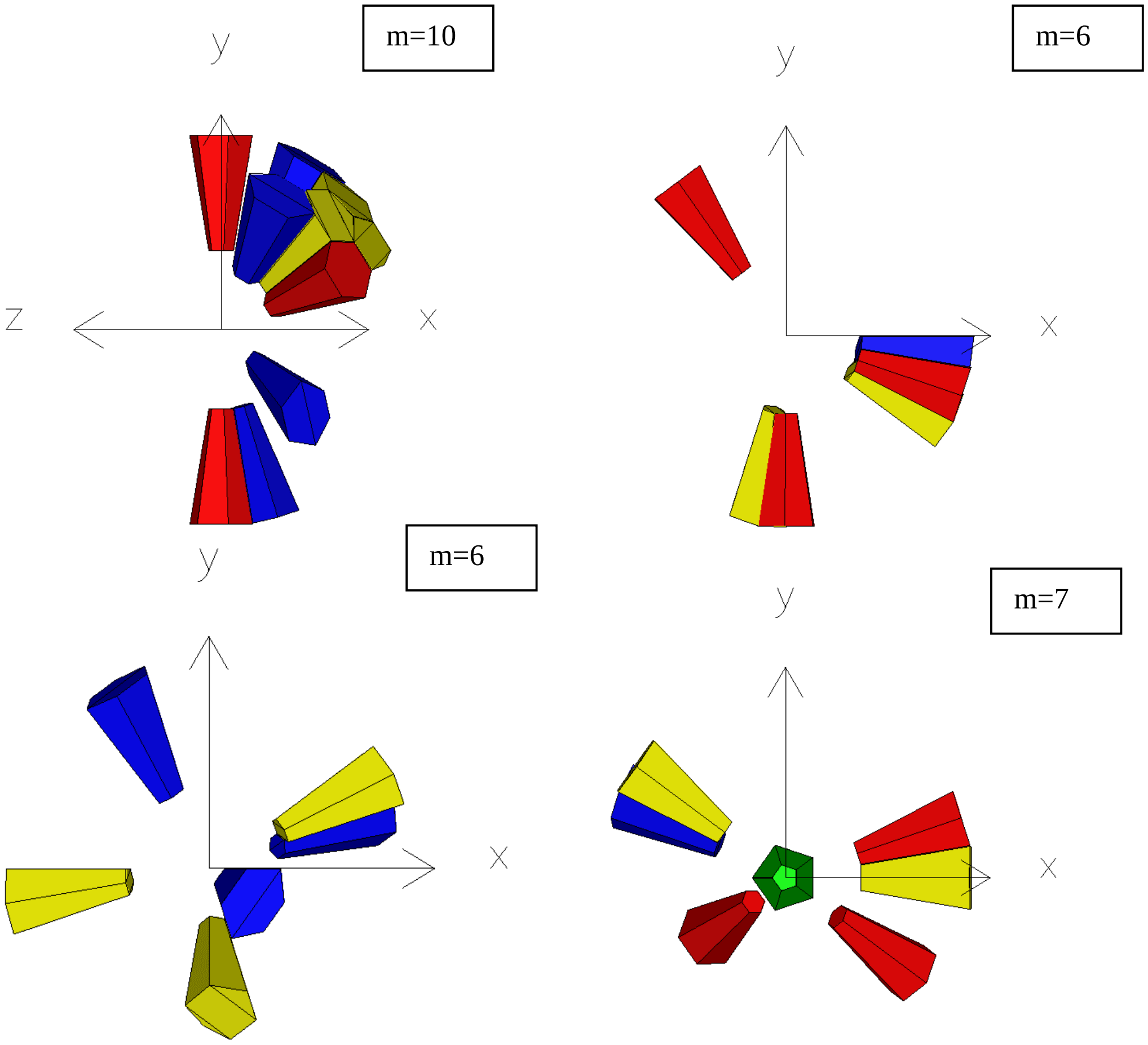

**Figure 12:** Representative hit patterns for events with different multiplicities ($E_\gamma$ = 6 MeV).

## 5.2 γ-cascades from neutron capture in $^{197}$Au

We also simulated the response of the detector for the $^{197}$Au(n,γ)$^{198}$Au reaction (Q = 6.512 MeV). For $^{198}$Au an isomer exists with a lifetime of 124 ns at 0.312 MeV. For neutron capture events populating this isomer the prompt energy is reduced to 6.2 MeV. The decay into the ground state as well as the decay of the ground state itself are not taken into consideration. For the simulation theoretical γ-cascades for capture of 100 keV neutrons calculated by Uhl [Uhl93a,Uhl93b] were used, which have multiplicities of at most 7. Again 20,000 events were simulated. The average multiplicity of detectors firing for the full crystal ball was 5.38. In order to see the effects of an incomplete crystal ball the response of the following geometrical setups was investigated:

- 151 crystals, where 1 pentagon (type A) and the 10 nearest irregular hexagons (5 of type B and 5 of type C) were left out.
- 140 crystals, where two opposite pentagons (type A) and the 20 nearest irregular hexagons (10 of type B and 10 of type C) were left out.
- 81 crystals, where all crystals of one hemisphere were left out.

Figure 13 shows a comparison of the resulting spectra. To get a better picture of the degradation of the peak, Figure 14 shows the percentage of events integrated above a given energy.

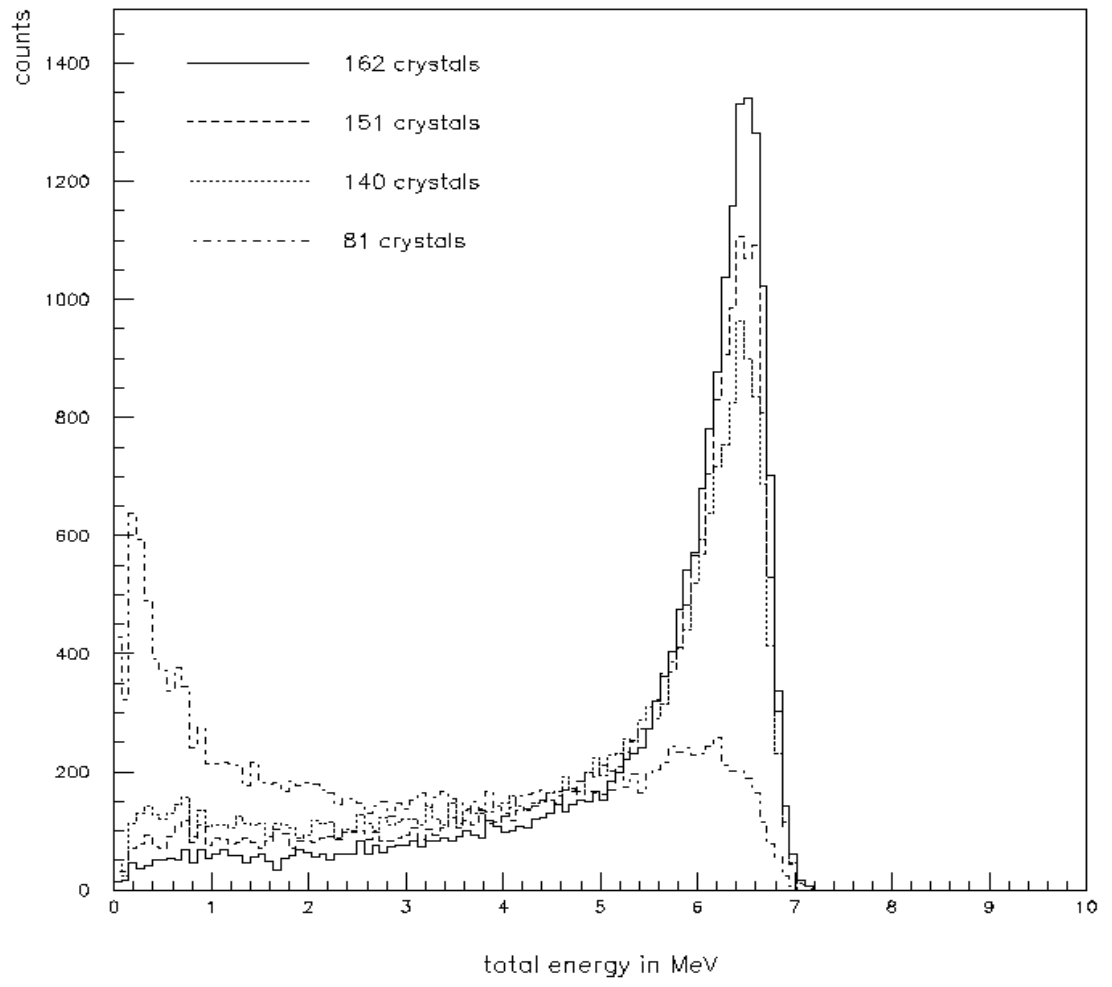

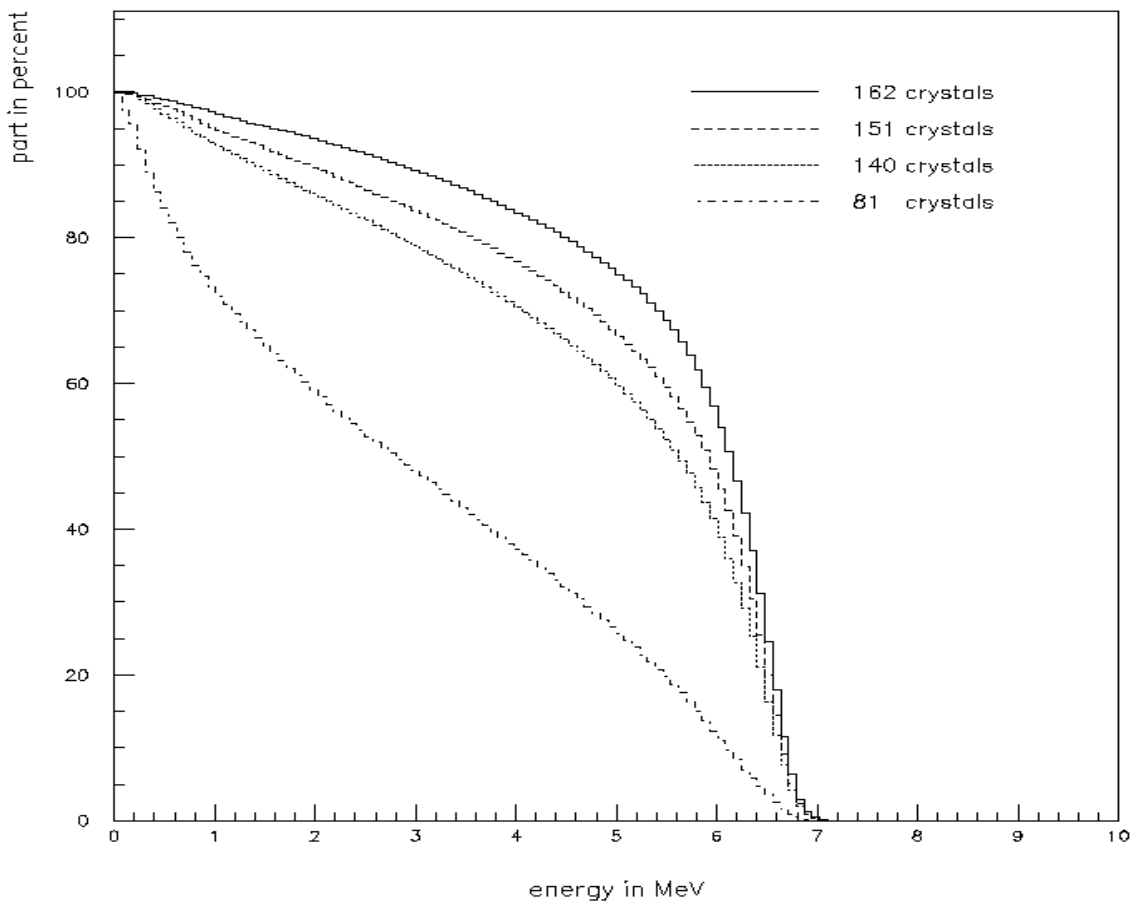

**Figure 13:** Sum energy spectra for the $^{197}$Au(n,γ) reaction using different numbers of crystals.

**Figure 14:** Percentage of counts above a given threshold energy for different numbers of crystals.

To get more realistic simulations 2 opposite crystals were left out for an aluminum beam pipe with 3 cm outer diameter and different wall thickness (Figure 15 and Figure 16). The beam pipe consists of aluminum and the effect of beam pipe thickness was also investigated. These results show that a significant degradation of the spectra occurs for aluminum beam pipes of more than 3 mm thickness.

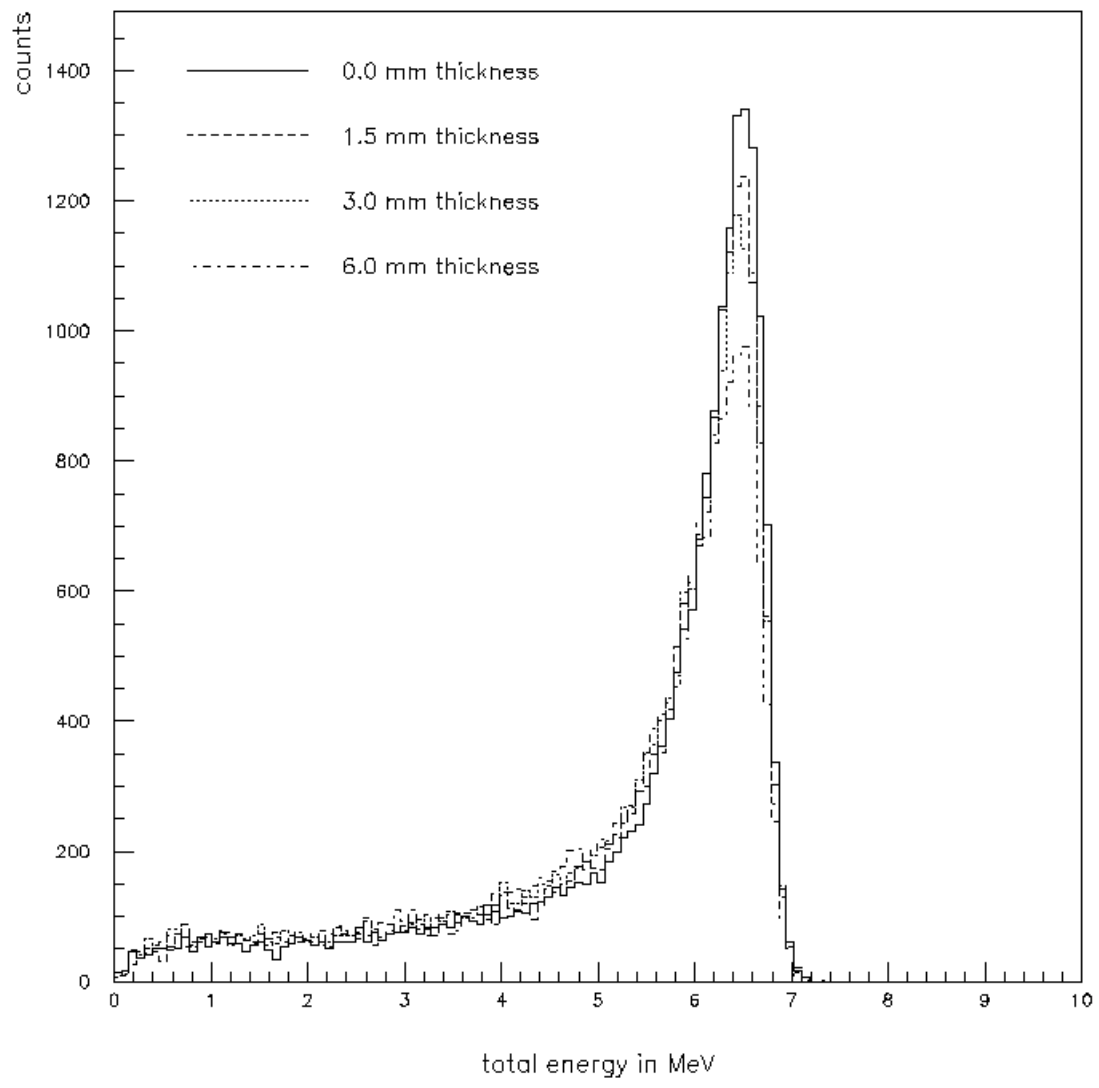

**Figure 15:** Effect of different beam pipe thickness on the energy deposited in the scintillators.

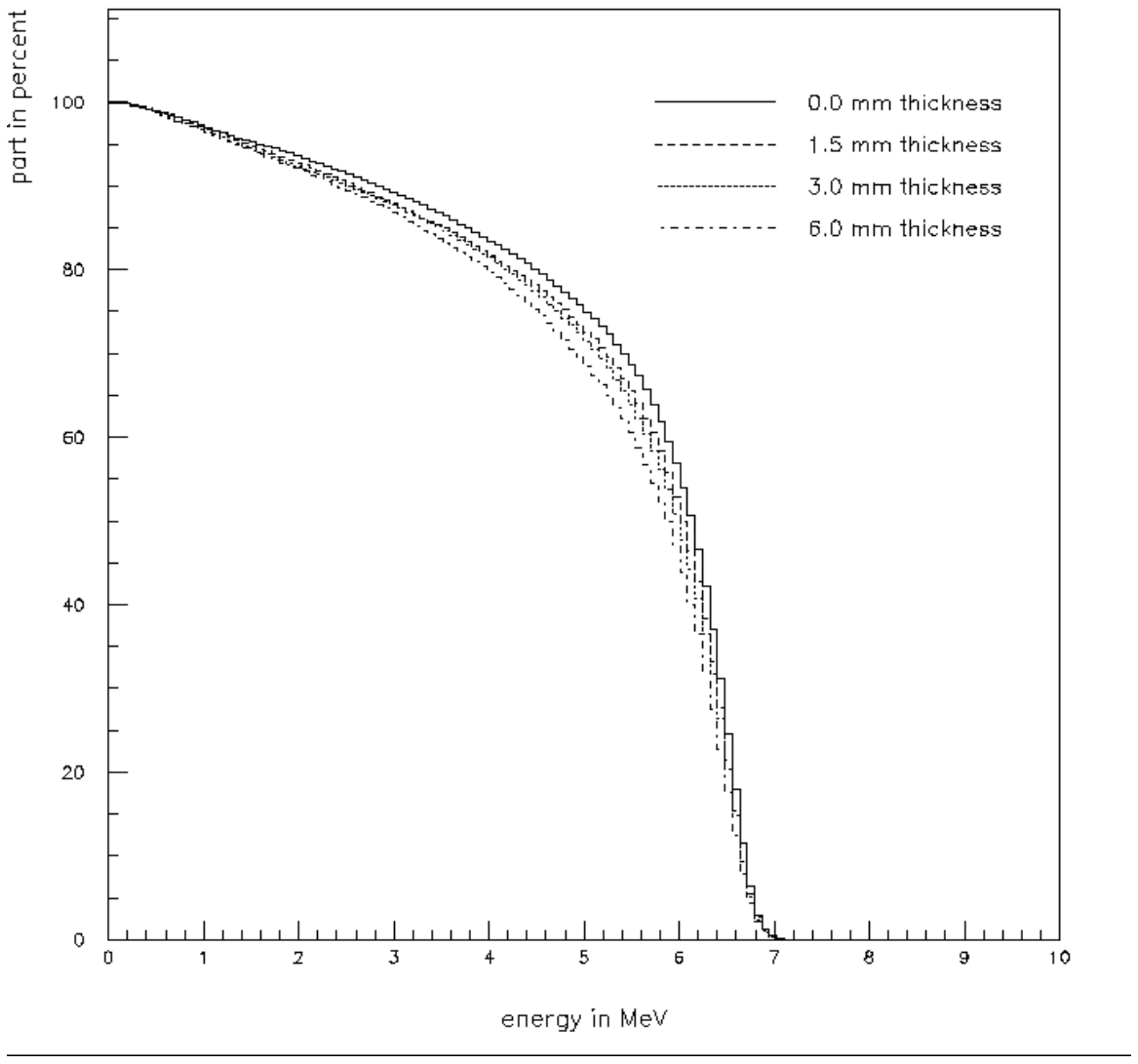

**Figure 16:** Percentage of counts above a given threshold energy for different beam pipe thickness.

## 5.3 Crystal thickness

The effect of different crystal thickness was also investigated. For this simulation we used the geometry with all 162 crystals and with the gaps filled with aluminum and Teflon. Simulations for $BaF_2$ crystals with a thickness of 12 cm, 15 cm and 18 cm were done for 6 MeV mono-energetic γ-rays and for Au cascades. No beam pipe was included in these calculations. The total efficiency for 6 MeV γ-rays was 84 % for a thickness of 12 cm and 93 % for a thickness of 18 cm. Of perhaps greater importance is the peak shape and the low energy tail. A significant degradation is calculated for 12 cm thick crystals. Figures 17 and 18 show the results for the Au cascades. These results need to be compared with the simulations for neutron capture in $BaF_2$ because a gain in the peak efficiency with thicker crystals needs to be compared with the increased background due to capture of scattered neutrons.

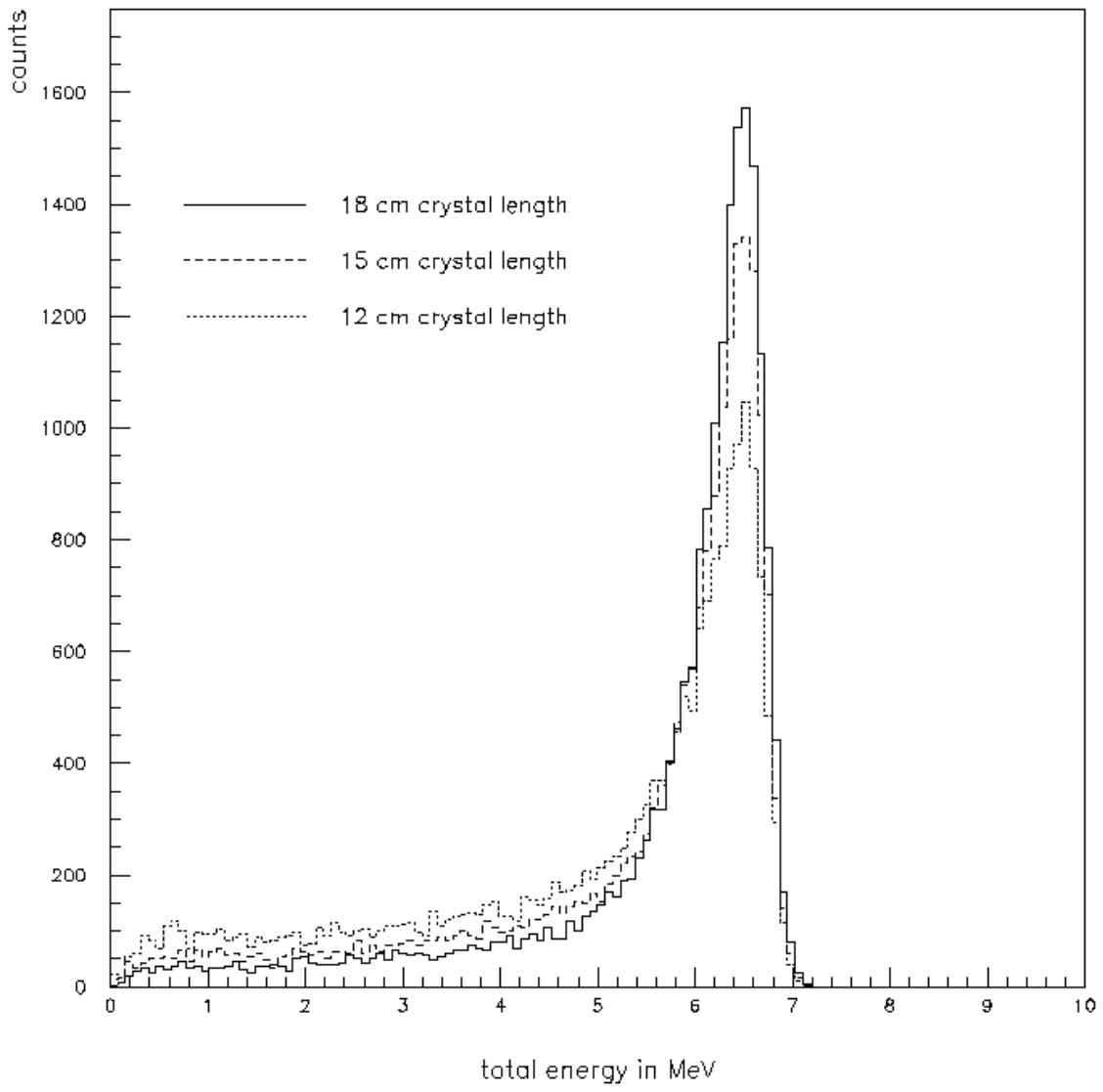

**Figure 18:** Sum energy spectra for Au cascades and different crystal thickness.

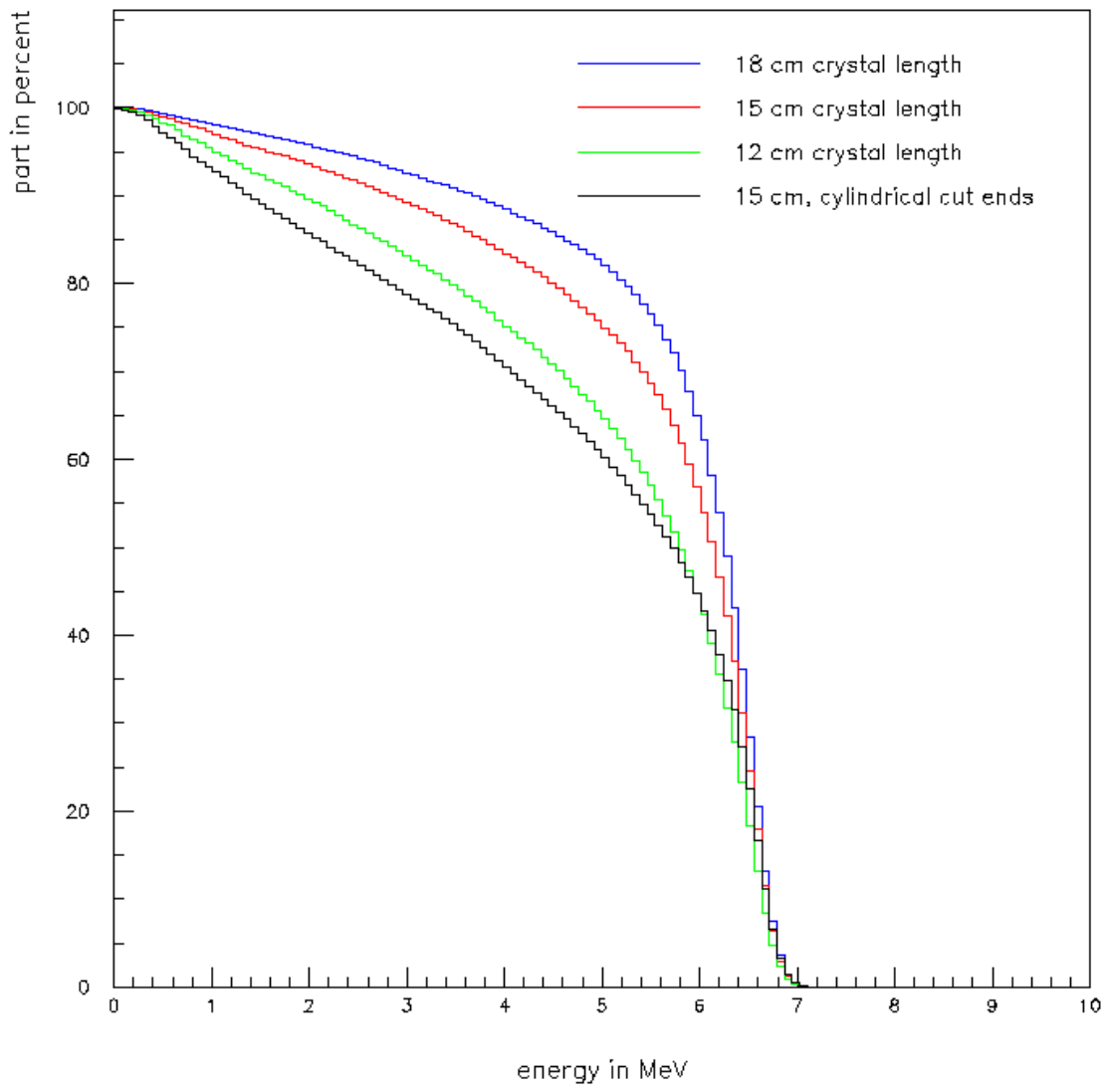

**Figure 17:** Percentage of counts above a given threshold energy for Au cascades and different crystal thickness.

In this context we also carried out a simulation with crystals cylindrically cut at the end. The cylindrical part is 5.08 cm in diameter, which would match a 5.08 cm photomultiplier tube. The sum spectra for the Au cascades are given in Figure 19 and a simulation with 6 MeV γ-rays showed that the total efficiency dropped from 92 % to 80 %.

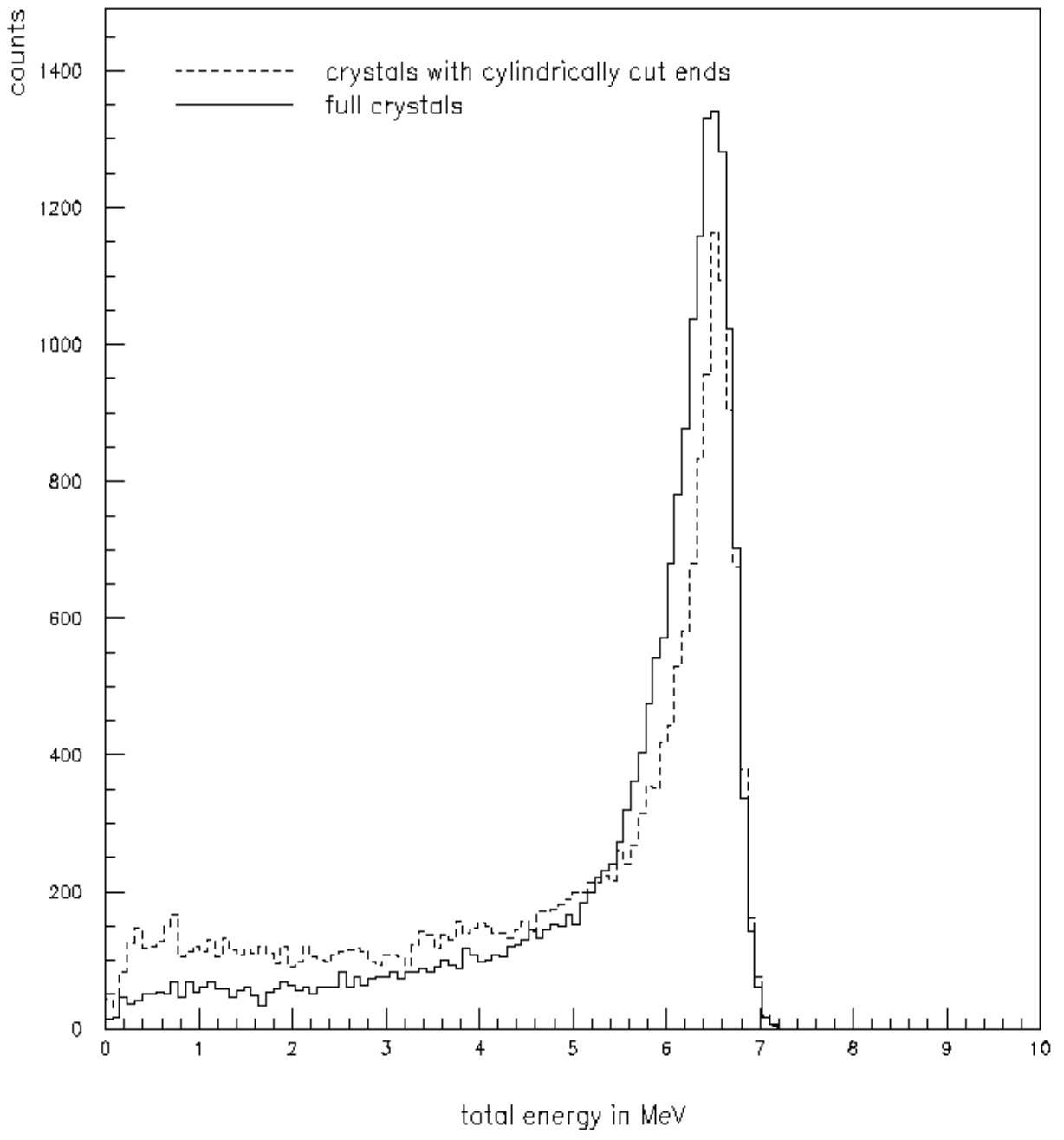

**Figure 19:** Sum spectra with full and cylindrically cut crystals.

## 5.4 Radiation from $^{60}Co$

Single elements of this array will be tested with radioactive sources. It was therefore thought useful to have on hand a calculated spectrum, which could be compared with the test data.

At present only a single $BaF_2$ crystal of form D (see section 2 of this report) is available for testing. Because of this, the response of a single crystal of type D (Figure 20) wrapped in 0.1 mm aluminum and 0.5 mm Teflon was simulated. The $^{60}Co$ source was positioned at a distance of 5 cm from the front surface of the crystal. The same calculation was carried out for a single crystal of type A (Figure 21) also. Since the different shapes are designed to cover the same solid angle no major changes are expected.

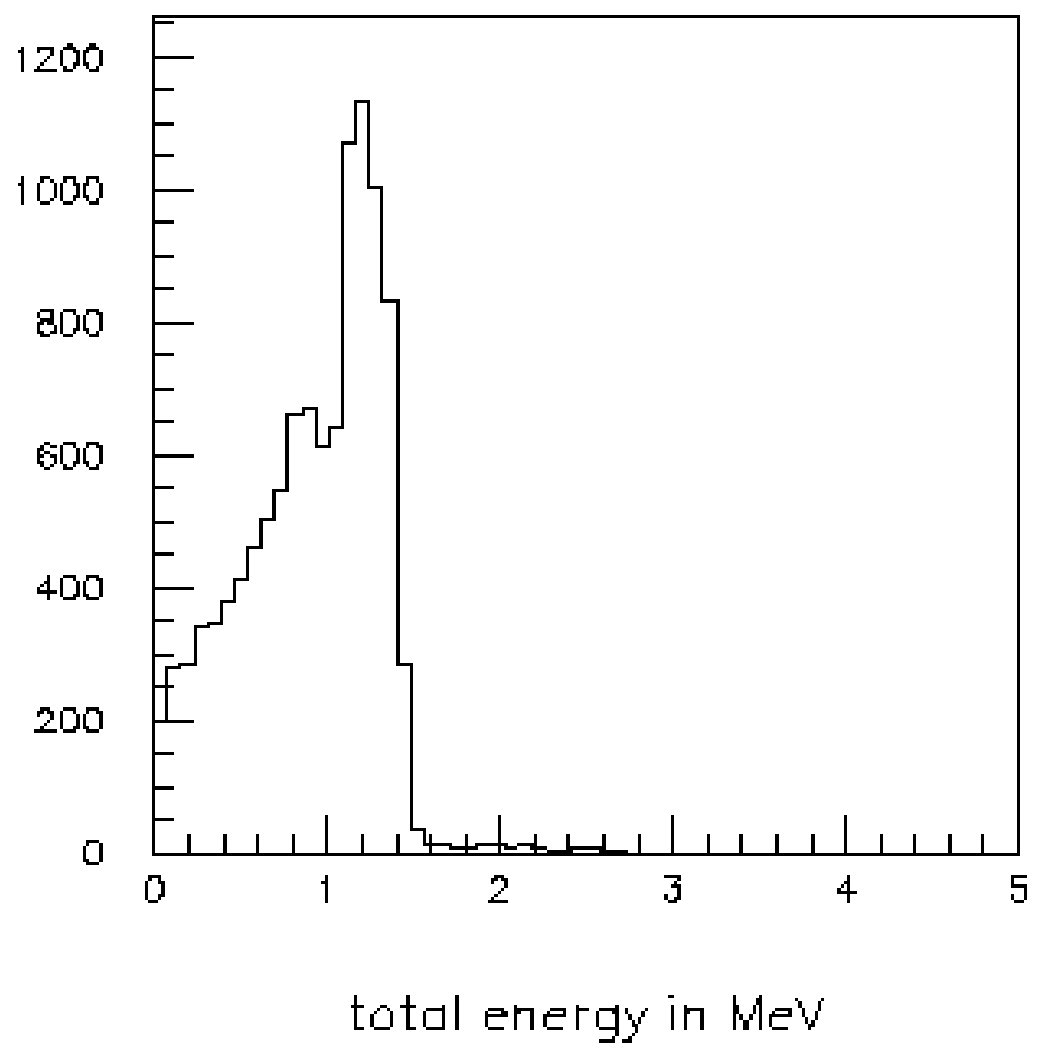

**Figure 20:** $^{60}Co$ spectrum with a single crystal of type D.

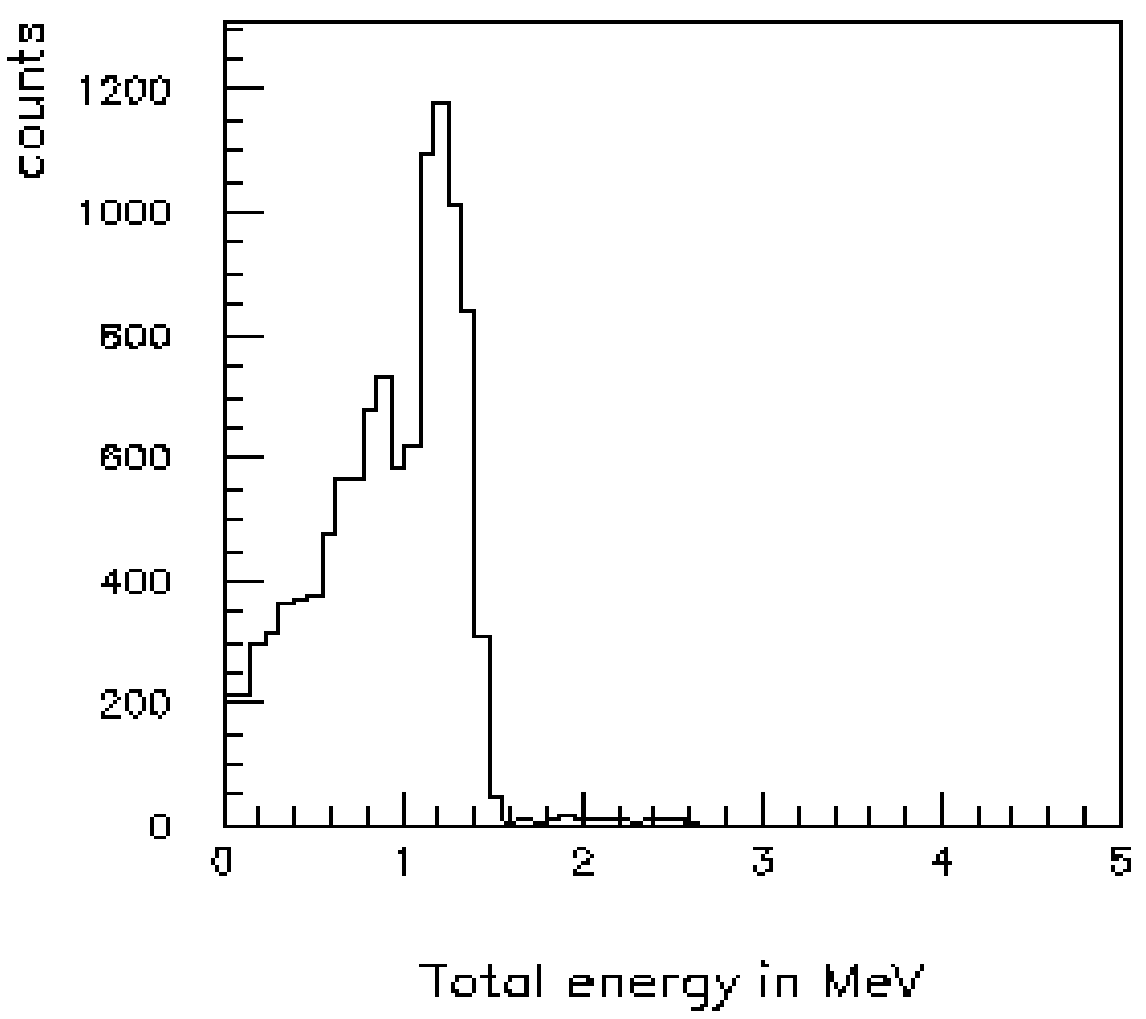

**Figure 21:** $^{60}Co$ spectrum with a single crystal of type A.

Figure 22 shows the sum energy spectrum of a $^{60}Co$ source located at the center of the detector with 162 crystals. The angle correlation between the 1.173 MeV and the 1.332 MeV γ-rays was taken into account. Electrons from β-decay were neglected.

A comparison of the resulting spectrum with the experimental spectrum could be used to assess the energy resolution. At the moment the energy resolution of the Karlsruhe setup [Wis90] is used (see Figure 23).

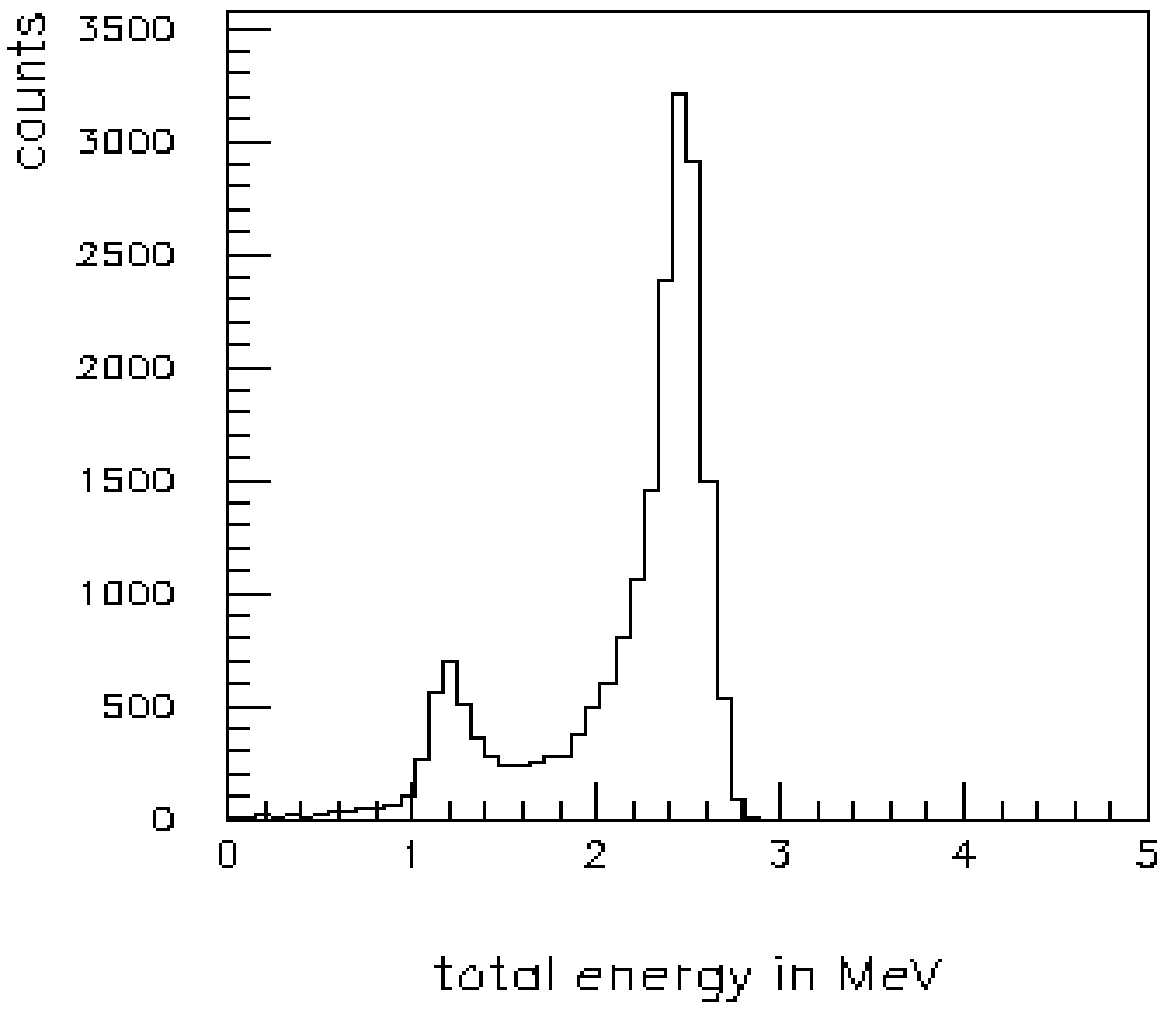

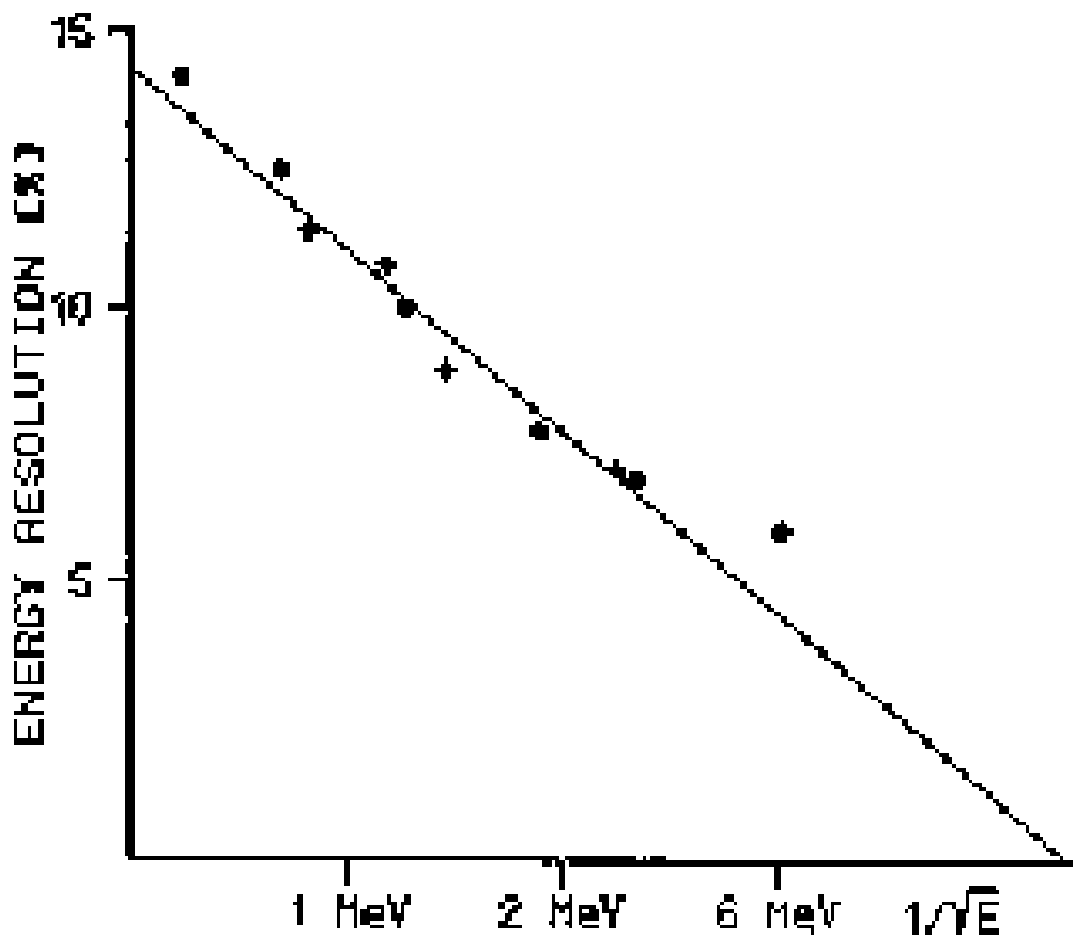

**Figure 22:** $^{60}Co$ spectrum with all 162 crystals.

**Figure 23:** Energy resolution of the Karlsruhe 4π $BaF_2$ detector in the energy range from 0.6 to 6.1 MeV.

# 6. Simulations with neutrons

## 6.1 Neutron spectrum

A fit of the measured data of Figure 24 gave a neutron spectrum of the form

$$\log_{10}(\text{flux})=3.07-0.948\,\log_{10}(E_n)$$

but for simplicity a spectrum following $1/E_n$ was assumed for the simulations. For neutron energies $E_n$ > 100 keV the spectrum is unknown. The neutron spectrum at a future flight path, FP14, is expected to have the same shape as is shown in Figure 25. The simulations showed that neutrons of energies between 8 MeV and 12 MeV could be important because the (n,2n) and the (n,3n) channels for $^{138}$Ba and $^{197}$Au open at 8 MeV and have cross sections of about 2 barns. Neutrons produced in these reactions have lower energies and are likely captured in the crystals, causing considerable background. In addition, such capture events appear in the TOF spectrum at later times and when primary neutrons of somewhat lower energy are just arriving at the sample. Simulations showed, however, that this effect is not significant for a neutron spectrum of the form $1/E_n$. So far, little is known about the true spectrum above 100 keV, but quantitative information would be required for assessing effects if the spectrum deviates from $1/E$ at these higher energies.

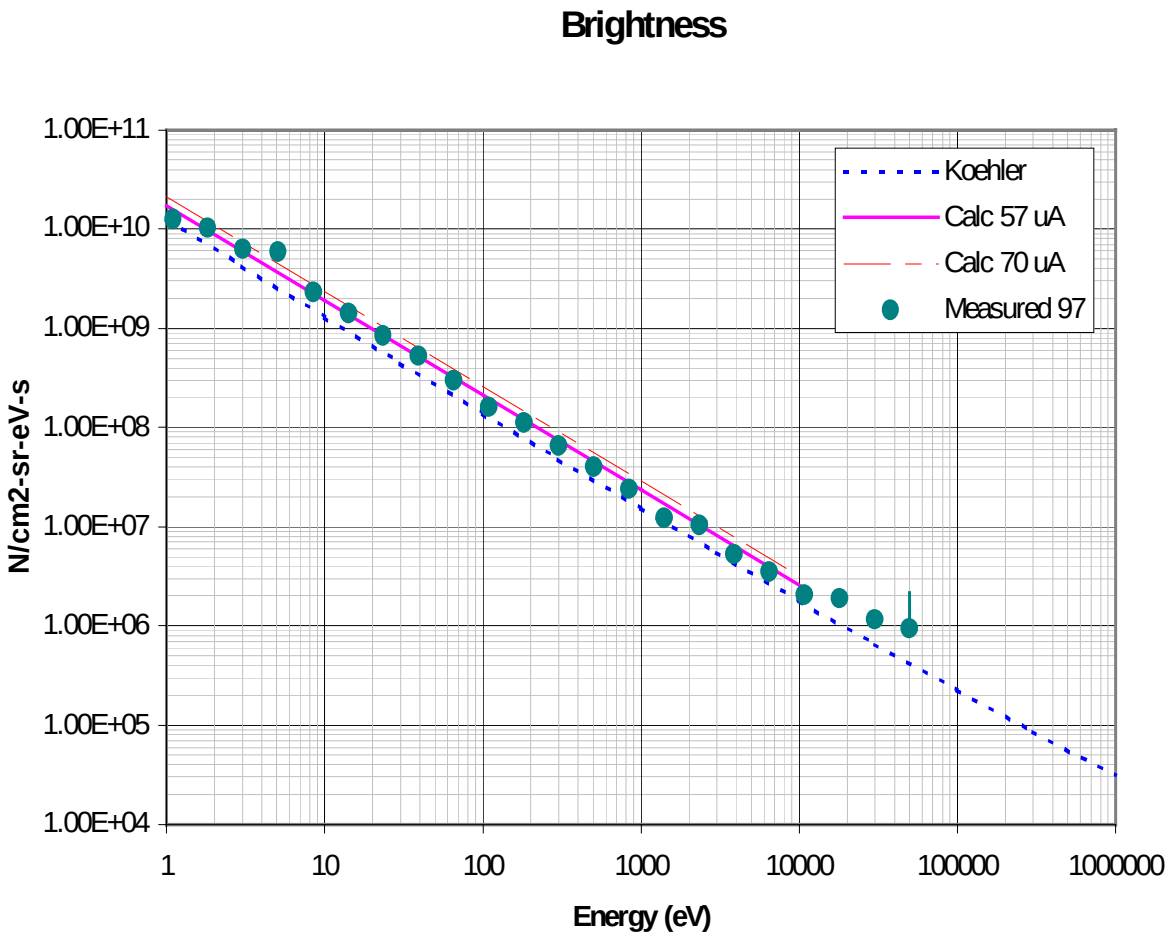

**Figure 24:** Calculated and measured brightness for the previous flight path (FP4).

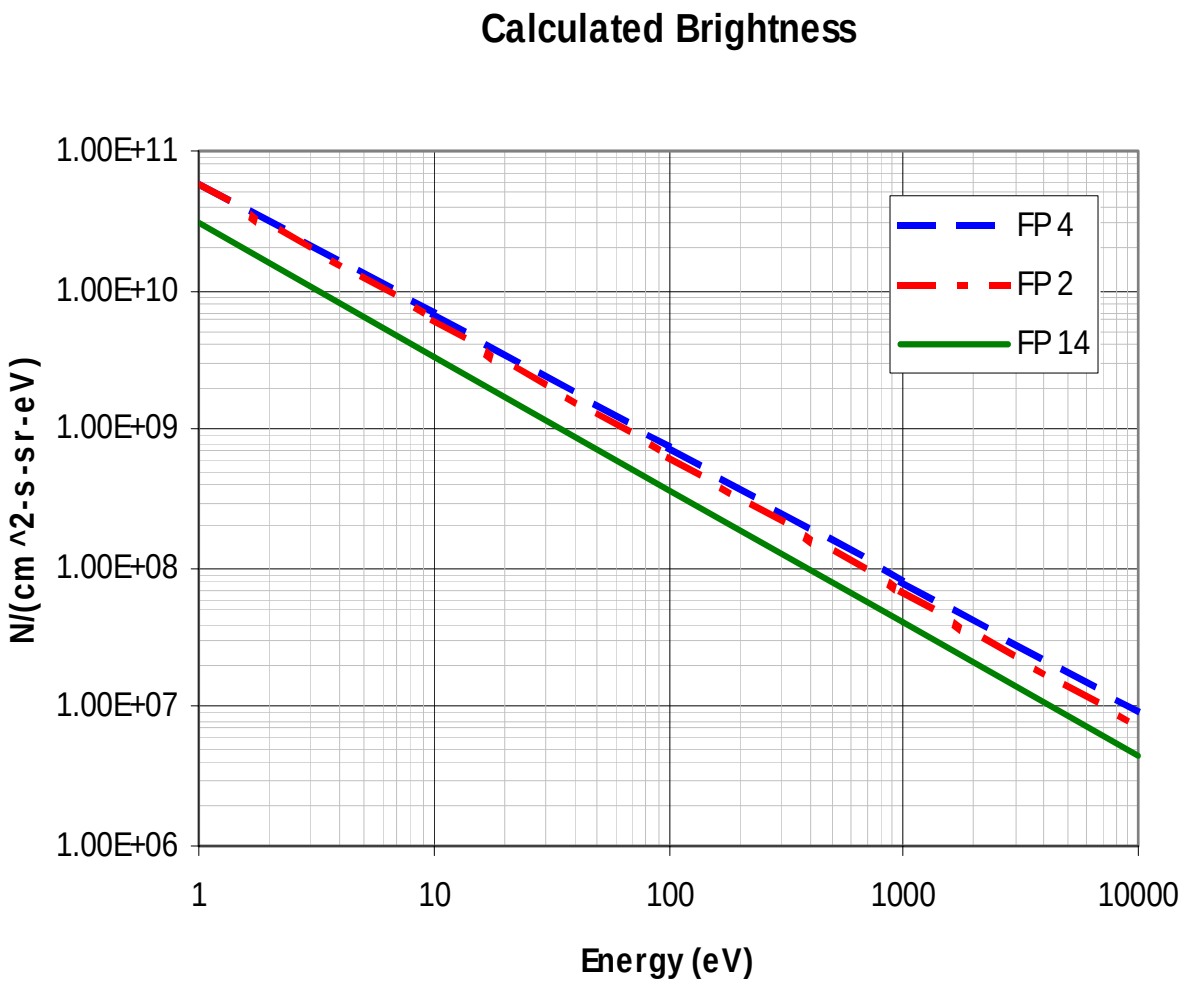

**Figure 25:** Calculated brightness on new flight paths (FP2 and FP14).

## 6.2 Time-of-flight (TOF) spectra

The following calculations were done with a flight path of 20 m and by neglecting the pulse width of the proton beam that drives the MLNSC spallation neutron source. To save computing time a rather thick gold sample (1 mm thick, 1 cm in diameter) was used for the simulations. This should not effect the ratio between scattered and captured neutrons of interest, while the γ-ray attenuation in the sample can still be tolerated. We used the geometry with 160 crystals of 15 cm length, aluminum and Teflon between the crystals and a beam pipe with an inner diameter of 2.4 cm and an outer diameter of 3.0 cm. Figures 26 and 27 show the recorded time of flight spectra for captured and scattered events respectively. One can clearly see the resonance structure of the (n,γ) cross sections for gold and the barium isotopes.

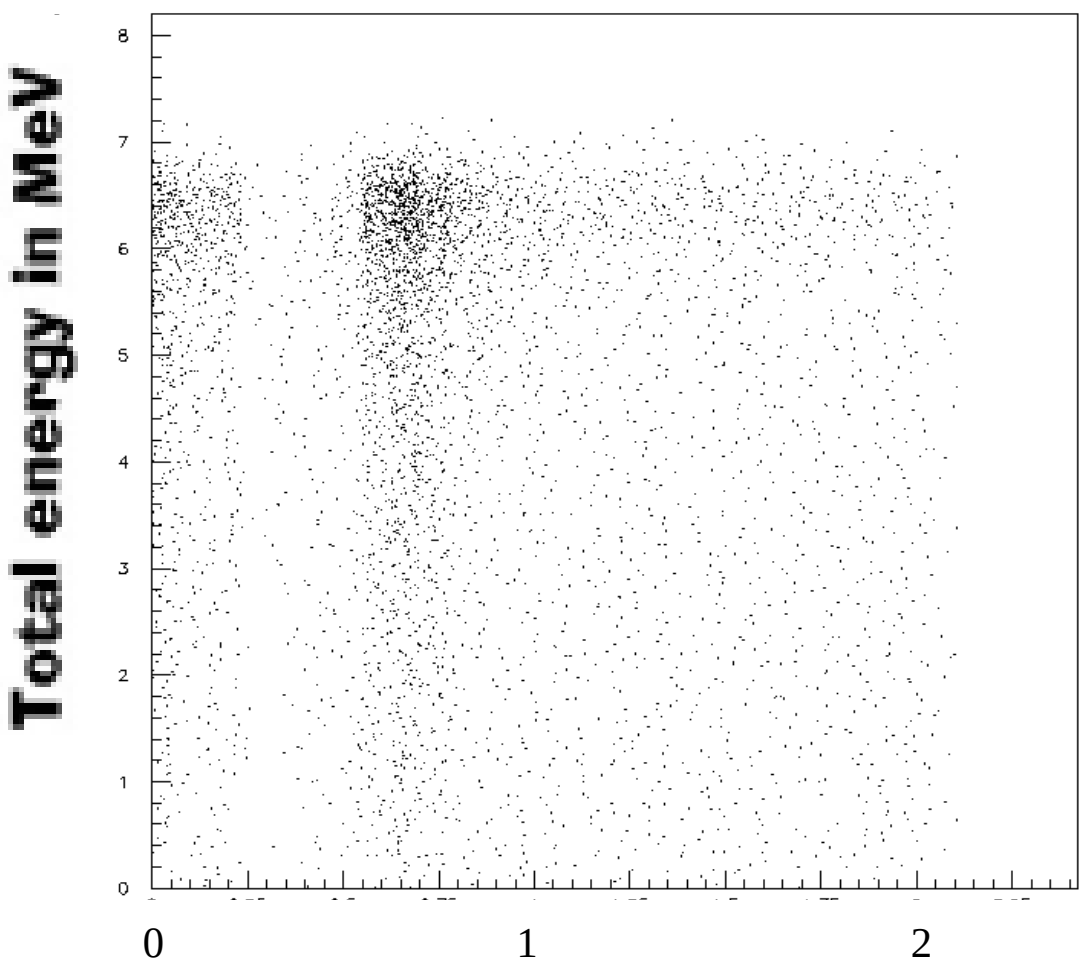

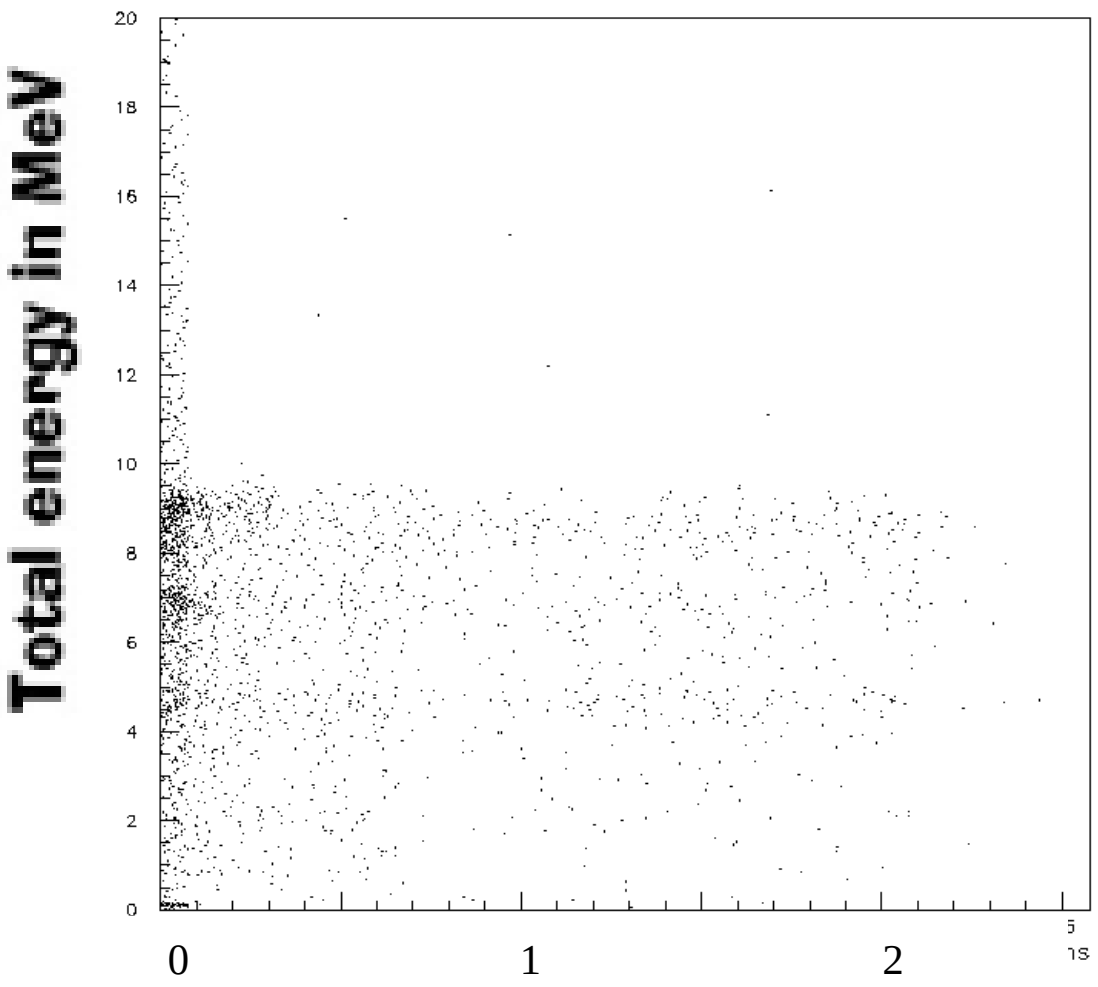

**Figure 27:** Total energy deposited in the scintillators versus TOF for capture events.

**Figure 26:** Total energy deposited in the scintillators versus TOF for scattered events.

Figure 28 and Figure 29 show γ-ray spectra for various cuts on TOF related to the inferred incident neutron energy. In the simulation the TOF was defined as the time difference between the beginning of the pulse and the first signal coming from the detector as in the real experiment. The first column shows spectra of events where neutrons are scattered from the gold sample and captured in barium afterwards. The second column shows spectra of events from neutron capture in the gold sample. The third column shows the sum of the two previous spectra (black line) and the spectrum for scattered neutrons (red line). For neutron energies up to 1 keV the large capture cross section of gold (see section 9) dominates the background from scattered neutrons. From 1 keV to 10 keV, however, the ratio between events from detected scattered and captured neutrons is about 1 and is further deteriorated for higher energies.

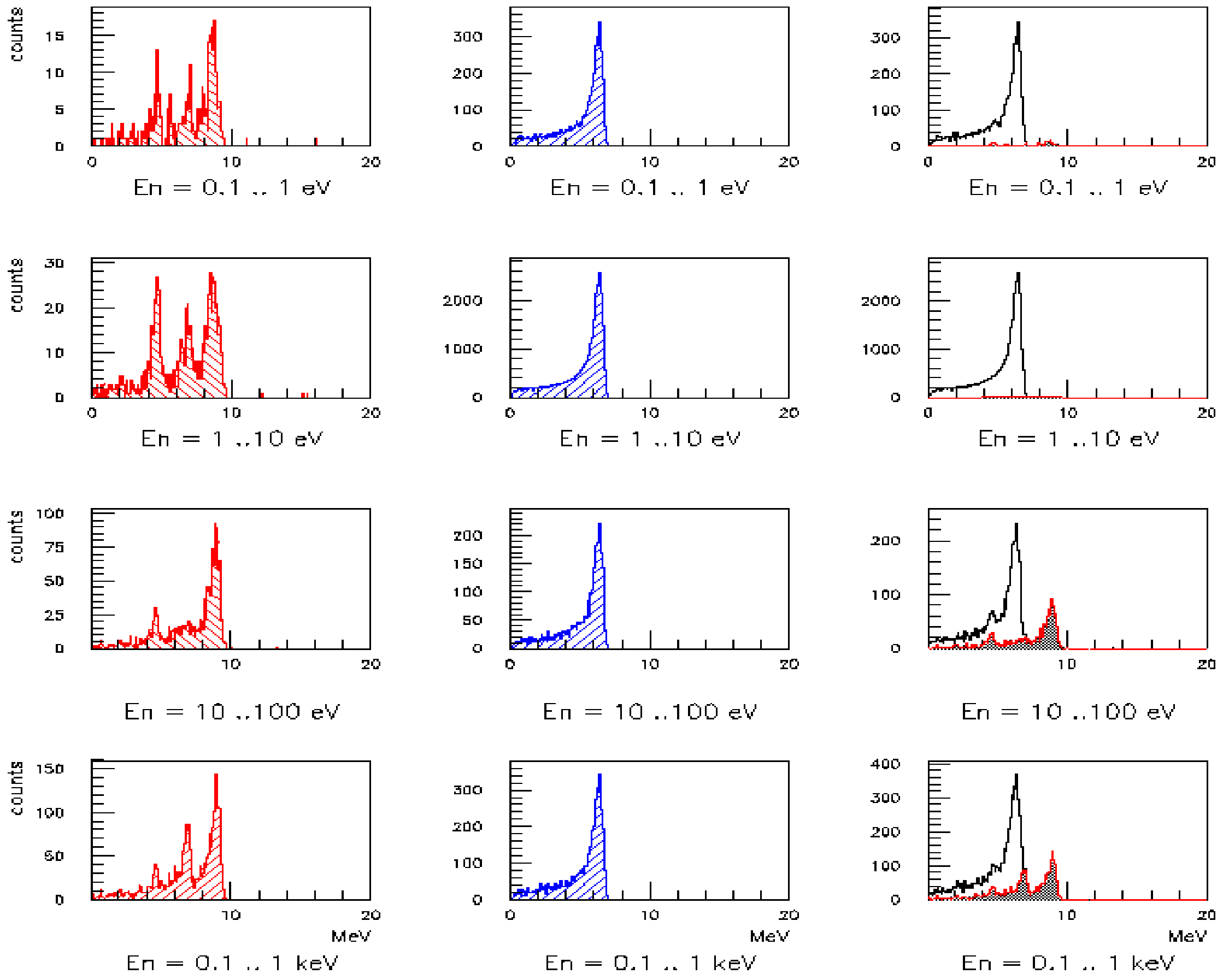

**Figure 28:** Projections of deposited energy spectra for scattered and captured neutrons for various TOF cuts. The background from capture of scattered neutrons in $BaF_2$ ($1^{st}$ column), the (n,γ) events on gold ($2^{nd}$ column) and the expected signal ($3^{rd}$ column -- plots of columns 1 and 2 on the same vertical scale) are compared for neutron energy intervals below 1 keV.

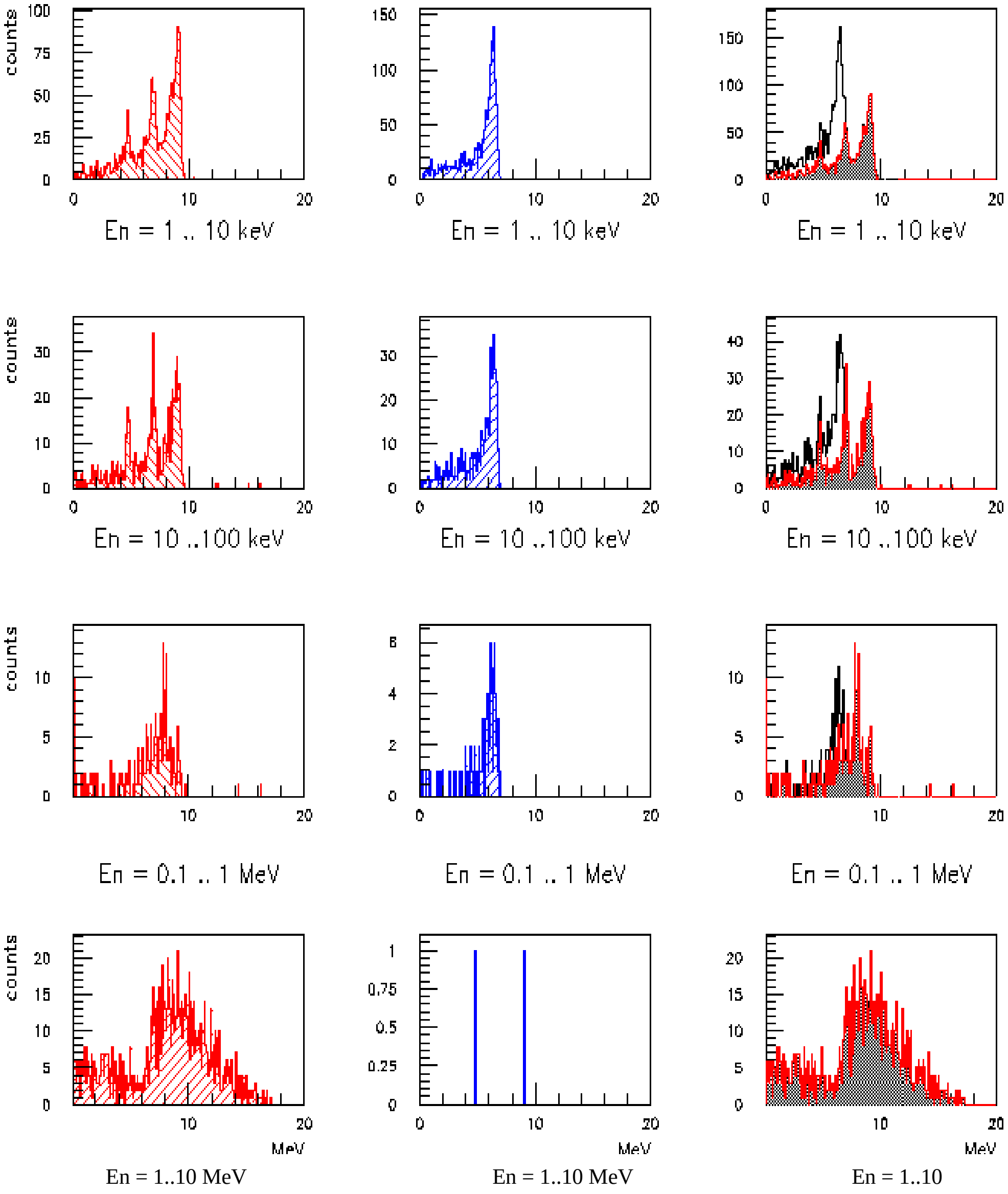

**Figure 29:** Projection of deposited energy spectra for scattered and captured neutrons and for various TOF cuts. The background from capture of scattered neutrons in $BaF_2$ (1st column), the (n,γ) events on gold (2nd column) and the expected signal (3rd column -- plots of columns 1 and 2 on the same vertical scale) are compared for neutron energy intervals between 1 keV and 10 MeV.

Calculations for crystals with 18 cm length gave a ratio of scattered to captured neutrons of 2.07 for the energy region from 10 keV to 100 keV, whereas this value reduces to 1.34 for crystals of 15 cm length. This means that the 10% gain in peak efficiency (see section 3.2) is offset by a 54% increase of the background from scattered neutrons.

As mentioned in section 5.4, the energy resolution of the Karlsruhe setup was used for the calculations. In order to avoid pile up effects for the higher neutron fluxes at the MLNSC neutron source it might be necessary to integrate only the fast component of the scintillator light, which contributes only about 10% to the total light output. Correspondingly one would expect a reduction in energy resolution by a factor of √10 as illustrated in Figure 30.

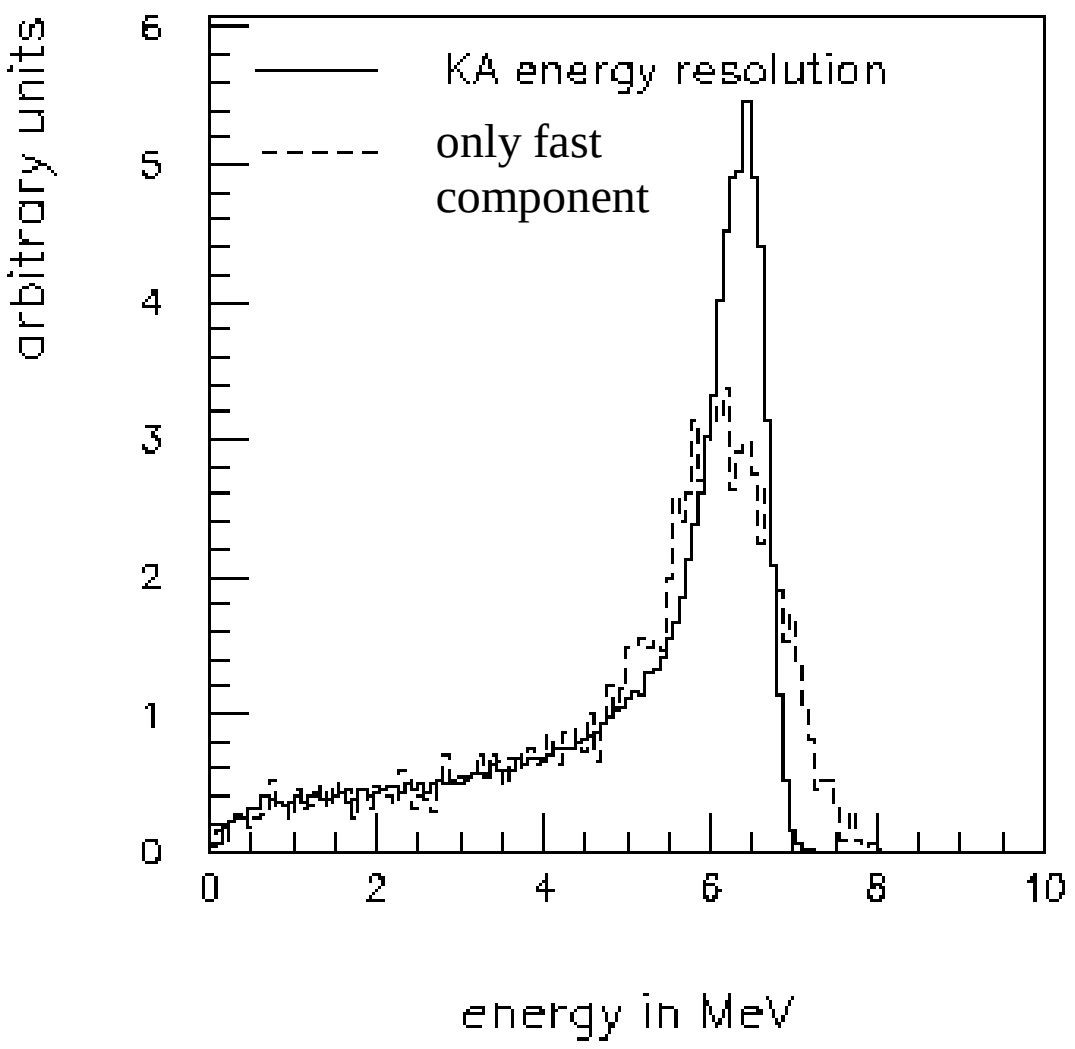

**Figure 30:** Comparison of the Karlsruhe energy resolution used for the calculations and the energy resolution one would expect by integrating only the fast component of the scintillation light.

The effect of such a reduced energy resolution with respect to the background from capture in barium is shown in Figure 31.

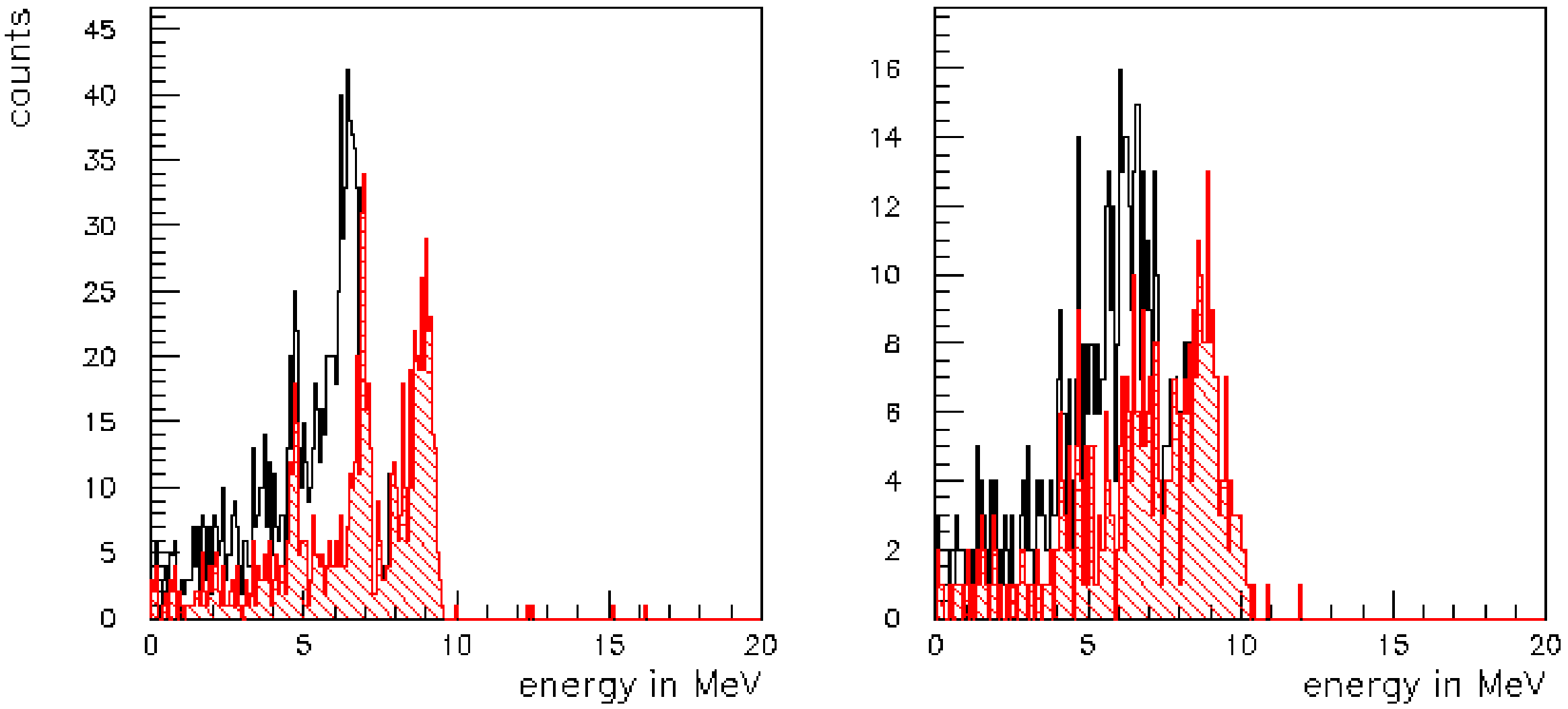

**Figure 31:** Comparison between the pulse height spectra for captured and scattered neutrons with the Karlsruhe energy resolution (left) and the resolution expected for the fast component (right). The neutron energy range was 10 keV to 100 keV. The pictures show the expected signal (black) as well as the background (red) caused by neutron capture in $BaF_2$.

## 6.3 Background from scattered neutrons

In this section, the possibilities to decrease the ratio of detected scattered to captured neutrons in the energy region from 1 keV to 1 MeV are discussed for different $BaF_2$ arrays. For comparison the last case considers a $C_6F_6$ detector. Descriptions of the different experimental setups are as follows:

1. Standard setup with 160 crystals of 15 cm length with aluminum and Teflon between the crystals as discussed above.
2. Setup with 160 crystals of 12 cm length with aluminum and Teflon between the crystals.
3. Setup with 160 crystals of 18 cm length with aluminum and Teflon between the crystals as discussed above.
4. Same as setup 1 but the gaps of 0.8 mm air between the crystals were filled with natural LiF of density 2.635 g/cm$^3$. The aluminum and Teflon were not removed. This has the advantage that the alpha particles and tritons from the $^6$Li(n,$\alpha$)t reaction do not make it into the crystals but are absorbed in the aluminum.
5. Same as setup 4 but with $^6$LiF.
6. Same as setup 4 but with crystals of 18 cm length.
7. Same as setup 4 but the gaps of 0.8 mm air between the crystals were filled with natural boron of density 2.0 g/cm$^3$.
8. Same as setup 7 but with $^{10}$B.
9. Same as setup 7 but the 0.5 mm Teflon layer was also replaced with boron. This means that the 2 mm gap between the crystals is filled with 0.2 mm aluminum and 1.8 mm natural boron.
10. Same as setup 9 but with $^{10}$B.
11. Same as setup 7 but sample moved by 9 cm in beam direction.
12. Same as setup 7 but sample moved by 9 cm against the beam direction.
13. Same setup as 11 with 5 crystals around the existing beam pipe removed. The idea of setup 11, 12 and 13 was to force the scattered neutrons from the sample to cross the boron layer between the crystals more often. Also for higher neutron energies one would expect small scattering angles and by moving the sample in beam direction those scattered neutrons could escape through the gap of the omitted crystals. It turned out that the angular distribution of the scattered neutrons for the energy region of interest is rather isotropic and the effects from moving the targets were small.
14. Same as setup 7 but beam pipe made out of a mixture of aluminum and natural boron.
15. Standard setup but with a sphere of natural LiH, 15 cm diameter around the sample, beam pipe intersecting this sphere.
16. Same as setup 15 but with a sphere of 19 cm diameter.
17. Same as setup 7 but with a sphere of 9 cm in diameter made of natural LiH.
18. Same as setup 8 but with a sphere of 15 cm in diameter made of $^6$LiH.
19. Same as setup 18 but with a sphere of 19 cm diameter.
20. Same as setup 14 but an inner sphere of 9 cm diameter made of polyethylene and an outer sphere of 15 cm diameter made of natural LiF.
21. Because of the higher price of isotopically enriched $^{10}$B, 1.8 mm natural boron was placed between the crystals. Additionally a spherical moderator/absorber with an outer diameter of 15 cm made of $^6$LiH was positioned around the beam pipe.
22. In order to check the influence of a possibly lower density of the $^6$LiH moderator/absorber, 15 cm outer diameter and 80 % of the density (0.68 instead of 0.85 g/cm$^3$) were assumed. The gaps between the 160 crystals were filled with $^{10}$B and aluminum.
23. To check the moderating effect of $^6$LiH, a moderator/absorber made of metallic $^6$Li with 15 cm diameter was used. Again $^{10}$B and aluminum were between the crystals.
24. Same as setup 18 but moderator/absorber made of BN. This simulation shows the importance of the moderator/absorber material for the neutron background.
25. Same as setup 18 but moderator/absorber made of a mixture of 5 % natural boron and polyethylene.
26. Standard setup but including a sphere with inner radius of 41.91 cm (16.5’’) and thickness of 3.81 cm (1.5’’) consisting of a mixture of aluminum, nickel, cobalt and iron. With this setup we investigated the influence of the supporting structure and the PMs. The supporting structure is made out of aluminum and the PMs are shielded with a mixture of nickel (78 %), copper (3.5 %) and iron (18.5 %). For the sphere an average density corresponding to the fractions of the different elements of the mixture was used (Al: 95.9 %, Ni: 3.2 %, Fe: 0.8 %, Cu: 0.1 %). Especially for the 10 to 100 keV region one can see more background from back-scattered neutrons. This is due to a large resonance in the scattering cross section of aluminum.

27. Same setup as before but including a Polyethylene (PE) shielding around the detector and a concrete floor. The shielding includes a lead block of 80 x 80 x 20 $cm^3$ (see Figure 32). The height of the box is 2.4 m, the length is 3 m and the width is 1.6 m. The center of the detector is 1.37 m above the floor. The PE walls on the side are 10 cm and the ceiling is 21 cm thick. The floor has the dimensions 3 x 3 x 0.5 $m^3$. The concrete is a mixture of 0.0056 mass fractions of hydrogen, 0.4983 oxygen, 0.0171 sodium, 0.0024 magnesium, 0.0456 aluminum, 0.3158 silicon, 0.0012 sulfur, 0.0192 potassium, 0.0826 calcium, and 0.0122 iron and has a density of 2.3 $g/cm^3$. As one can see from Table 6, the effect of the PE walls and the concrete for scattered neutron from the sample is negligible but this situation might change when neutrons scattered from the collimator are considered also.

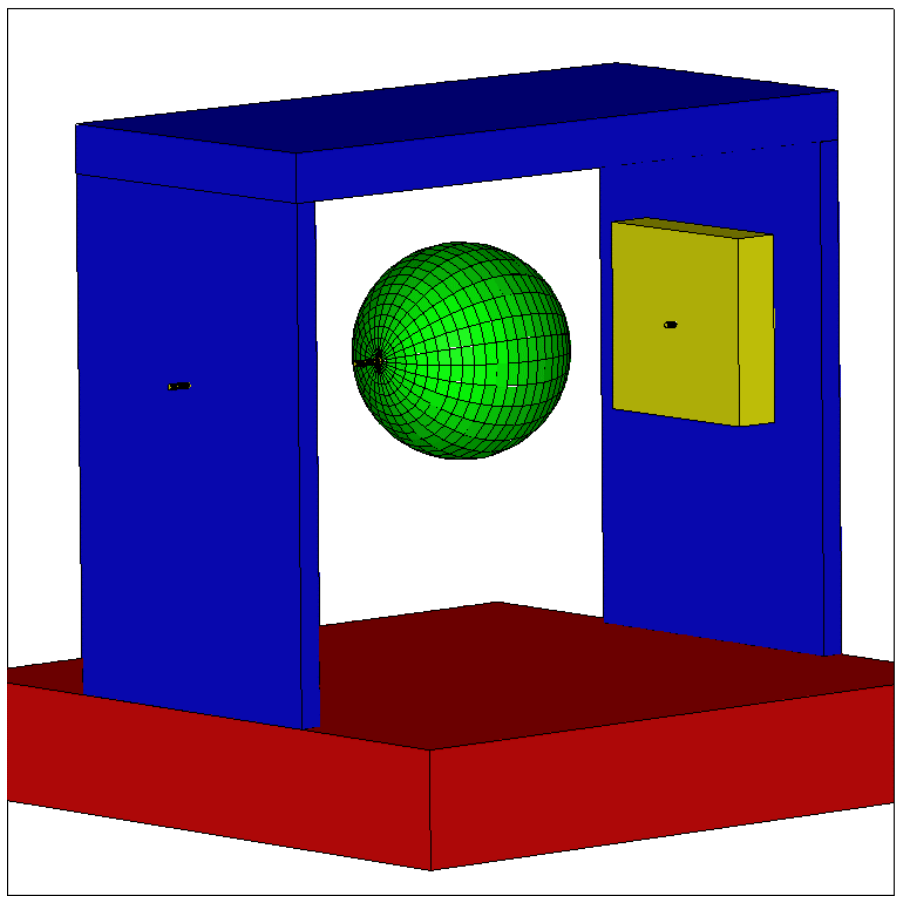

**Figure 32:** View of the setup. Side walls are left out. Concrete (red), PE (blue), aluminum supporting structure (green) and lead (yellow) are shown.

28. Same as standard setup but crystal ball with inner radius of 5 cm and outer radius of 20 cm.
29. Same as setup before but gaps between the crystals filled with $^{10}B$ and aluminum and with a spherical moderator/absorber with 9 cm in diameter (maximum size) made of $^{6}LiH$.
30. 160 modules made of $C_6F_6$ with an inner radius of 10 cm and an outer radius of 70 cm (60 cm scintillator length) and a beam pipe like in setup 1.

The results are summarized in Table 6.

| Setup | Ratio between scattered and captured events for different neutron energy regions. | | | |
|---|---|---|---|---|
| | 0.1 .. 1 keV | 1 ..10 keV | 10 .. 100 keV | 0.1 .. 1 MeV |
| 1 standard | 0.522±0.02 | 1.09±0.04 | 1.34±0.10 | 3.19±0.60 |
| 2 crystals of 12 cm length | 0.374±0.02 | 0.752±0.05 | 1.09±0.13 | 4.11±1.18 |
| 3 crystals of 18 cm length | 0.657±0.02 | 1.47±0.06 | 2.07±0.14 | 4.44±0.78 |
| 4 LiF, Al and Teflon between crystals | 0.519±0.02 | 0.954±0.04 | 1.34±0.10 | 3.63±0.62 |
| 5 $^{6}$LiF, Al and Teflon between crystals | 0.345±0.017 | 0.723±0.005 | 1.10±0.13 | 2.68±0.67 |
| 6 LiF between crystals of 18 cm length | 0.644±0.02 | 1.27±0.05 | 2.03±0.14 | 3.60±0.60 |
| 7 B, Al and Teflon between crystals | 0.282±0.01 | 0.697±0.03 | 1.28±0.10 | 3.12±0.57 |
| 8 $^{10}$B, Al and Teflon between crystals | 0.126±0.008 | 0.35±0.03 | 0.88±0.11 | 3.66±1.02 |
| 9 B and Al between crystals | 0.203±0.01 | 0.512±0.03 | 1.01±0.08 | 3.54±0.65 |
| 10 $^{10}$B and Al between crystals | 0.068±0.006 | 0.234±0.021 | 0.599±0.085 | 3.41±0.86 |
| 11 B and Al; sample +9 cm moved | 0.169±0.01 | 0.484±0.03 | 1.02±0.11 | 3.80±0.80 |
| 12 B and Al; sample –9 cm moved | 0.177±0.01 | 0.487±0.03 | 1.01±0.10 | 3.85±0.86 |
| 13 B and Al; sample +9 cm, 5 crystals left out | 0.130±0.01 | 0.410±0.03 | 0.98±0.12 | 3.62±0.98 |
| 14 B and Al; beam pipe made out of B and Al | 0.144±0.01 | 0.447±0.03 | 1.01±0.13 | 4.11±1.16 |
| 15 6 cm $^{6}$LiH around beam pipe (sph) | 0.013±0.002 | 0.058±0.009 | 0.243±0.045 | 2.12±0.63 |
| 16 8 cm $^{6}$LiH around beam pipe (sph) | 0.007±0.002 | 0.023±0.005 | 0.12±0.03 | 1.59±0.49 |
| 17 B and Al; 3 cm LiH around beam pipe (sph) | 0.0401±0.004 | 0.253±0.02 | 0.903±0.12 | 3.21±0.86 |
| 18 $^{10}$B and Al; 6 cm $^{6}$LiH around beam pipe (sph) | 0.0027±0.001 | 0.0084±0.003 | 0.044±0.015 | 1.83±0.53 |
| 19 $^{10}$B and Al; 8 cm $^{6}$LiH around beam pipe (sph) | 0.0007±0.0005 | 0.005±0.002 | 0.013±0.008 | 1.22±0.33 |
| 20 B and Al; 3 cm PE , 3cm LiH around beam pipe | 0.067±0.006 | 0.202±0.02 | 0.491±0.07 | 2.36±0.63 |
| 21 B and Al; 6cm $^{6}$LiH around beam pipe | 0.037±0.012 | 0.027±0.006 | 0.78±0.21 | 1.91±0.56 |
| 22 $^{10}$B and Al; 6cm $^{6}$LiH ($\rho$ = 0.8 $\rho_0$) moderator | 0.0030±0.0010 | 0.012±0.004 | 0.061±0.02 | 2.61±0.83 |
| 23 $^{10}$B and Al; 6cm $^{6}$Li (metallic) around beam pipe | 0.008±0.002 | 0.093±0.012 | 0.37±0.06 | 3.15±0.79 |
| 24 $^{10}$B and Al; 6cm $^{10}$BN around beam pipe | 0.132±0.012 | 0.423±0.044 | 1.08±0.18 | 3.59±1.17 |
| 25 $^{10}$B and Al; 6cm 5% B + PE around beam pipe | 0.127±0.009 | 0.342±0.027 | 0.497±0.07 | 2.29±0.62 |
| 26 Al (Ni, Co, Fe) sphere around | 0.529±0.02 | 1.07±0.04 | 1.70±0.13 | 3.29±0.54 |
| 27 Al sphere, PE walls, concrete floor, lead block | 0.557±0.02 | 1.12±0.04 | 1.68±0.13 | 3.39±0.56 |
| 28 crystals with inner / outer radius of 5 cm / 20cm | 0.586±0.03 | 0.975±0.074 | 1.54±0.2 | 3.68±1.12 |
| 29 crystals of 5 cm / 20cm; $^{10}$B and Al; 3 cm $^{6}$LiH | 0.007±0.002 | 0.045±0.008 | 0.320±0.056 | 3.15±0.86 |
| 30 $C_6F_6$ modules; 60 cm length | 0.0349±0.004 | 0.0181±0.0045 | 0.0364±0.014 | 1.28±0.37 |

**Table 6:** Ratio of events from scattered neutrons and true capture for different setups.

From item 16 of this table, it is apparent that including a $^{6}$LiH shell around the sample results in about a factor of 10 improvement in the scattered-to-capture ratio in the neutron energy range of 10- 100 keV, of great interest for astrophysics. This material may cause some degradation of the capture peak, however. This effect is illustrated in Figure 33 for different moderator/absorber thickness. Obviously the effect is rather small since LiH consists of low-Z elements.

In summary, the best solution is obtained by filling the gaps between the crystals with $^{10}$B and by using a moderator/absorber made of $^{6}$LiH around the beam pipe.

Figure 34 shows a comparison of the highlighted cases of Table 6 for the critical neutron energy range from 10-100 keV.

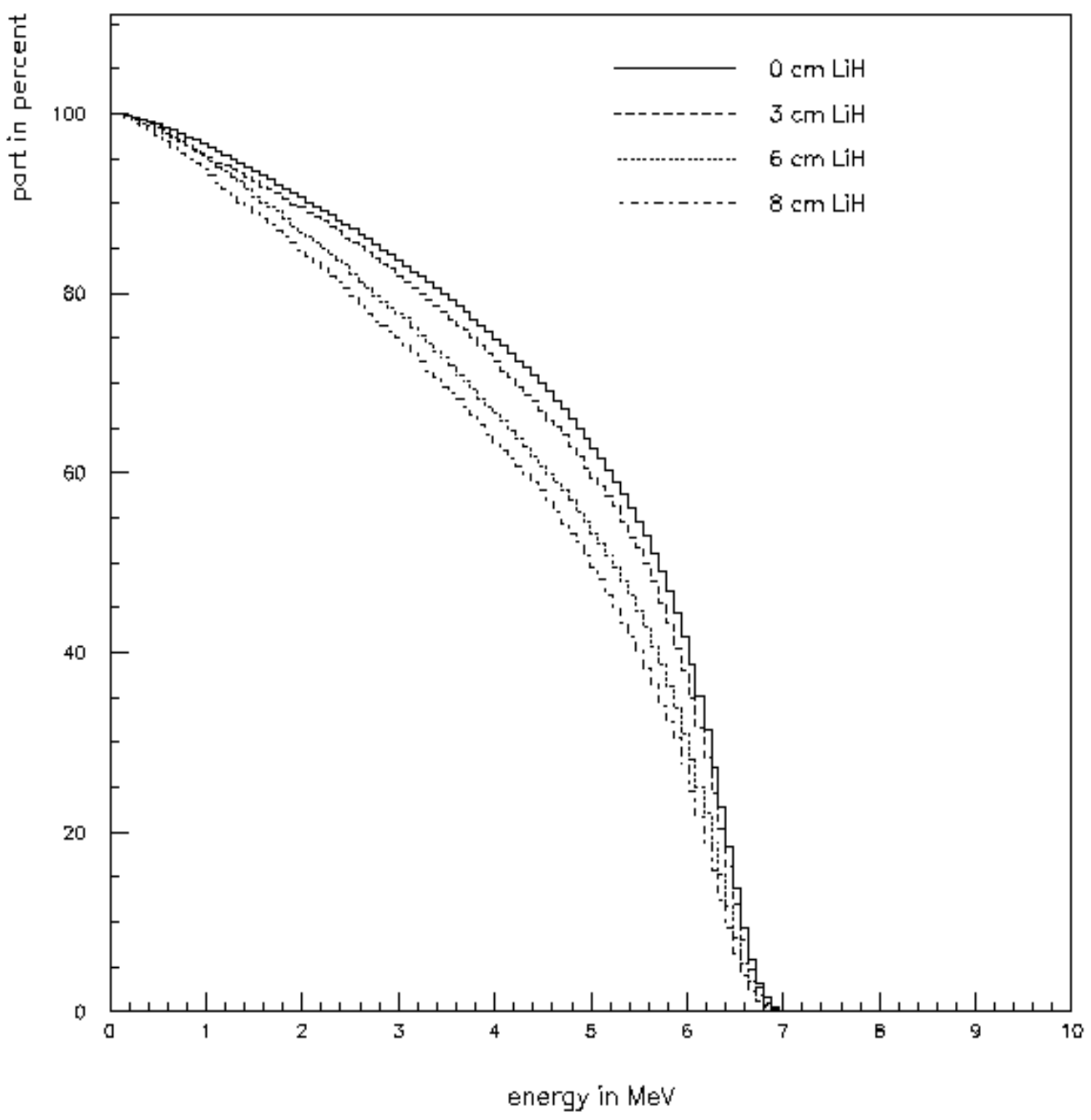

**Figure 33:** Influence of moderator/absorber thickness on the shape of the gold peak.

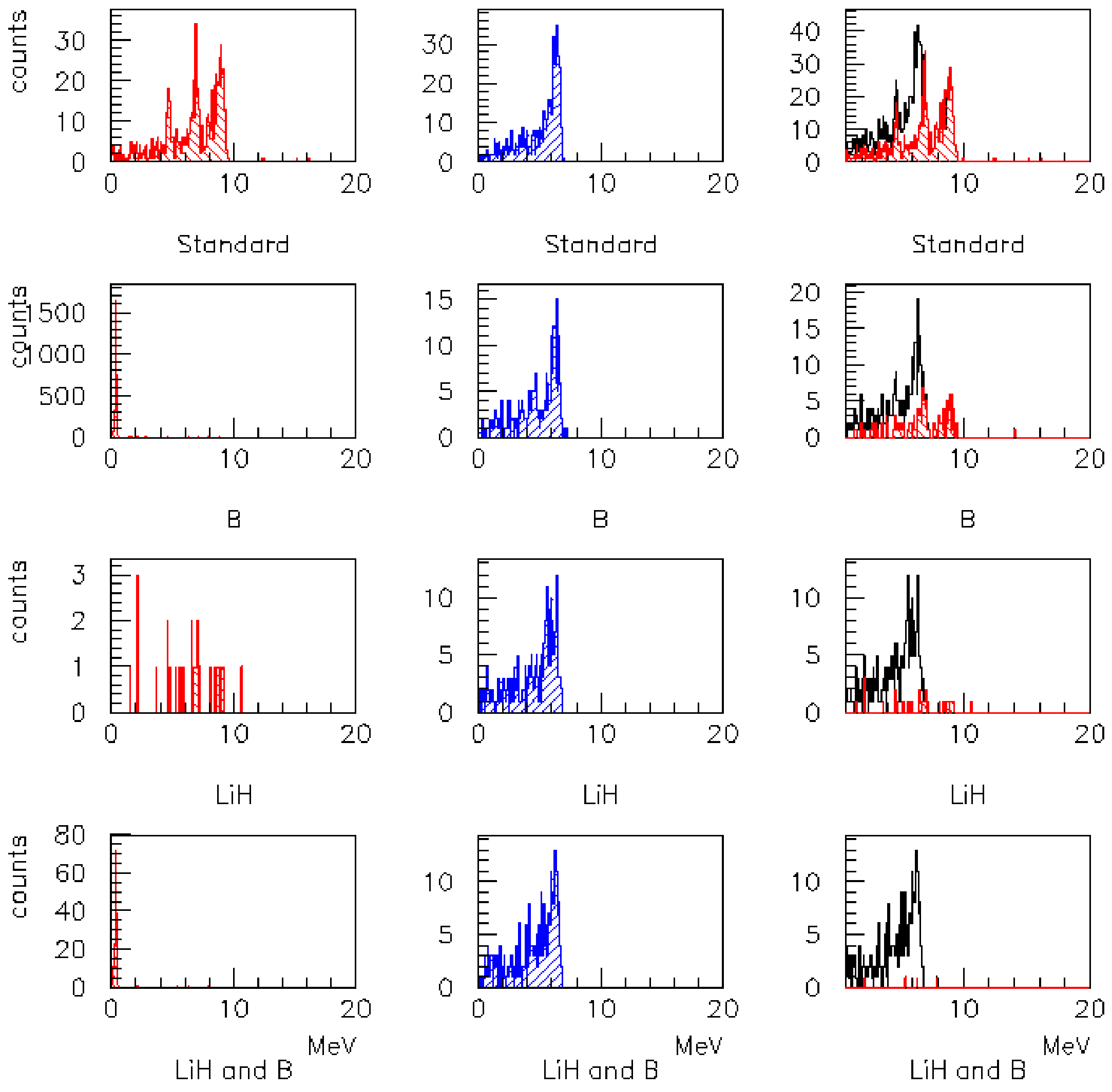

**Figure 34:** Comparison of setups 1, 8, 16 and 19 in the neutron energy range from 10-100 keV. First column shows the response to neutrons scattered from the gold sample, the middle column the response to neutron capture in the gold sample and the right column the sum of these two processes compared to the background (shaded area). Case 1 ($1^{st}$ row) is the standard setup, in case 8 ($2^{nd}$ row) the gaps between the crystals are filled with $^{10}B$, in case 16 ($3^{rd}$ row) a moderator/absorber of $^{6}LiH$ around the beam pipe was used and in case 19 ($4^{th}$ row) both, the $^{10}B$ between the crystals and the moderator/absorber of $^{6}LiH$ were combined.

## 6.4 Target backings

In this section the influence of possible backing materials for target preparation is discussed.

To investigate the background caused by scattered or captured neutrons in the backing, a disk-shaped backing of various materials 1 cm in diameter and 1 mm thick was positioned in the center of the crystal ball and irradiated with neutrons. The geometrical shape and size of the backing was the same as for the gold sample in calculations before. An aluminum beam pipe of 3 mm thickness and the standard setup with 160 $BaF_2$ crystals were considered (see section 6.3).

The neutron energy distribution was the same as described above (1/E from 100 eV up to 20 MeV).

Table 7 shows the ratio between the number of background events due to scattering in backing and the number true capture events in the gold sample for carbon and beryllium backings and various neutron energy intervals. Scattering from the gold is analyzed in detail in section 6.3 and is not considered here.

| Backing | Ratio between backing events (scatter followed by capture in the scintillators) and capture events in a gold sample for different neutron energy regions. The geometrical size of the backing is the same as that of the gold sample. | | | |
|---|---|---|---|---|
| | 0.1 .. 1 keV | 1 ..10 keV | 10 .. 100 keV | 0.1 .. 1 MeV |
| Carbon © | 0.332±0.016 | 0.499±0.035 | 1.13±0.13 | 1.21±0.36 |
| Beryllium (Be) | 0.508±0.022 | 0.796±0.050 | 1.77±0.18 | 1.43±0.40 |

**Table 7:** The ratio of events due to scattering from the backing and true capture events for a gold sample with the same dimensions as the backing.

Assuming a gold sample with 1 mg/cm$^2$ and a density of 20 g/cm$^3$ (which corresponds to a thickness of 0.5 µm) and a backing made of 1 µm beryllium foil, one would expect a ratio of 3.54 between the background caused by the backing and the signal in the most interesting energy region between 10 and 100 keV. Hence, the calculated ratio of 1.34 for the standard setup would increase to 4.88 because of the backing.

Since the cross section data file used by GCALOR has no entries for titanium, it was not possible to investigate this backing material as well. However the comparison between the elastic scattering cross sections of gold and titanium (see section 9) indicates similar backgrounds for most of the energy intervals, but somewhat larger effects may be expected in the 10 – 100 keV region due to a 20 keV resonance in the scattering cross section of Ti as well as in the resonance region, where Ti exhibits significant capture resonances.

## 6.5 Hit pattern

The hit pattern from captured and scattered events may also carry important information. Could this provide a way to distinguish events caused by capture in the $BaF_2$ from the 'true' capture events in the sample? The first attempt is to look at the multiplicity distribution of events from scattered and captured neutrons. The results obtained with the standard setup (Figure 35) provide certainly no means for such a distinction. No significant feature which would allow one to distinguish between the captured and scattered events was found.

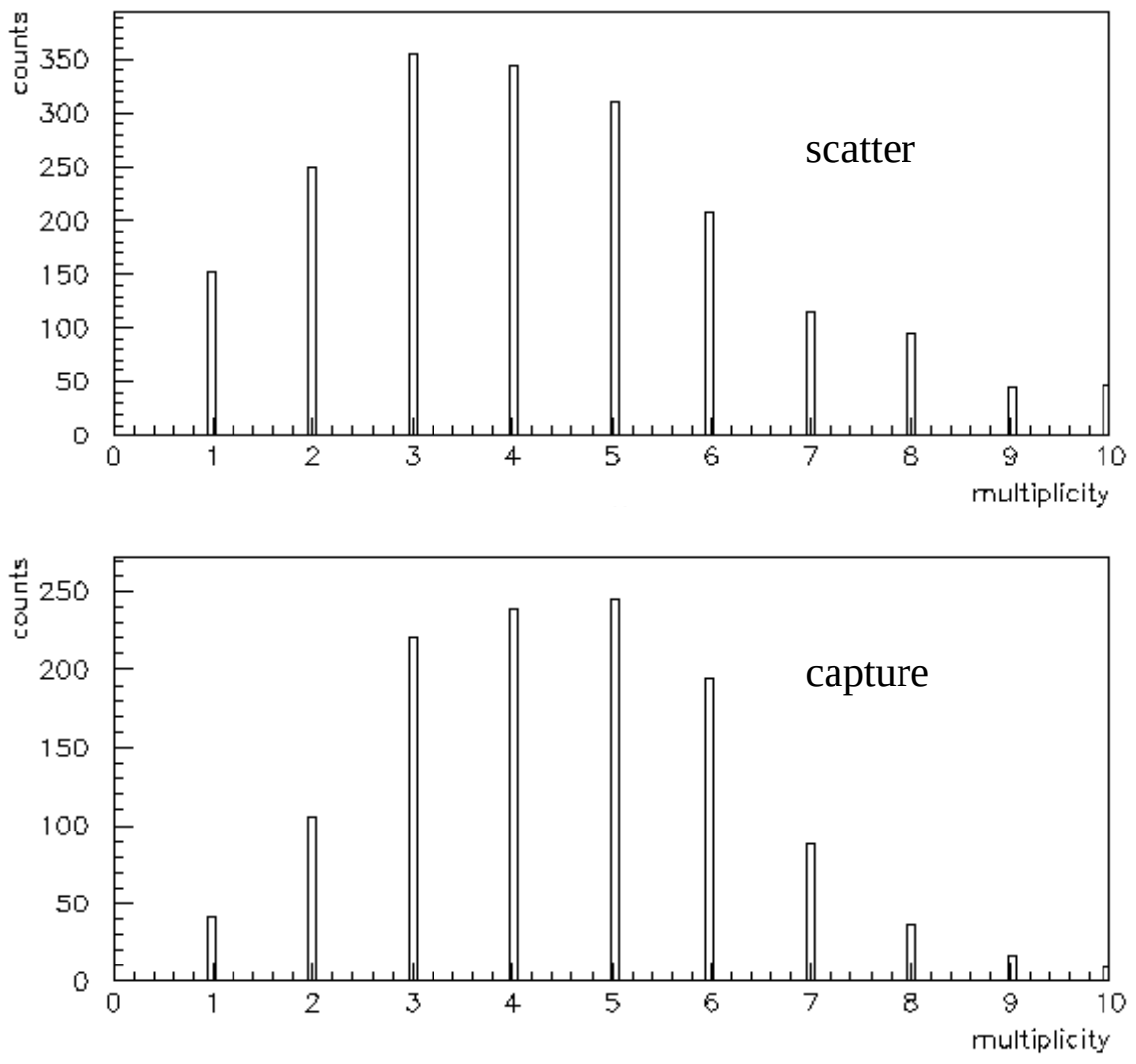

**Figure 35:** Multiplicity distribution for events from scattered neutrons captured in $BaF_2$ (top) and for true capture events in a gold sample (bottom).

The next attempt was to relate each crystal to a unit vector according to its position and add up all vectors multiplied by the energy deposit in the hit crystals for one event. The length of these vectors is plotted in the diagrams of Figure 36 for captured and scattered neutrons separately. Figure 37 shows the same diagrams, but without multiplying the length of the vector with the deposited energy. Only in this approach, there are clear peaks at a length of 2, 3, 4 and 5 units due to events where only neighboring crystals were hit, suggesting that one could improve the ratio between scattered and captured events by analyzing the non-energy-weighted vector sum.

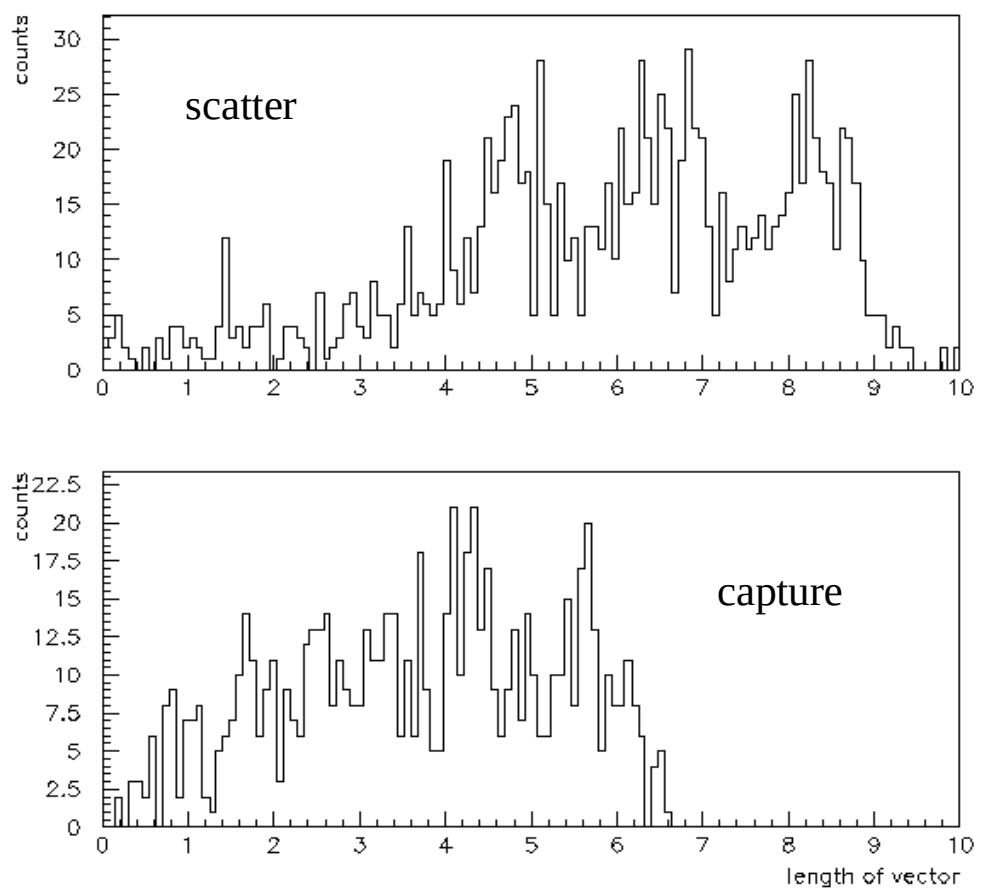

**Figure 36:** Each crystal was related to a unit vector according to its position in the three dimensional space. The diagrams show the length of the vectors after summing all unit vectors multiplied by the energy deposit in the crystals.

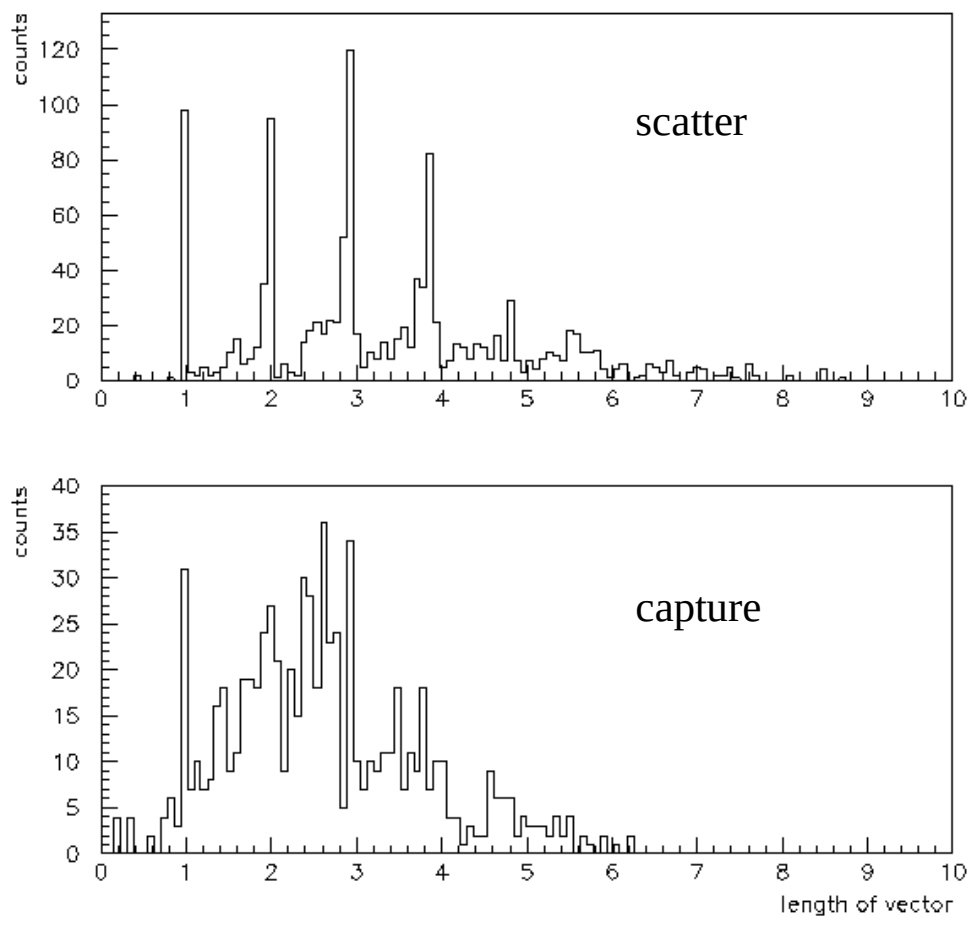

**Figure 37:** Vector sum of all hits per event.

Indeed, analysis in the neutron energy region from 10 to 100 keV proved that the events due to scattered neutrons could be reduced by a factor of 2 by eliminating all events where only neighboring crystals have fired, whereas for the “true” capture events, only 10 % were sorted out by this procedure.

It seems that background events from neutron capture in barium have a tendency to build big clusters. For the special geometry of 162 crystals, 2 crystals are neighbored if and only if the angle between the two rays starting in the center of the ball and going through the middle of the crystals is lower then 20 degrees (or if the scalar product between the corresponding unit vectors is greater than 0.94).

Since the randomly distributed γ-rays caused by a neutron capture event in barium start all in one crystal it is plausible that it interacts most likely in this or one of the neighboring crystals. In contrast, for two γ-rays emitted in the center of the crystal ball (e.g. after neutron capture in the sample) it is most unlikely to interact in neighboring crystals. Therefore the number of hit crystals in the clusters due to genuine captures and those due to background from scattered neutrons are significantly different.

Figure 38 and Figure 39 show that captures in $BaF_2$ due to scattered neutrons tend to form large clusters with a 50 % probability to from only one cluster. The energy deposited per cluster is investigated in Figure 40, where we see a large difference between many of the scattered and captured events.

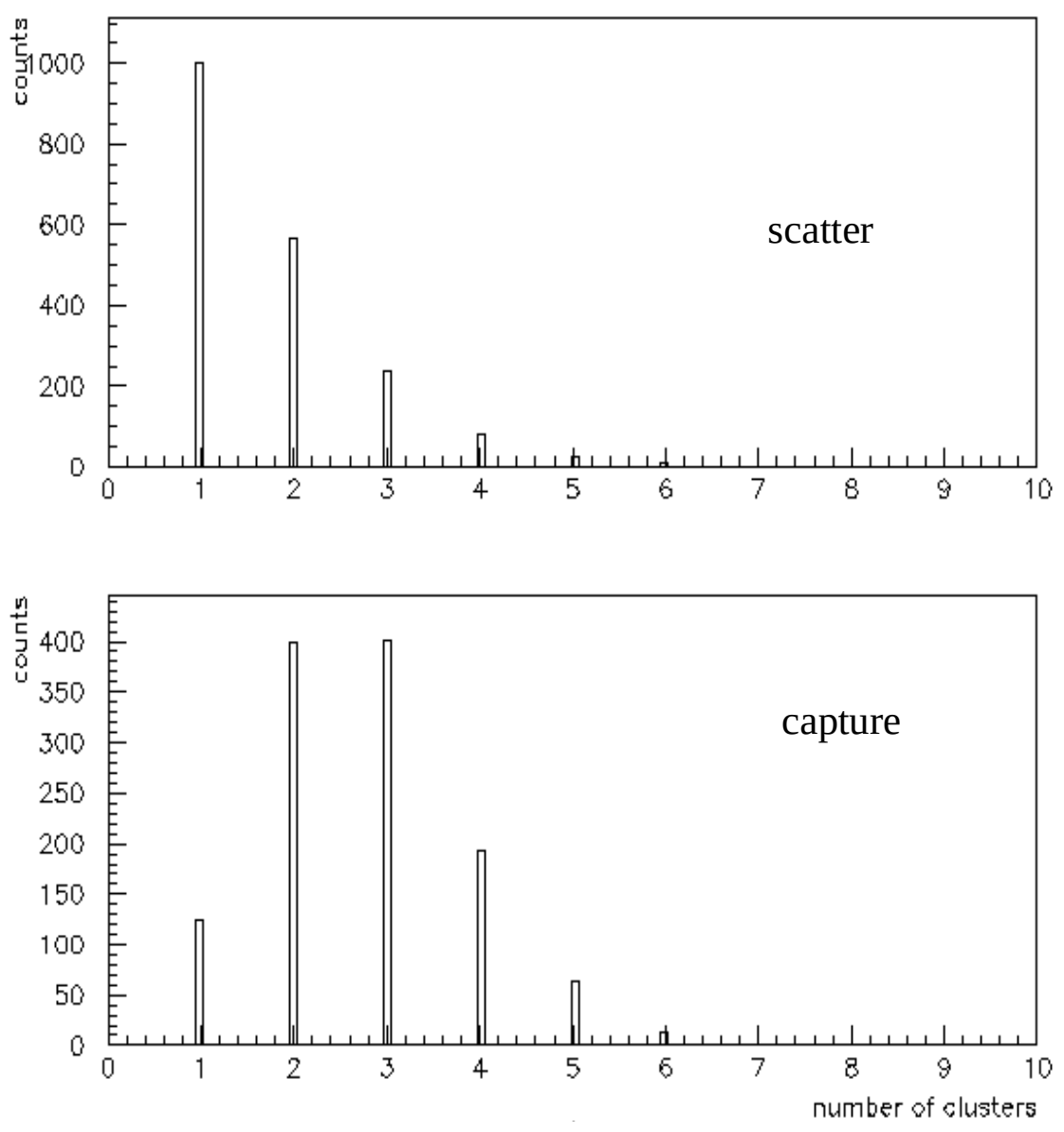

**Figure 38:** Number of clusters for capture-events (bottom) and for scattered events (top).

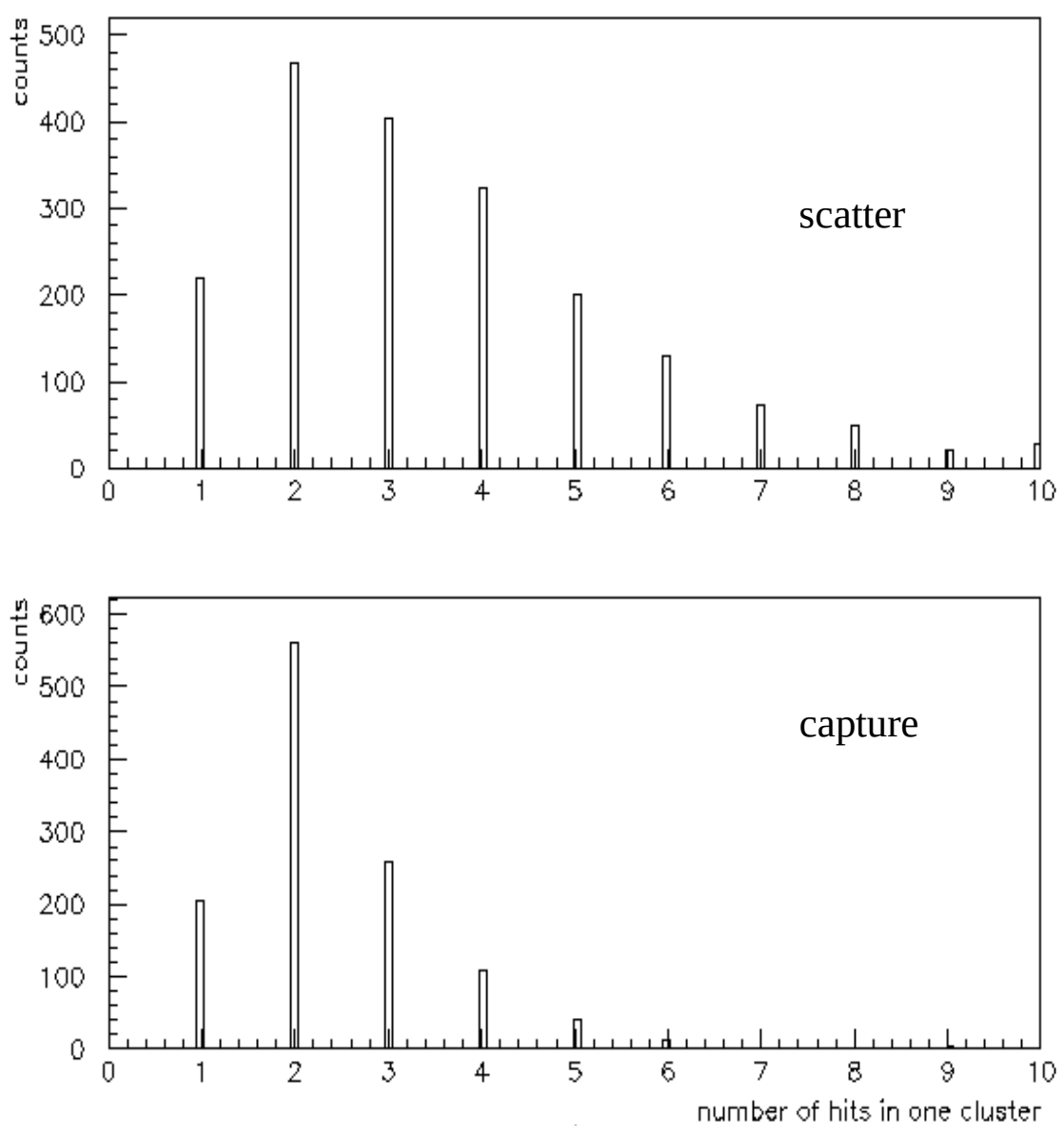

**Figure 39:** Number of hit crystals in the largest cluster of capture-events (bottom) and of events due to scattered neutrons (top).

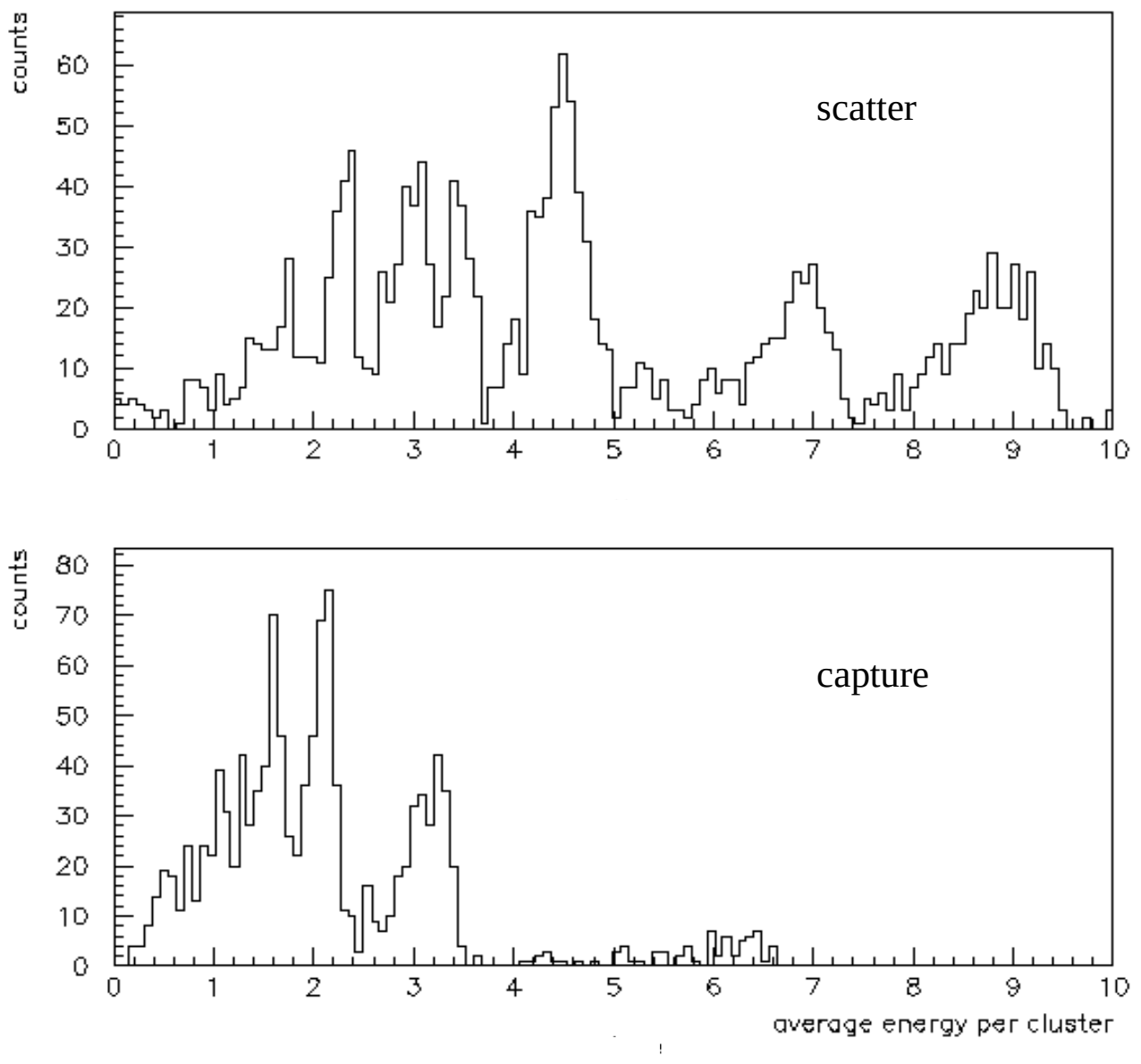

**Figure 40:** Sum energy divided by the number of clusters for scattered (top) and captured (bottom) neutrons.

Therefore, the best separation between captures and background events due to scattered neutrons could be achieved by combining two cuts: If one requires the formation of more than one cluster and in addition restricts the energy per cluster to be less than 3.8 MeV, the ratio between scattered and captured events could be reduced by a factor of about 3.

It should be noted that applying additional cuts in the sum energy could reduce this ratio even more, but this was avoided because it would be specific for the gold sample. Also, if only the fast component of the scintillator light is used, cuts in energy might not be possible due to the poorer energy resolution.

# 7. Radioactive background from the sample

Background caused by the decay of a radioactive sample may present a particular challenge. Usually, a decay deposits not more than about 1 MeV energy in the crystals and the capture event deposits between 5 and 7 MeV. Therefore the problem is not deciding between single background and capture events but rather is related to the high radioactive count rate of a 1 mg sample with a lifetime between a few days and several years, resulting in massive pile up effects and dead time losses.

Let $t_{1/2}$ be the half life, $m$ the sample mass , $M$ the molar mass of the investigated isotope and $N_A$

$$A(t) = -\frac{\mathrm{d}N}{\mathrm{d}t}(t) = \frac{\ln 2}{t_{1/2}} \cdot \frac{m(t=0)}{M} N_A \cdot \mathrm{e}^{-\frac{\ln(2)}{t_{1/2}} \cdot t}$$

Avogadros number, the activity is:

With $t_{1/2} >> t$ this simplifies to

$$A(t) = \frac{\ln 2}{t_{1/2}} \cdot \frac{m}{M} N_A \quad .$$

The number of decays in a given time follows the Poisson distribution. If

$$a = A(t) \cdot t_c$$

is the average number of events during the (coincidence) time $t_c$, the probability $P(n)$ for $n$ decays during this time is:

$$P(n) = \mathrm{e}^{-a} \frac{a^n}{n!} \quad .$$

The probability for at least one background event during the coincidence time of a capture event can be written as:

$$P_{coincident} = \sum_{n=1}^{\infty} P(n) = 1 - P(0) = 1 - \mathrm{e}^{-a} \quad .$$

A number of isotopes of astrophysical interest is considered in Table 8. For a sample mass of 1 mg the average number of counts in the detector system for an integration time of 150 ns (more than enough for the fast component) and 2 µs (slow component) is given in column 2 if no shielding against radiation from the sample is used . In this case the probability of an accidental coincidence with true capture events would be unity almost all isotopes.

A lead shielding of 5 mm thickness would be sufficient to solve this background problem (columns 4 and 5) except for a few cases. The corresponding probabilities were calculated for the standard setup of 160 crystals assuming a spherical lead shielding of 5 mm radius. Gamma rays from the sample in the center have to pass at least 5 mm lead. In a real experiment this shielding has to be modified for avoiding direct interaction with the neutron beam. Radioactive background is not significantly affected by the neutron moderator/absorber around the beam pipe. Only for a few isotopes a 5 mm thick lead shielding is not yet sufficient. Some of these could still not be measured even if the lead shield is replaced by a 20 mm thick gold sphere (column 4). Gold was chosen as an extreme case for its high atomic number and, compared with lead, its high density.

The respective end point energy of the $\beta^{-}/\beta^{+}$-decay as well as the energy of the highest (but still probable) γ-ray are given in the last column.

| Isotope | no shielding | | 5 mm Pb | | 20 mm Au | | Radiation |
|---|---|---|---|---|---|---|---|
| | 150 ns | 2 μs | 150 ns | $P_{coincident}$ | 150 ns | $P_{coincident}$ | |
| $^{151}$Sm | 141 | $2{\cdot}10^3$ | 0 | 0 | | | γ: 20 keV; $\beta^-$: 80 keV |
| $^{170}$Tm | $3.3{\cdot}10^4$ | $4.4{\cdot}10^5$ | 0.7 | 0.5 | 0 | 0 | γ: 70 keV; $\beta^-$: 1000 keV |
| $^{63}$Ni | 314 | $4.2{\cdot}10^3$ | 0 | 0 | | | $\beta^-$: 70 keV |
| $^{79}$Se | 0.38 | 5.2 | 0 | 0 | | | $\beta^-$: 150 keV |
| $^{85}$Kr | 8.7 | 115 | 4.6 | 0.99 | $8.7{\cdot}10^{-3}$ | $8.6{\cdot}10^{-3}$ | γ: 514 keV; $\beta^-$: 700 keV |
| $^{90}$Sr | 770 | $1{\cdot}10^4$ | 0 | 0 | | | $\beta^-$: 550 keV |
| * $^{90}$Y | 770 | $1{\cdot}10^4$ | 6.2 | 0.998 | 0.4 | 0.32 | γ: 1760 keV; $\beta^-$: 2.3 MeV |
| $^{94}$Nb | 1.1 | 14 | 1.0 | 0.62 | 0.16 | 0.15 | γ: (870+700) keV; $\beta^-$: 470 keV |
| $^{106}$Ru | $1.8{\cdot}10^4$ | $2.4{\cdot}10^5$ | 0 | 0 | | | $\beta^-$: 40 keV |
| * $^{106}$Rh | $1.8{\cdot}10^4$ | $2.4{\cdot}10^5$ | $3{\cdot}10^3$ | 1 | 178 | 1 | γ: 511 keV; $\beta^-$: 3.5 MeV |
| $^{135}$Cs | $7{\cdot}10^{-3}$ | $1{\cdot}10^{-1}$ | 0 | 0 | | | $\beta^-$: 200 keV |
| $^{137}$Cs | 480 | $6.4{\cdot}10^3$ | 0 | 0 | | | $\beta^-$: 500 keV |
| $^{147}$Pm | 13 | 180 | $1.3{\cdot}10^{-5}$ | 0 | | | γ: 120 keV; $\beta^-$: 220 keV |
| $^{155}$Eu | 1400 | $1.9{\cdot}10^4$ | $1.4{\cdot}10^{-3}$ | $1.4{\cdot}10^{-3}$ | | | γ: 100 keV; $\beta^-$: 250 keV |
| $^{153}$Gd | $2{\cdot}10^4$ | $2.6{\cdot}10^6$ | $1.5{\cdot}10^4$ | 1 | 314 | 1 | γ: 100 keV; $\beta^+$: 500 keV |
| $^{163}$Ho | 2.6 | 35 | 2.0 | 0.87 | $4.2{\cdot}10^{-2}$ | $4.2{\cdot}10^{-2}$ | $\beta^+$: 3 keV |
| $^{169}$Er | $4.5{\cdot}10^5$ | $6.0{\cdot}10^6$ | $6{\cdot}10^{-6}$ | 0 | | | γ: 120 keV; $\beta^-$: 350 keV |
| $^{171}$Tm | $6.9{\cdot}10^3$ | $80{\cdot}10^3$ | $6{\cdot}10^{-6}$ | 0 | | | γ: 70 keV; $\beta^-$: 100 keV |
| $^{175}$Yb | $9.4{\cdot}10^4$ | $1.2{\cdot}10^6$ | $2.4{\cdot}10^4$ | 1 | 52 | 1 | γ: 400 keV; $\beta^-$: 450 keV |
| $^{182}$Hf | $1.2{\cdot}10^{-3}$ | $1.6{\cdot}10^{-2}$ | 0 | 0 | | | γ: 270 keV; $\beta^-$: 160 keV |
| * $^{182}$Ta | $1.2{\cdot}10^{-3}$ | $1.6{\cdot}10^{-2}$ | $1.2{\cdot}10^{-3}$ | $1.2{\cdot}10^{-3}$ | | | γ: 1.3 MeV; $\beta^-$: 1.8 MeV |
| $^{179}$Ta | $6.1{\cdot}10^3$ | $81{\cdot}10^3$ | $4.6{\cdot}10^3$ | 1 | 97 | 1 | $\beta^+$: 110 keV |
| $^{185}$W | $52{\cdot}10^3$ | $7.0{\cdot}10^5$ | $1{\cdot}10^{-5}$ | 0 | | | γ: 125 keV; $\beta^-$: 430 keV |
| $^{191}$Os | $2.5{\cdot}10^5$ | $3.3{\cdot}10^6$ | $7.2{\cdot}10^{-2}$ | $6.9{\cdot}10^{-2}$ | | | γ: 130 keV; $\beta^-$: 140 keV |
| $^{193}$Pt | 0 | 0 | 0 | 0 | | | No |
| $^{204}$Tl | 66 | 880 | 0 | 0 | | | $\beta^-$: 350 keV |
| $^{210m}$Bi | $3.1{\cdot}10^{-3}$ | $4.2{\cdot}10^{-2}$ | 0 | 0 | | | γ: 300 keV; α: 5 MeV |
| $^{210}$Bi | $6.9{\cdot}10^5$ | $9.2{\cdot}10^6$ | 587 | 1 | 0 | 0 | γ: 300 keV; α: 5 MeV |

**Table 8:** Expected background from some radioactive isotopes of astrophysical interest. Isotopes marked by an asterisk are decay products of the isotope listed before with lifetimes shorter than the lifetime of the present nucleus. Due to decay equilibrium the decay rates of the daughter and present isotope is equal.

Shields of 5 mm lead and especially of 20 mm gold influence significantly the γ-ray resolution of the detector. This is simulated for the detector response to $^{197}Au(n,\gamma)$ cascades, representative for all other neutron capture events, was simulated also (see Figure 41).

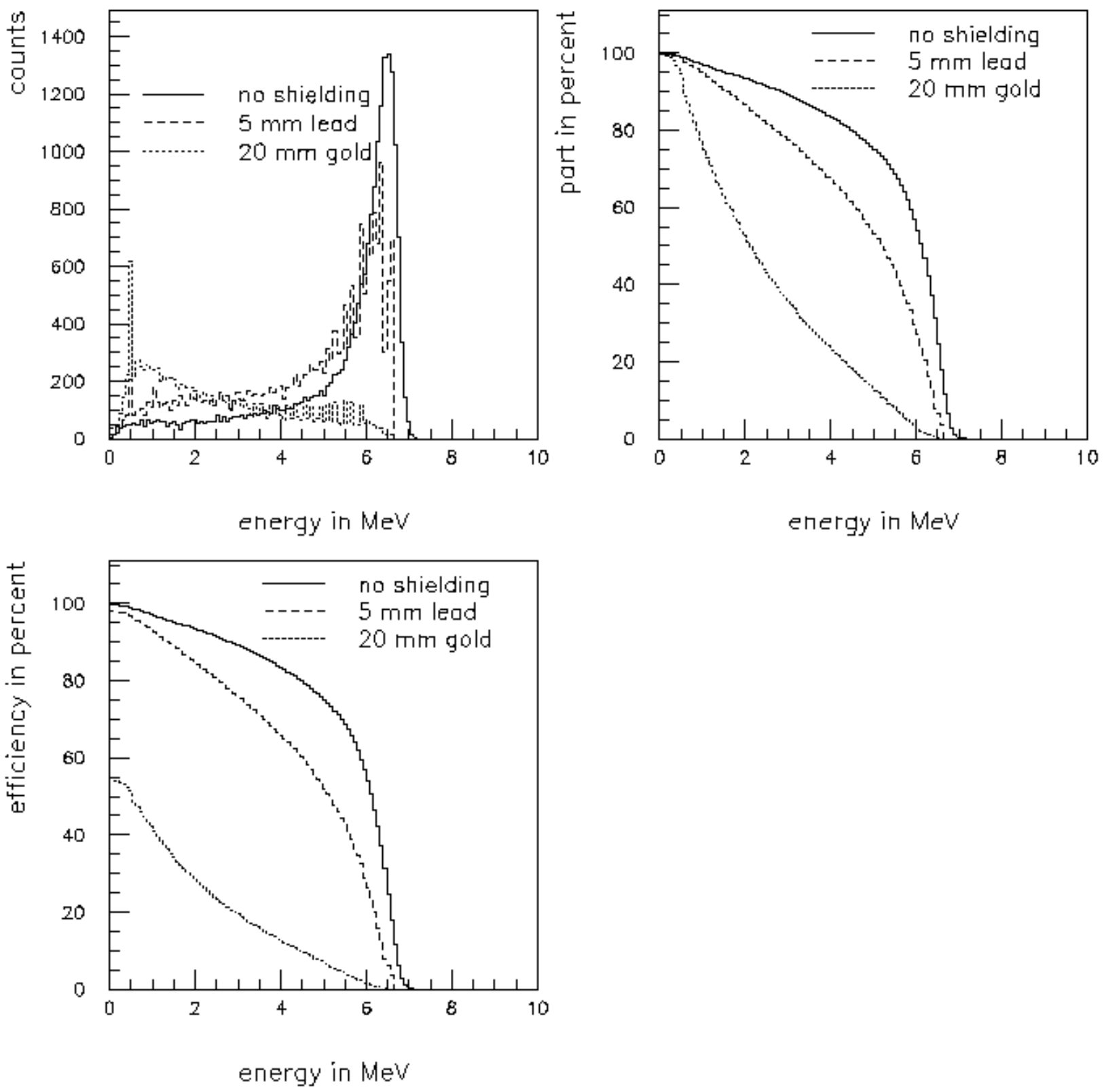

**Figure 41:** Influence of various γ-shieldings on the $^{197}Au(n,\gamma)$ spectrum.

The total γ-efficiency drops from 100 % (without shielding) to 98.5 % (0.5 mm lead) and 54.1 % (20 mm Au) respectively. If the 20 mm gold shielding has to be used the full energy peak would disappear. But even in this case there are still enough events at high sum energies to allow for discrimination between background and capture events.

# 8. Conclusions

- The collection of the scintillator light would probably be better using crystals that transition into circular cylinders to match the diameter of the photomultiplier tubes. But for a crystal ball with inner radius of 10 cm and outer radius of 25 cm cutting the crystal ends to a diameter of 5.08 would result in a loss of efficiency of more than 10 %. In our opinion it would be better to use PMs with a larger diameter.
- The beam pipe made of aluminum has an effect on the $\gamma$-rays as can be seen in Figures 15 and 16. This effect could be even worse for capture cascades with higher multiplicities and smaller $\gamma$-energies than gold especially if one uses a neutron moderator/converter also. Selection of a beam pipe that attenuates the $\gamma$-rays minimally is therefore essential.
- Using only the fast light component results in an energy resolution about a factor of 3 worse then the energy resolution assumed in the calculations. This would reduce the signal-to-background ratio. Also energy cuts may be not as efficient as with the higher resolution. In addition, signals from the decay of isomeric states (e.g. $^{198}$Au) would not be integrated which would result in a loss in peak efficiency.
- Using a moderator/absorber of $^{6}$LiH around the beam pipe and filling the gaps between the crystals with $^{10}$B was found to be very efficient in reducing the background from scattered neutrons. About a factor of 30 could be achieved for neutrons of energies from 10 keV to 100 keV.
- The supporting structure for the PMs made of aluminum could cause higher background at neutron energies from 10 keV to 1 MeV. This is due to a neutron scattering resonance for aluminum at a neutron energy of 35 keV (see section 9). It would be better to have less aluminum for the supporting structure. A honeycomb structure, which is known to be very stable, could reduce the amount of aluminum needed for the supporting structure.
- The results of section 7 have shown that it is possible to reduce the count rate from many of the radioactive samples to an acceptable amount by using a lead shield. For most isotopes one could even integrate over the long component of the scintillator light.

# 9. Cross sections for various isotopes

In this section the total, elastic and (n,γ) cross section for various materials used in the setups are shown. The cross sections were evaluated with the computer program JEF-PC [Nor92]. Materials include structural materials (Al), scintillator materials (Ba-isotopes, Ce-isotopes, Bi), samples (Au, Bi), backings (Be, Ti, C).

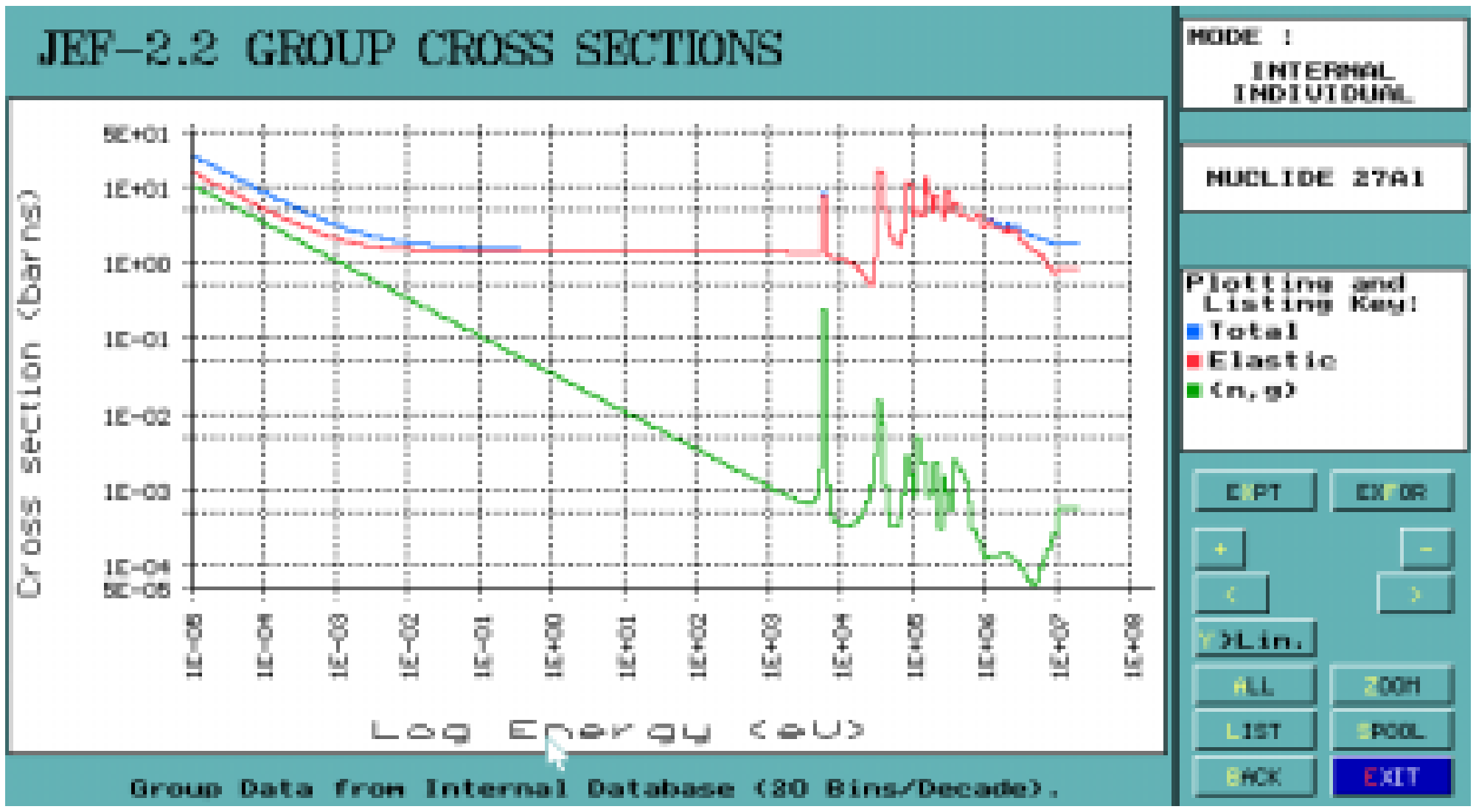

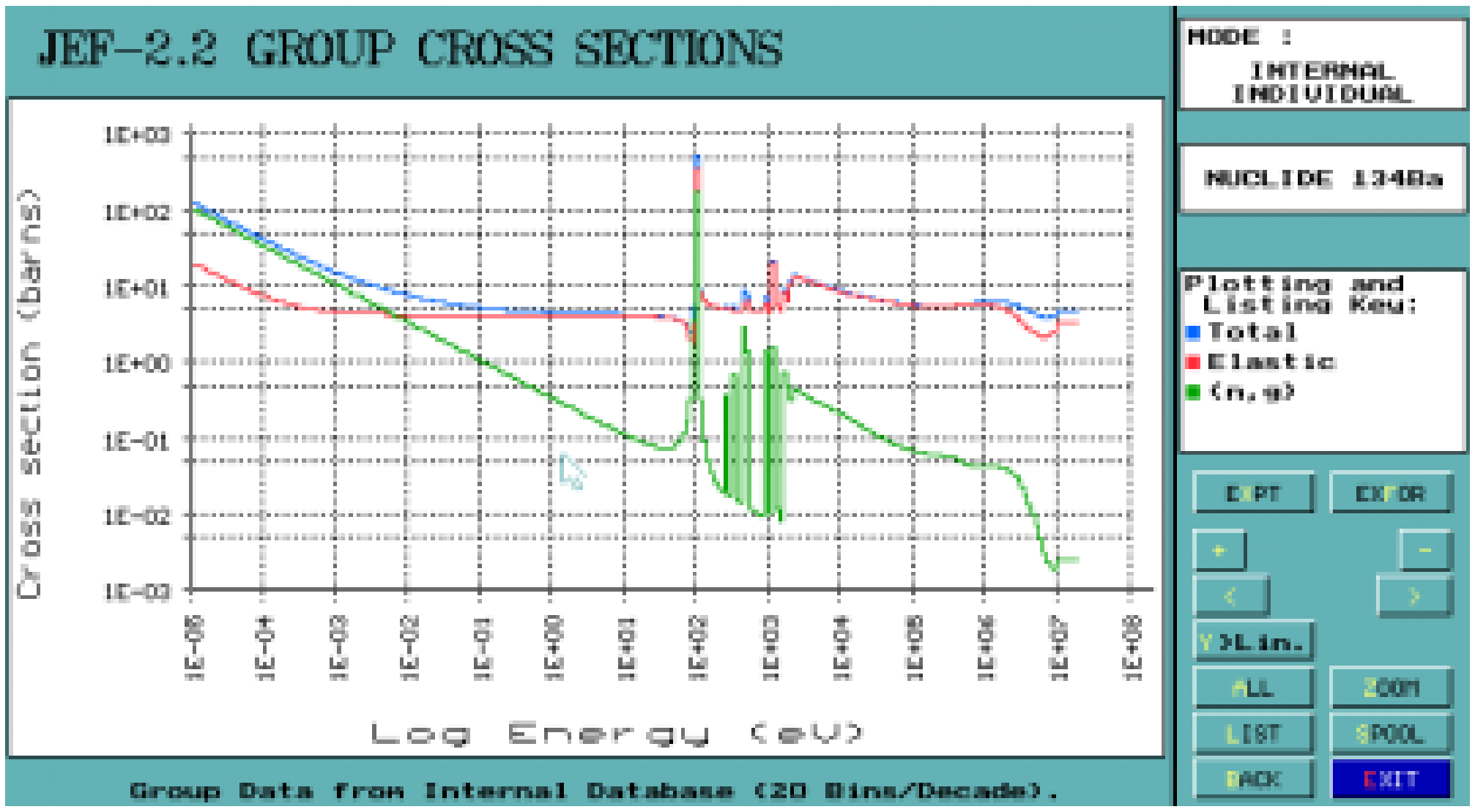

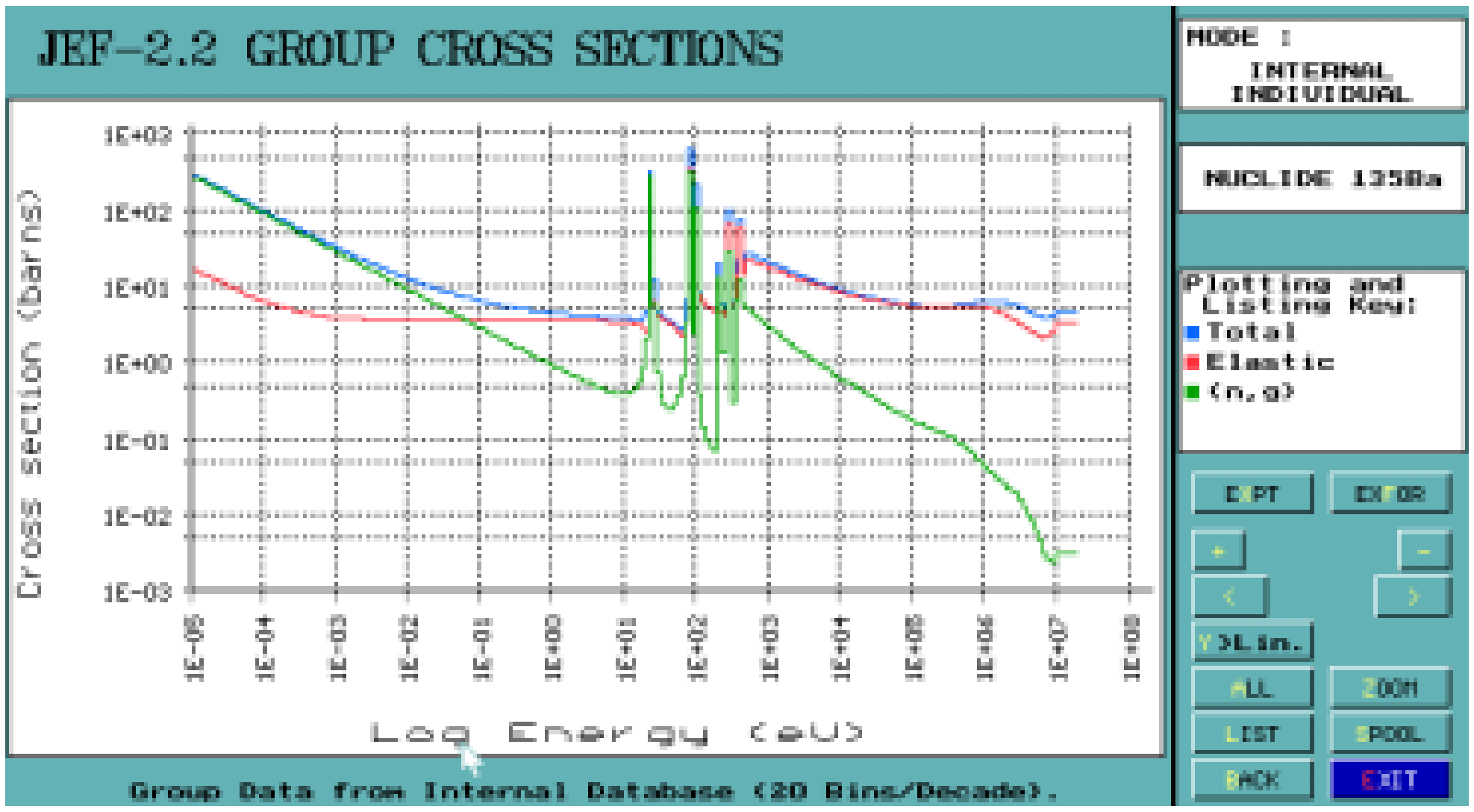

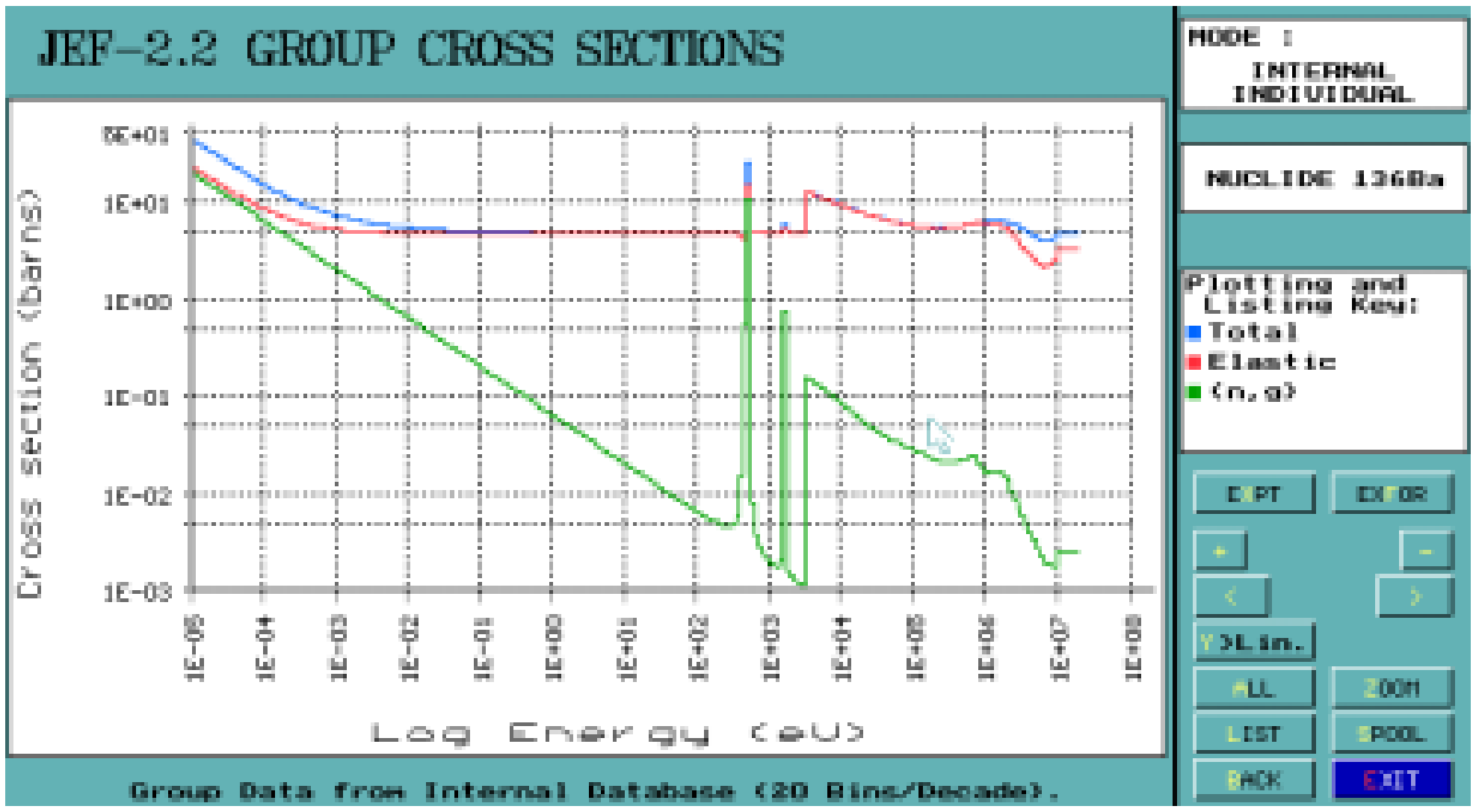

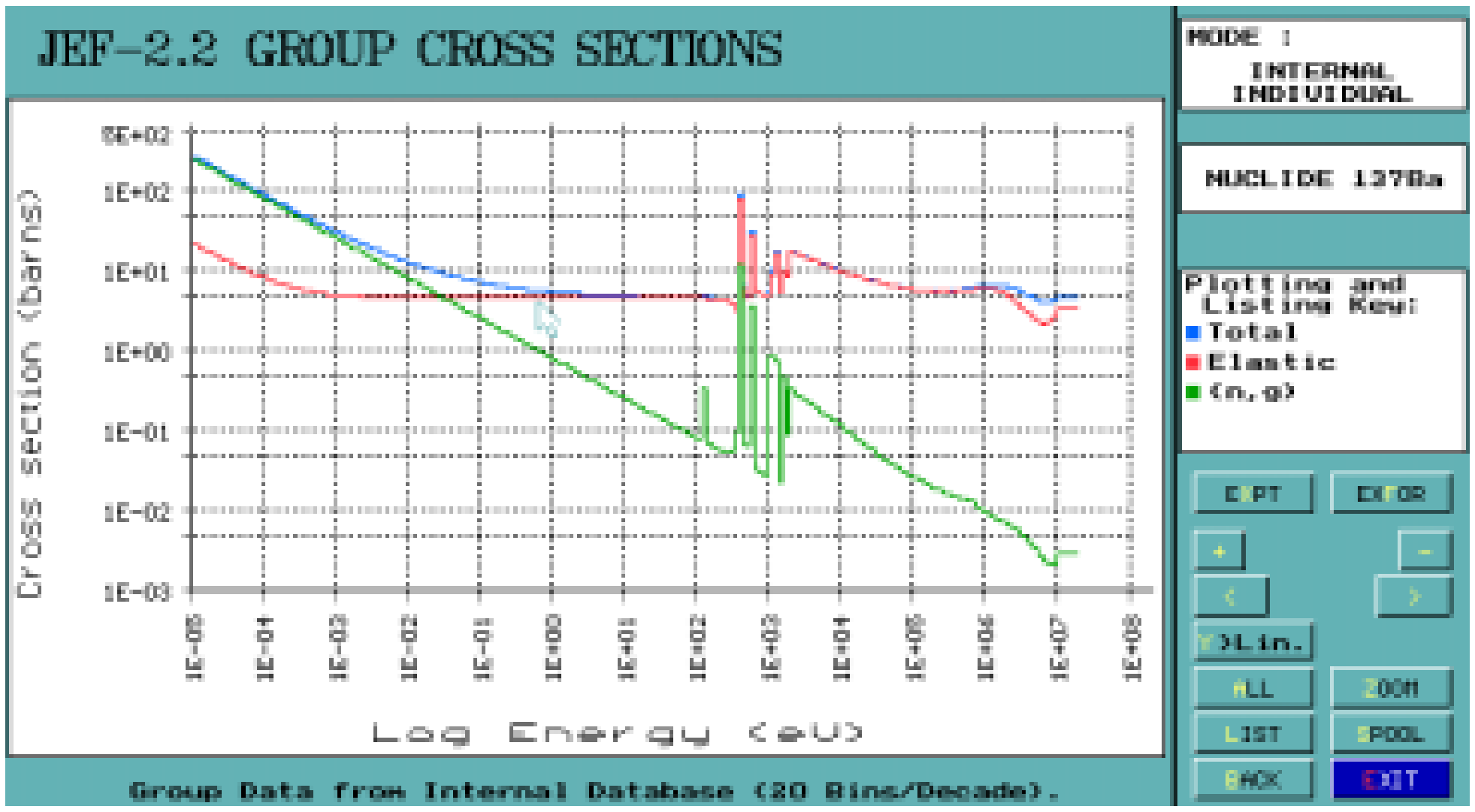

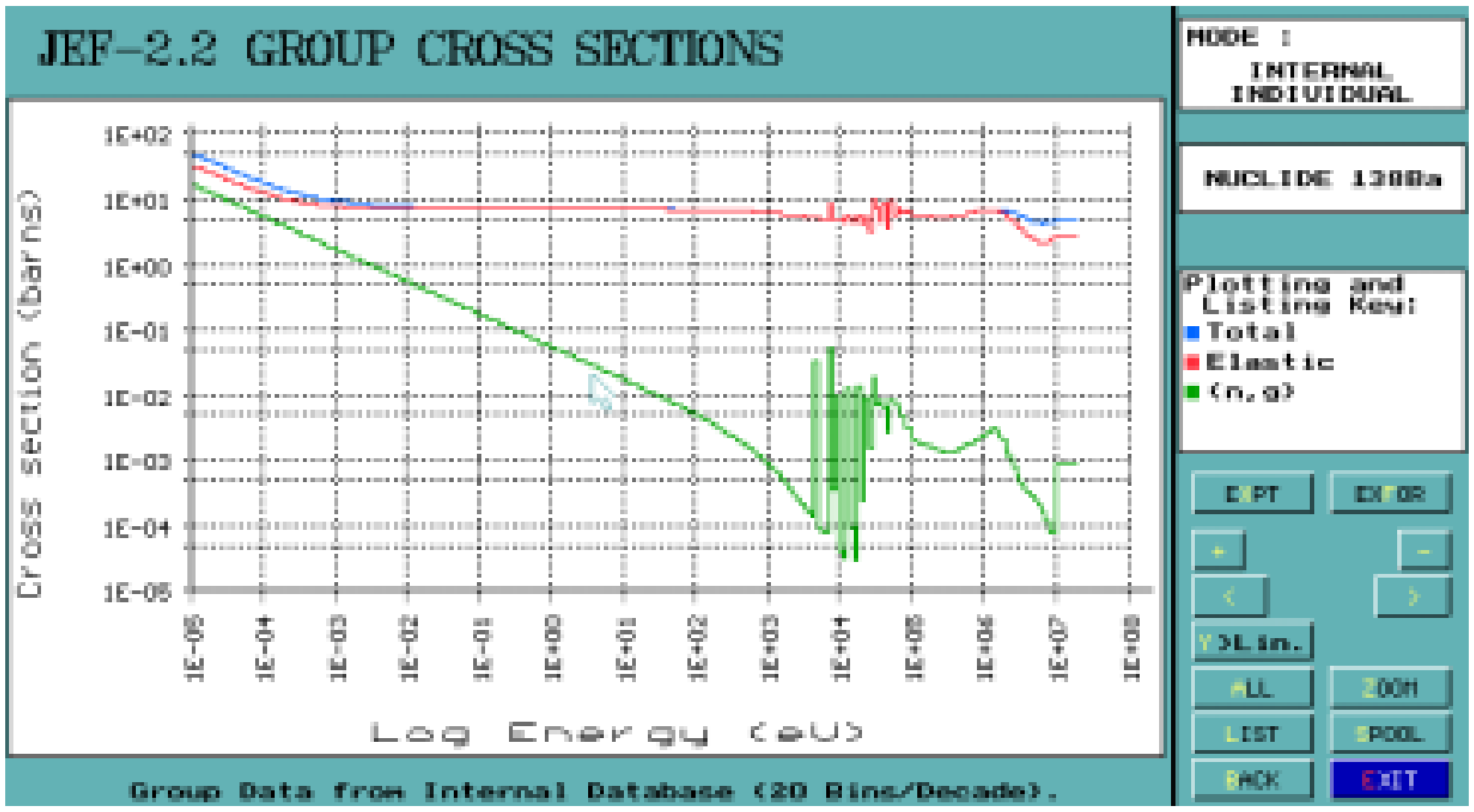
JEF-2.2 GROUP CROSS SECTIONS
MODE : INTERNAL INDIVIDUAL
NUCLIDE 138Ba
Plotting and Listing Key:
Total
Elastic
(n,g)
Cross section (barns)
Log Energy (eV)
Group Data from Internal Database (20 Bins/Decade).

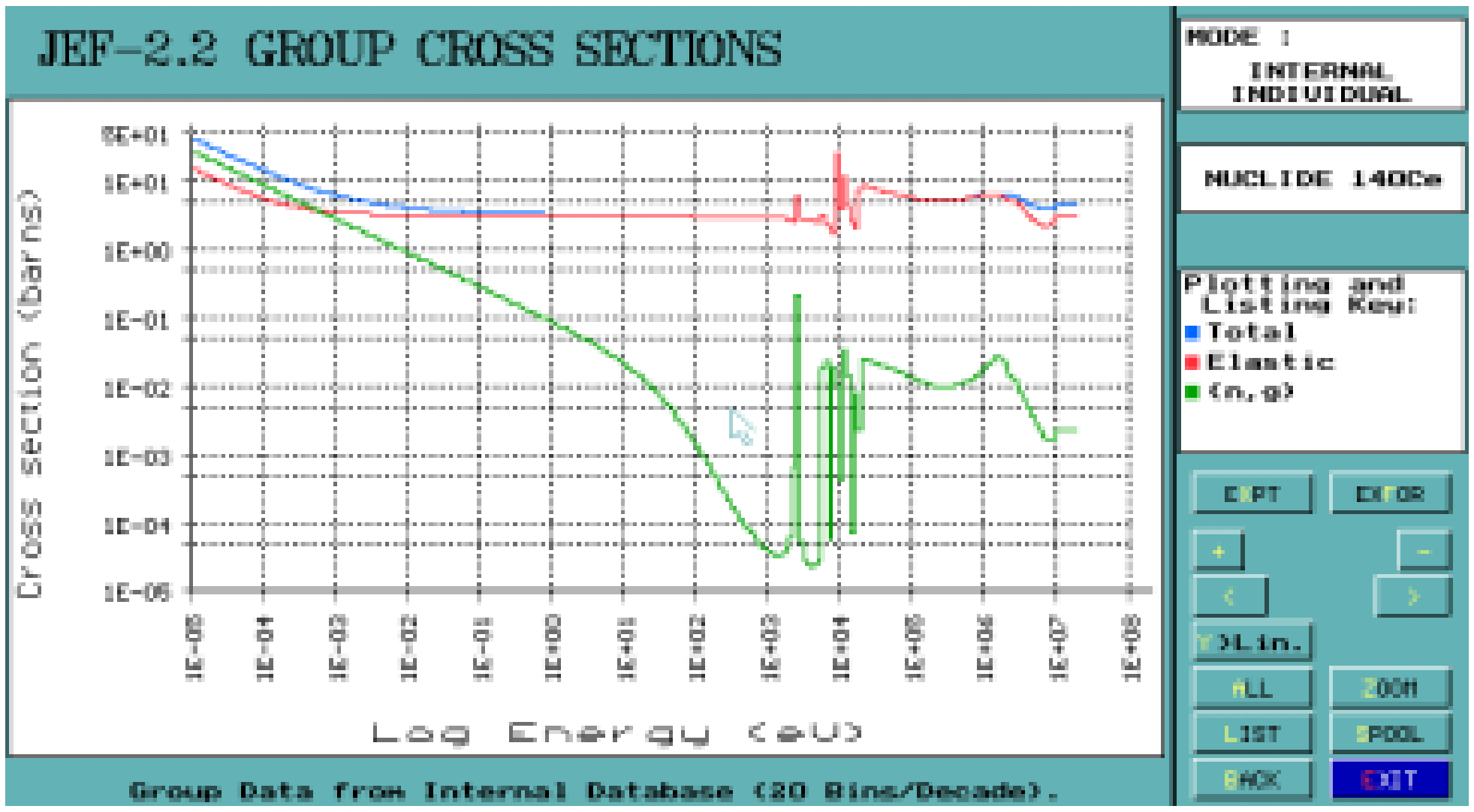
JEF-2.2 GROUP CROSS SECTIONS
MODE : INTERNAL INDIVIDUAL
NUCLIDE 140Ce
Plotting and Listing Key:
Total
Elastic
(n,g)
Cross section (barns)
Log Energy (eV)
Group Data from Internal Database (20 Bins/Decade).

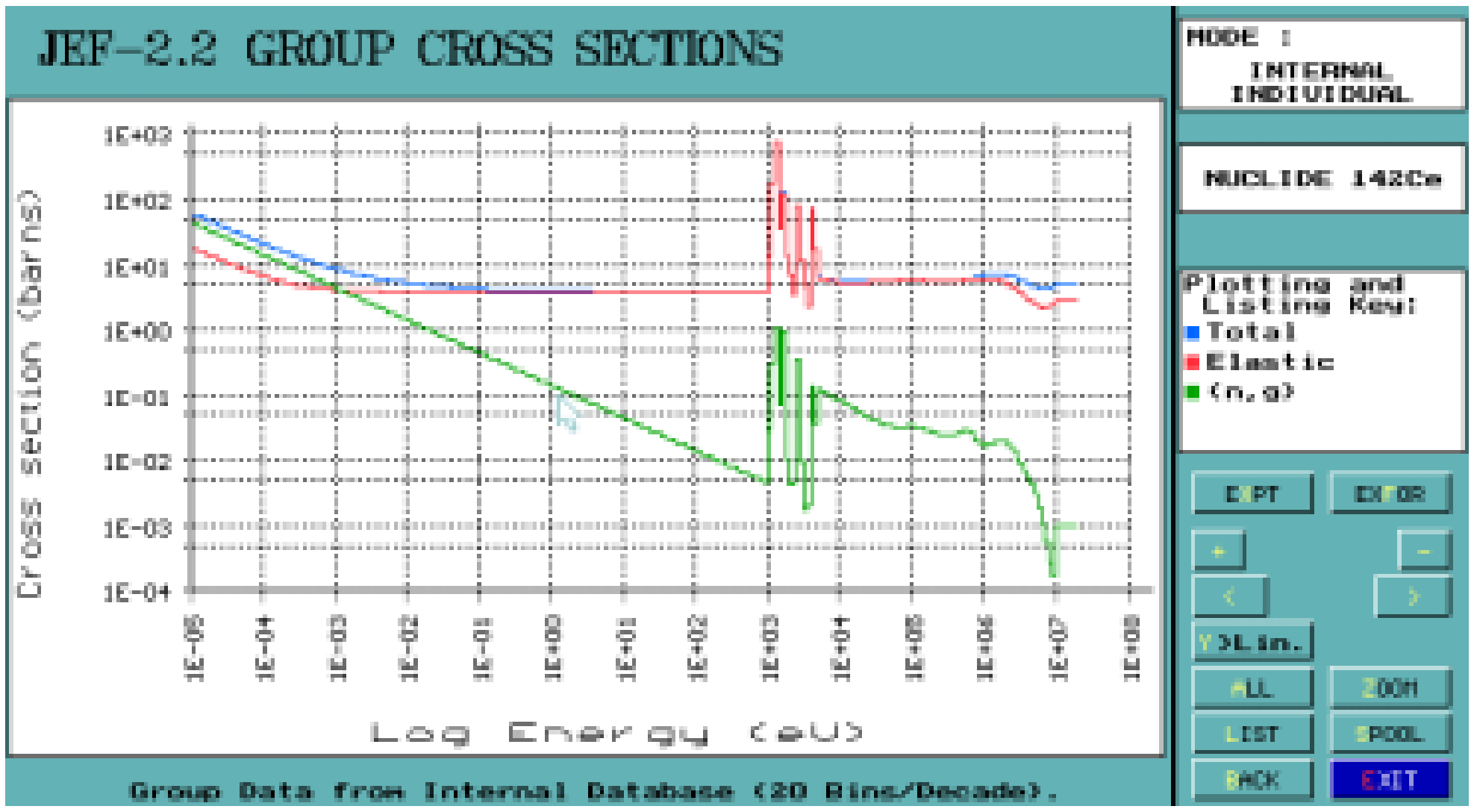
JEF-2.2 GROUP CROSS SECTIONS
MODE : INTERNAL INDIVIDUAL
NUCLIDE 142Ce
Plotting and Listing Key:
Total
Elastic
(n,g)
Cross section (barns)
Log Energy (eV)
Group Data from Internal Database (20 Bins/Decade).

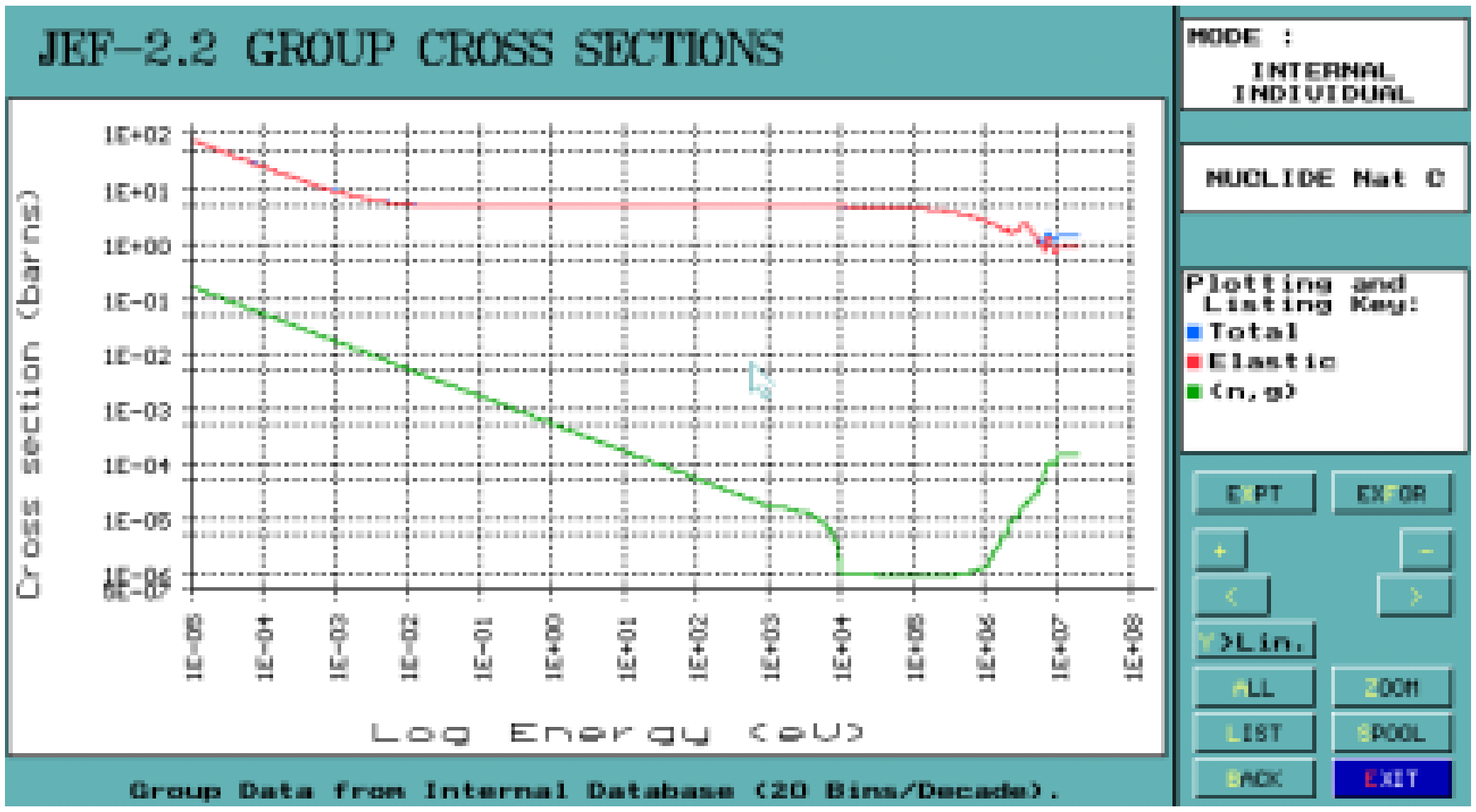
JEF-2.2 GROUP CROSS SECTIONS
MODE : INTERNAL INDIVIDUAL
NUCLIDE Nat C
Plotting and Listing Key:
Total
Elastic
(n,g)
1E+02
1E+01
1E+00
1E-01
1E-02
1E-03
1E-04
1E-05
Cross section (barns)
1E-05 1E-04 1E-03 1E-02 1E-01 1E+00 1E+01 1E+02 1E+03 1E+04 1E+05 1E+06 1E+07 1E+08
Log Energy (eV)
Y>Lin.
ALL
ZOOM
LIST
SPOOL
BACK
EXIT
Group Data from Internal Database (20 Bins/Decade).

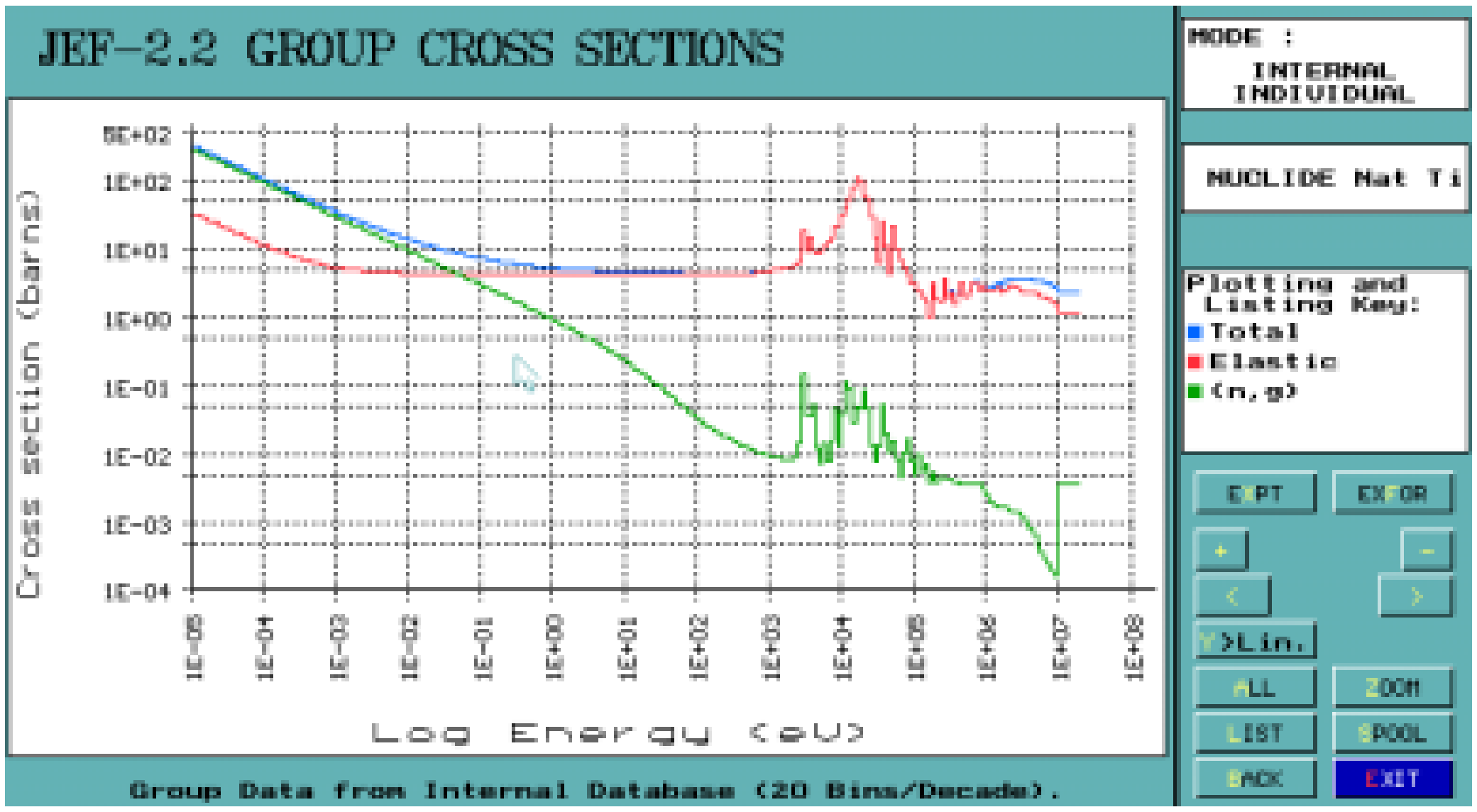
JEF-2.2 GROUP CROSS SECTIONS
MODE : INTERNAL INDIVIDUAL
NUCLIDE Nat Ti
Plotting and Listing Key:
Total
Elastic
(n,g)
5E+02
1E+02
1E+01
1E+00
1E-01
1E-02
1E-03
1E-04
Cross section (barns)
1E-05 1E-04 1E-03 1E-02 1E-01 1E+00 1E+01 1E+02 1E+03 1E+04 1E+05 1E+06 1E+07 1E+08
Log Energy (eV)
Y>Lin.
ALL
ZOOM
LIST
SPOOL
BACK
EXIT
Group Data from Internal Database (20 Bins/Decade).

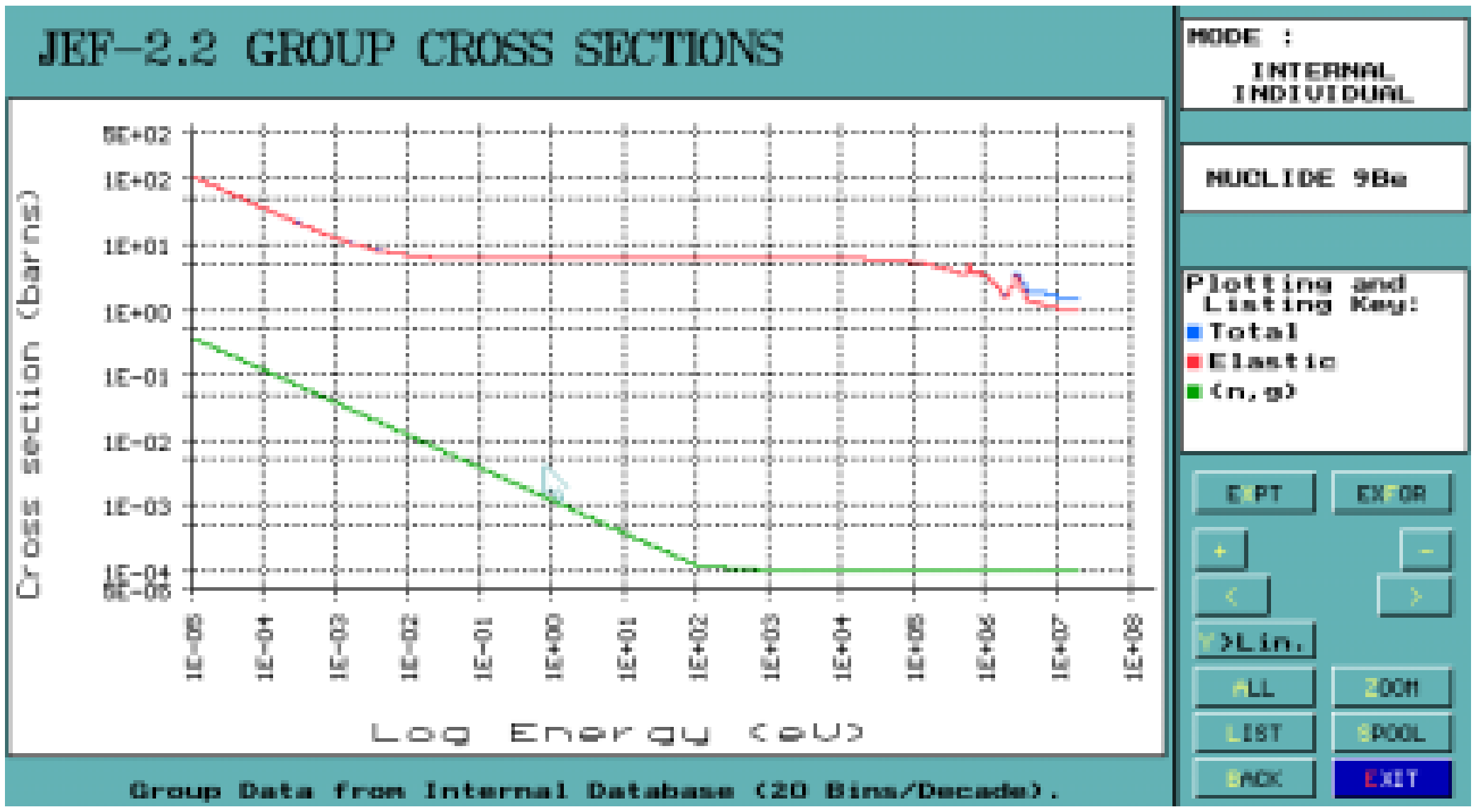
JEF-2.2 GROUP CROSS SECTIONS
MODE : INTERNAL INDIVIDUAL
NUCLIDE 9Be
Plotting and Listing Key:
Total
Elastic
(n,g)
5E+02
1E+02
1E+01
1E+00
1E-01
1E-02
1E-03
1E-04
Cross section (barns)
1E-05 1E-04 1E-03 1E-02 1E-01 1E+00 1E+01 1E+02 1E+03 1E+04 1E+05 1E+06 1E+07 1E+08
Log Energy (eV)
Y>Lin.
ALL
ZOOM
LIST
SPOOL
BACK
EXIT
Group Data from Internal Database (20 Bins/Decade).

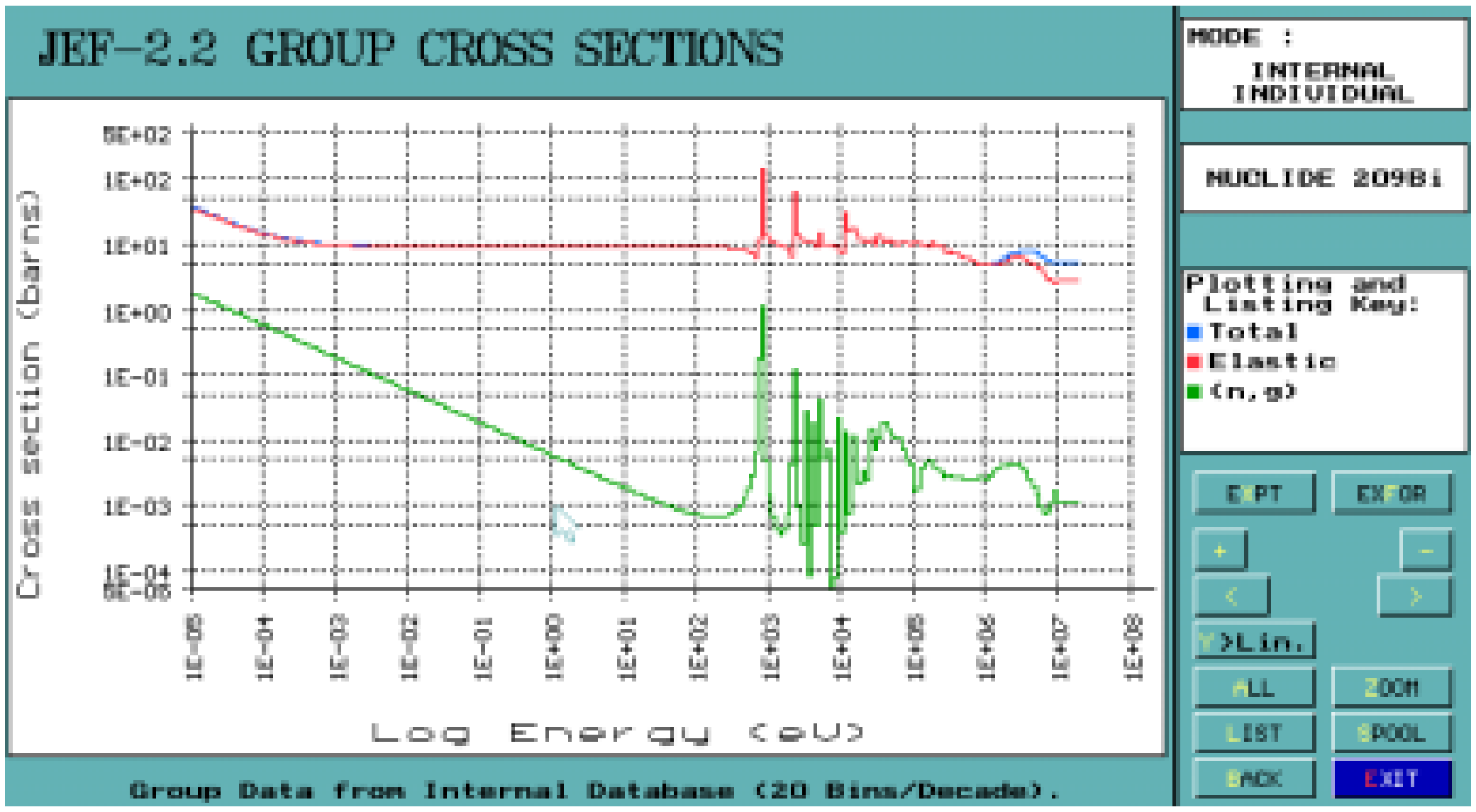

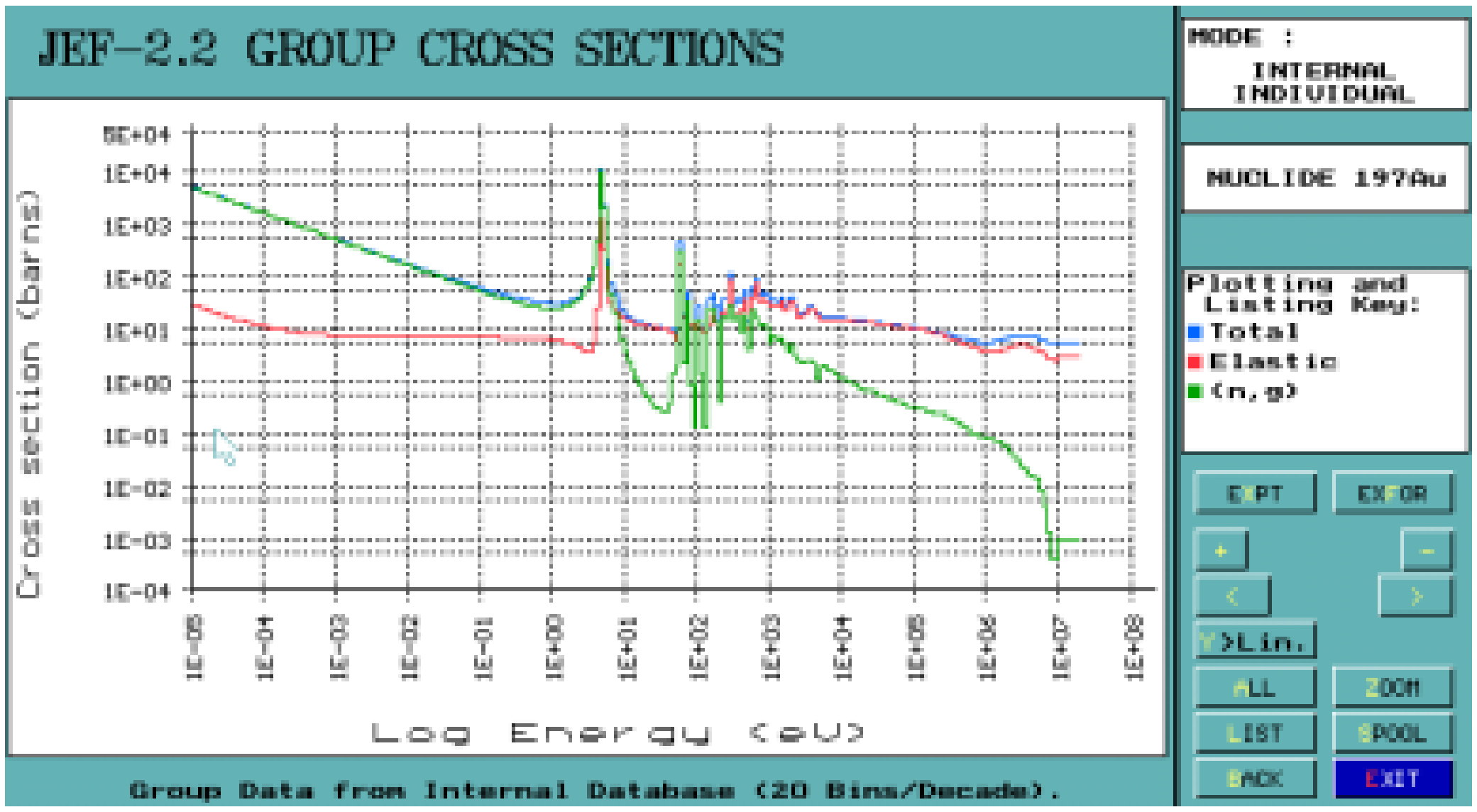